\documentclass[aps, prl, amsmath, amssymb, preprintnumbers, twocolumn, ,noshowpacs, superscriptaddress]{revtex4-1}
\usepackage{amsmath}
\usepackage{amssymb}
\usepackage{graphics}
\usepackage{graphicx}
\usepackage{bm}
\usepackage{color}

\usepackage[breaklinks]{hyperref}
\newcommand{\be}{\begin{equation}}
\newcommand{\ee}{\end{equation}}
\newcommand{\bas}{\begin{eqs}}
\newcommand{\eas}{\end{eqs}}

\newcommand{\bq}{\mathbf{q}}

\newcommand{\bk}{\mathbf{k}}

\newcommand{\bw}{\begin{widetext}}
\newcommand{\ew}{\end{widetext}}

\begin{document}

\title{Superradiant phase transition in electronic systems and emergent topological phases}
\author{Daniele Guerci}
\affiliation{Universit{\'e} de Paris, Laboratoire Mat{\'e}riaux et Ph{\'e}nom{\`e}nes Quantiques, CNRS, F-75013 Paris, France.} 
\affiliation{Universit{\'e} Paris-Saclay, CNRS, Laboratoire de Physique des Solides, 91405, Orsay, France.} 
\author{Pascal Simon}
\affiliation{Universit{\'e} Paris-Saclay, CNRS, Laboratoire de Physique des Solides, 91405, Orsay, France.} 
\author{Christophe Mora}
\affiliation{Universit{\'e} de Paris, Laboratoire Mat{\'e}riaux et Ph{\'e}nom{\`e}nes Quantiques, CNRS, F-75013 Paris, France.}

\begin{abstract}
  We derive a general criterion for determining the onset of superradiant phase transition in electronic bands coupled to a cavity field, with possibly electron-electron interactions.
  For longitudinal superradiance in 2D or genuine 1D systems, we  prove that it is always prevented, thereby extending existing no-go theorems. Instead, a superradiant phase transition can occur to a nonuniform transverse cavity field and we give specific examples in non-interacting models, either through Fermi surface nesting or parabolic band touching. Investigating the resulting time-reversal symmetry breaking superradiant states, we find in the former case Fermi surface lifting down to four Dirac points on a square lattice model, with topologically protected zero-modes, and in the latter case topological bands with non-zero Chern number on an hexagonal lattice.
\end{abstract}

\date{\today}

\maketitle

The study of the  quantum-mechanical interaction between light and matter has been a driving field in physics in the past century with its application in different research fields, such as laser cooling \cite{Aspect_1988,Philipps_1998}, quantum information and quantum computing \cite{Monroe_1995,Cirac_1997,Duan_2004}.
Experimental advances in cavity quantum electrodynamics (CQED) \cite{Haroche_2001,Wineland_2003} have made
it possible to  integrate solid state materials with optical
 cavities \cite{Faist_2012,Maissen_2014,Smolka_2014,Liu_2015,Basov_2016}, thus paving the way for CQED  at the micrometer and even nanometer scale.
These recent tremendous progresses open the door to harness electronic properties of solid state materials \cite{Forst_2011,Subedi_2014,Jaksch_2019} and eventually to emulate new exotic collective phases \cite{Mazza_2019,Claassen_2019}.

In this context, the phenomenon of superradiance  plays a pivotal role \cite{Haroche_1982}. It was  originally predicted in the Dicke model \cite{Dicke,Lieb_1973,Wang_1973}, where a single cavity mode is coupled to an ensemble of  two-levels systems (dipoles). The collective and coherent interaction can lead to a so-called superradiant phase in which the dipoles emit light at high intensity, {\it i.e.} macroscopically populate the cavity. This phase transition has been observed first in optically pumped gas \cite{Feld_1973}, in 
photo-excited semiconducting quantum dots \cite{Scheibner_2007,Timothy_2012} and in pumped ultra-cold gases trapped in an ultrahigh-finesse optical cavity \cite{Baumann_2010}. These experiments involve however an external drive and no equilibrium version of superradiance has yet been experimentally demonstrated.

Indeed, in realistic systems, the linear light-matter coupling of the Dicke model is supplemented by a diamagnetic term, quadratic in the potential vector and  detrimental to a superradiant phase transition. The relative balance between the two competing terms is generally fixed by the Thomas-Reiche-Kuhn (TRK) sum rule and prevents a superradiant state to occur through no-go theorems~\cite{Rzazewski_1975,Rzazewski_1979,Rzazewski_1981} in most systems. Suggestions to bypass no-go theorems have been made, involving for instance magnetic dipolar interactions~\cite{EMELJANOV1976366,Keeling_2007,Rabl_2018} or electron-electron interactions~\cite{Pellegrino_2016}, but the proper account or not of the TRK sum rule have led to mistakes and controversies in past studies~\cite{Nataf_2010,Marquardt_2011,Ciuti_2012,Hagenmuller_2012,Brandes_2012,Chirolli_2012,Bamba_2014,Mazza_2019}. In the case of electronic systems, a no-go theorem for photon condensation (or equilibrium superradiance) has been recently proven~\cite{Andolina_2019}, seemingly closing the door to equilibrium exotic polaritonic phases. It holds even in the presence of (strong) electron-electron interactions but requires a uniform cavity field. Incidentally, a crossing of Landau levels in a 2D electron gas has been predicted to induce a superradiant instability in a spatially varying cavity field~\cite{Basko_PRL2019}.

The purpose of this letter is to provide a general framework for predicting superradiant phase transitions in electronic systems, thereby connecting the above studies. Building on a lattice model, which automatically exhibits gauge invariance and the associated TRK sum rule, we derive a general criterion for the occurrence of superradiance. The no-go theorem of Ref.~\cite{Andolina_2019} is circumvented by taking into account the finite momentum exchanged between the photon mode and the electron gas, extending the findings of Ref.~\cite{Basko_PRL2019}. We find that superradiance can occur only for a transverse cavity field. For a longitudinal field~\footnote{even in the Coulomb gauge considered in this work, the projection of the 3D cavity field to a 2D material may have a longitudinal component.}, we derive an extended TRK sum rule at finite momentum which definitely prevents photon condensation and superradiance in one-dimensional settings. We explore several explicit non-interacting models in which a superradiant phase transition takes place, either through nesting or quadratic band touching. We detail the resulting superradiant phases, where the magnetic flux gives rise to a spatially-modulated orbital current order. Interestingly, the symmetry broken phases bear non-trivial topological properties.

\textit{The model ---} Without loss of generality, we consider a lattice, or tight-binding, model to describe the crystalline band structure of a solid-state material. The Hamiltonian $H_{el} = H_0+H_{\text{int}}$ includes a kinetic term
\begin{equation}
\label{Matter_Hamiltonian}
 H_0=-\sum_{\bm{j}, \bm{\delta}} \sum_{\alpha\beta}\,t^{\bm{\delta}}_{\alpha\beta} \, c^\dagger_{\bm{R}_{\bm{j},\alpha}}\, c_{\bm{R}_{\bm{j},\beta}+\bm{\delta}}
\end{equation}
with the cell index $\bm{j}$,  the  orbital indices $\alpha/\beta$ and $\bm{R}_{\bm{j},\alpha}$ the corresponding site positions. The resulting Bloch Hamiltonian $h_{\alpha\beta} (\bk)  =-\sum_{\bm{\delta}} t^{\bm{\delta}}_{\alpha\beta}\,e^{i\,\bk\cdot(\bm{R}_{\bm{j},\beta}+\bm{\delta}-\bm{R}_{\bm{j},\alpha})}$  can be written in terms of the hopping amplitudes $t^{\bm{\delta}}_{\alpha\beta}$.  $H_{\text{int}}$ is assumed to contain only density-density interactions. 
The electronic system is either embedded into a three-dimensional cavity or coupled to free space photons described by the quantum potential vector $\hat{\bm{A}} (\bm{r})$. The light-matter coupling is performed in the Coulomb gauge through Peierls substitution in Eq.~\eqref{Matter_Hamiltonian}, $t^{\bm{\delta}}_{\alpha\beta} \to t^{\bm{\delta}}_{\alpha\beta} e^{-i e \lambda /c}$ with
\begin{equation}\label{peierls}
  \lambda = \int_{\bm{R}_{\bm{j},\alpha}}^{\bm{R}_{\bm{j},\beta}+\bm{\delta}} d \bm{r} \cdot \hat{\bm{A}} (\bm{r}),
\end{equation}
changing the hopping terms but leaving the interaction part $H_{\text{int}}$ invariant. $-e$ is the electron charge and $c$ the speed of light.

Inherited from the original minimal coupling, the Peierls substitution entails an associated gauge invariance. It is best described by replacing $\hat{\bm{A}} (\bm{r})$ with a classical uniform and time-independent vector potential $\bm{A}_0$. Eq.~\eqref{peierls} becomes $\lambda = \bm{A}_0 \cdot (\bm{R}_{\bm{j},\beta}+\bm{\delta} - \bm{R}_{\bm{j},\alpha})$. The resulting phase factor is readily absorbed by the gauge transform $c_{\bm{R}_{\bm{j},\alpha}} \to e^{i e\bm{A}_0 \cdot \bm{R}_{\bm{j},\alpha}/c} c_{\bm{R}_{\bm{j},\alpha}}$ and Eq.~\eqref{Matter_Hamiltonian} is  recovered. This is expected on physical ground as a constant vector potential is associated with vanishing electric and magnetic fields. As discussed below, this gauge invariance ensures the TRK sum rule. The  Bloch Hamiltonian is modified as $h_{\alpha\beta} (\bk - e \bm{A}_0 /c )$, {\it i.e.} a simple momentum shift removes $\bm{A}_0$. The momentum shift is harmless as the Brillouin zone is a compact space, in contrast with continuous or ${\bf k} \cdot {\bf p}$ approximations which often violate sum rules and incorrectly predict superradiance.
The generic lattice model~\eqref{Matter_Hamiltonian} offers a powerful antidote to enforce gauge invariance and protect sum rules.

We are interested at the onset of superradiance and therefore expand the phase factors~\eqref{peierls} to second order to obtain the Hamiltonian 
$H_0(\bm{A})=H_0+H_A+H_{A^2}$ with
\begin{subequations}
\begin{align}
\label{HA_linear}
H_A & = \frac{e}{c}\sum_{\bq}\,\hat{\bm{A}}(\bq)\cdot\bm{J}_p(-\bq), \\[2mm]
H_{A^2} & = -\frac{e^2}{2c^2}\sum_{\bq_1,\bq_2}\,\hat{A}_i (\bq_1) \mathcal{T}^{i,j} (-\bq_1,-\bq_2)  \hat{A}_j (\bq_2).
\end{align}
\end{subequations}
We thereby introduce the paramagnetic current $\bm{J}_p(\bq)$ and the diamagnetic tensor $\mathcal{T}^{i,j}(\bq_1,\bq_2)$. As detailed in the Supplemental Material (SM), they can be written solely in terms of the Bloch Hamiltonian $h_{\alpha\beta} (\bk)$.
Together, they define the current operator
\begin{equation}
J_i (\bq)=  J_{p,i}(\bq)-\frac{e}{c}\sum_{\bq^\prime,j} \mathcal{T}^{i,j} \left(\bq,-\bq^\prime\right) \hat{A}_j \left(\bq^\prime\right) 
\ee
where $i=x,y,z$ ($x,y$) in three (two) dimensions. For a classical potential vector $\bm{A} (\bm{r})$, the average current $\bm{j} = \langle \bm{J} \rangle/V$ follows from linear response theory
\begin{equation}\label{current-response}
j_i (\omega,\bq)=\frac{e}{c} \sum_{j} Q^{i,j} (\omega,\bq)\, A_j(\omega,\bq),
\ee
with the current susceptibility $Q^{i,j}(\omega,\bq)$. The above-mentioned gauge invariance implies that the current response Eq.~\eqref{current-response} to the uniform field $\bm{A}_0$ must vanish in the static limit, and therefore 
\begin{equation}\label{TRKsum}
\lim_{\bm{q} \to 0}  Q^{i,j} (0,\bm{q}) =  0.
\end{equation}
This is the TRK sum rule expressing the cancellation of paramagnetic and diamagnetic responses at long wavelength.


\textit{Condition for superradiance ---} We turn to the electromagnetic cavity in which, for the sake of simplicity, we keep only two modes with wavevectors $\bm{q}$ and  $-\bm{q}$. The potential vector takes the form
\begin{equation}\label{vectorpotential}
  \hat{\bm{A}} (\bm{r})  = \bar{A} \,  \bm{u} \,  e^{i \bm{q} \cdot \bm{r}} (a_{\bm q} + a_{-\bm q}^\dagger) + {\rm h.c.}~,
\end{equation}
where the direction is determined by the unit vector $\bm{u}$ and $\bar{A}$ sets the strength of light-matter interaction.
The light-matter Hamiltonian is then $H_{el} +H_A+H_{A^2} + H_{cav}$ with the cavity energy $H_{cav} = \hbar \omega_q (a_{\bm q}^\dagger a_{\bm q} + a_{-\bm q}^\dagger a_{-\bm q})$. In the thermodynamic limit, the light-matter ground state factorizes and one can show that the photon state is a coherent state. This justifies the replacement of bosonic operators $a_{\pm \bm q} \to \alpha_{\pm \bm q}$ by classical fields which must be chosen to minimize the ground state energy. We use linear response theory and the stiffness theorem to arrive at the ground state energy to leading order in $\alpha_{\pm \bm q}$  
\begin{equation}\label{expansion}
  E (\alpha_{\bm q}) - E(0) = {\cal N}_q \left[ \hbar \omega_q + 2\gamma^2Q_{T/L} (0,\bm{q}) \right] |\alpha_{\bm q}|^2,
\end{equation}
with $\alpha_{- \bm q} =  \alpha_{ \bm q}^*$, $\gamma=e\,|\bar{A}|/c$ and $T/L$ depends on whether Eq.~\eqref{vectorpotential} is a transverse ($\bm{q} \cdot \bm{u}=0$) or longitudinal field. 
${\cal N}_q$ is a  positive factor given in the SM.  $E (\alpha_{\bm q})$ is the ground state energy with a coherent state of photons of amplitude $\alpha_{\bm q}$ (and $\alpha_{-\bm q}$). When $\alpha_{\bm q} \ne 0$, it describes the superradiant state and the phase transition occurs when the term inside the bracket in Eq.~\eqref{expansion} changes sign. The derivation leading to Eq.~\eqref{expansion}, detailed in the SM, follows from Ref.~\cite{Andolina_2019} but extends it to finite $\bm{q}$. In the uniform case $\bq =0$, the TRK sum~\eqref{TRKsum} and Eq.~\eqref{expansion} prove the so-called no-go theorem~\cite{Andolina_2019} which prevents any photon condensation (superradiance) to a uniform cavity field.

However, Eq.~\eqref{expansion} at finite $\bm{q}$ goes beyond the TRK sum rule and predicts a superradiant state if the following condition is achieved
\begin{equation}\label{criterion}
  Q_{T} (0,\bm{q}) < - \frac{\hbar \omega_q c^2}{2 e^2 |\bar{A}|^2}.
\end{equation}
This criterion can alternatively be obtained from computing the pole of the photon Green's function at vanishing frequency.
The longitudinal response function behaves quite differently from the transverse one at finite $\bm{q}$. We identify a second sum rule, called the $f$-sum rule (see SM),
\begin{equation}\label{ql}
  Q_{L} (0,\bm{q}) = 0,
\end{equation}
stemming from charge conservation
Inserting Eq.~\eqref{ql} into the ground state energy~\eqref{expansion}, we find no phase transition to a longitudinal potential vector, fully excluding superradiance in one-dimensional electron lattice systems, ladder models aside.



Our analysis has shown that the transverse current susceptibility $Q_T$ determines the onset of superradiance. For non-interacting electrons, we consider the eigenstates $|\bm{k},n \rangle$ of the Bloch Hamiltonian $h_{\alpha\beta} (\bk)$ with energies $\epsilon_{\bk,n}$. For convenience, we label the states with $n<0$ ($>0$) for negative (positive) energies. 
We introduce the notation $|\bm{k},\bm{q}\rangle_{n,m}$ for an electron-hole excitation on top of the ground state $| 0 \rangle$, where the hole (electron) is in state $|\bm{k},n \rangle$ ($|\bm{k+q},m \rangle$). At zero temperature, the susceptibility is given by
$Q_T(\bm{q}) = K_T (\bm{q}) - u_i \langle {\cal T}^{i,j}_{\bm{q},-\bm{q}} \rangle  u_j /V$ with the paramagnetic response ($d$ is the space dimension)
\begin{equation}\label{paramagnetic}
K_T (0,\bm{q}) =  \int_{BZ} \frac{d^d\bk}{(2 \pi)^d}  \sum_{n<0<m} \sum_{\pm} \frac{| g^{n,m}_{ \bm{k},\pm \bm{q}} |^2  }{\epsilon_{\bk,n}-\epsilon_{\bk \pm \bq,m}},
\end{equation}
where the denominator is the energy of the electron-hole excitation. The numerator depends on the dipole couplings $g^{n,m}_{ \bm{k},\bm{q}} =   \langle 0 |J^T_{\bm q} |\bm{k},\bm{q} \rangle_{n,m}$. Interestingly, the corresponding dipole for the longitudinal response vanishes with $\epsilon_{\bk,n}-\epsilon_{\bk+\bq,m}$ which prevents any divergence in the integral. The absence of such cancellation for the transverse part is crucial and opens the way for a diverging paramagnetic response~\eqref{paramagnetic}. There are various ways to obtain a singularity, either by having two lines of points in the Fermi surface connected by a single momentum $\bm{q}$ (nesting) in two dimensions, or if the density of states at a Fermi point becomes infinite, such as quadratic band touching or Landau level crossing~\cite{Basko_PRL2019}. Since the paramagnetic susceptibility $K_T$ is negative, its divergence signals a superradiant phase transition, no matter how weak light-matter interaction is,  since the criterion~\eqref{criterion} is always satisfied.

\textit{Superradiant phase ---}  We illustrate the above criterion~\eqref{criterion} for superradiance with concrete examples of tight-binding models and discuss the resulting superradiant phases. 

The first model that we consider is the textbook two-dimensional square lattice with nearest-neighbor hopping. The Bloch Hamiltonian is $h (\bk) = - t ( \cos k_x + \cos k_y)$, with unit lattice spacing for simplicity. At half-filling, the electronic ground state exhibits a square Fermi surface shown as a solid line in Fig.~\ref{fig1}, and a nesting between two segments of the Fermi surface by the wavevector $\bm{q}^* = (\pi,\pi)$. The whole band structure can be arbitrarily separated into a valence band  and a conduction band depending on the sign of $\epsilon_{\bk}$.
\begin{figure}[t]
\includegraphics[width=0.9\columnwidth]{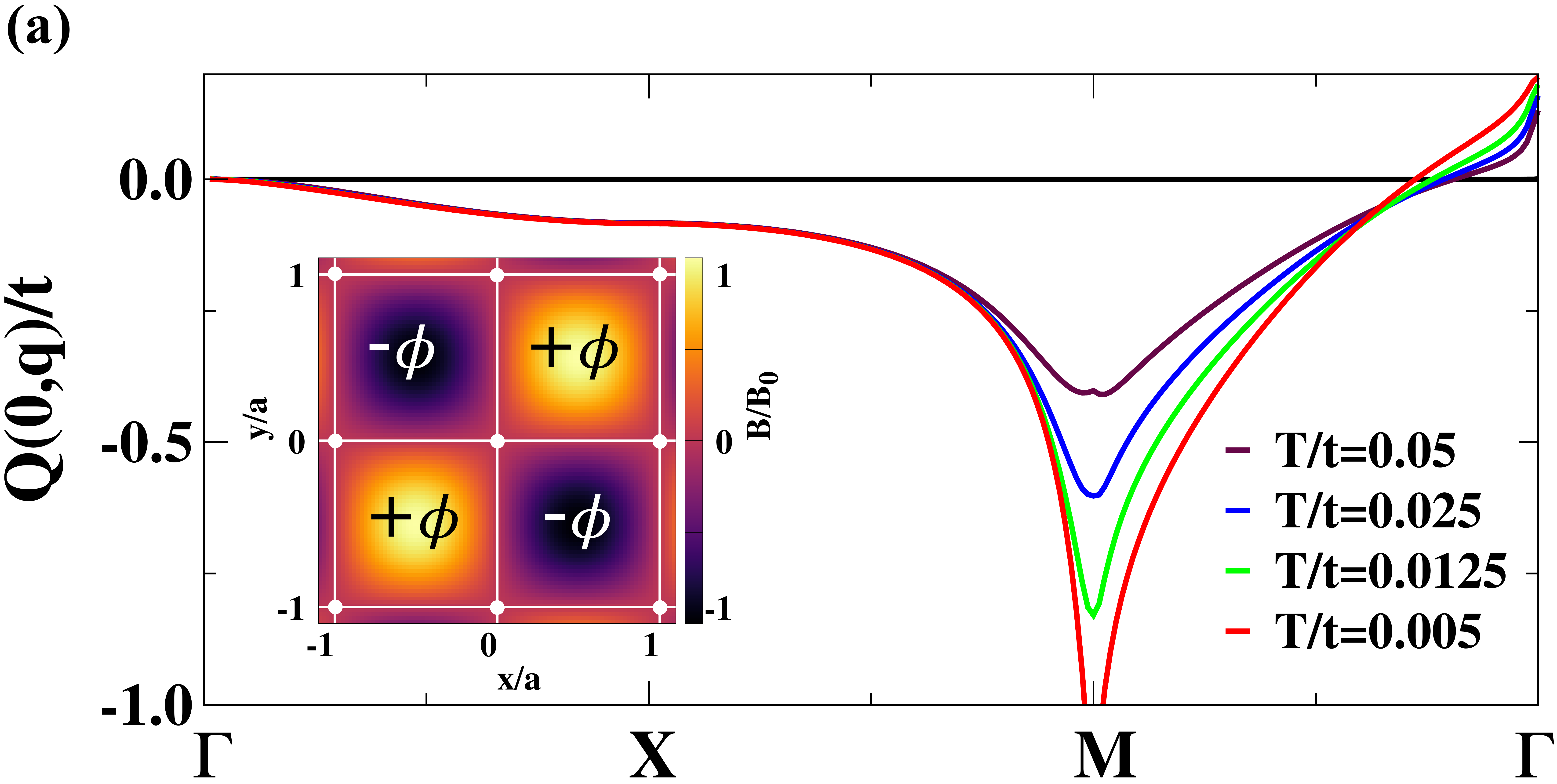}
\includegraphics[width=0.9\columnwidth]{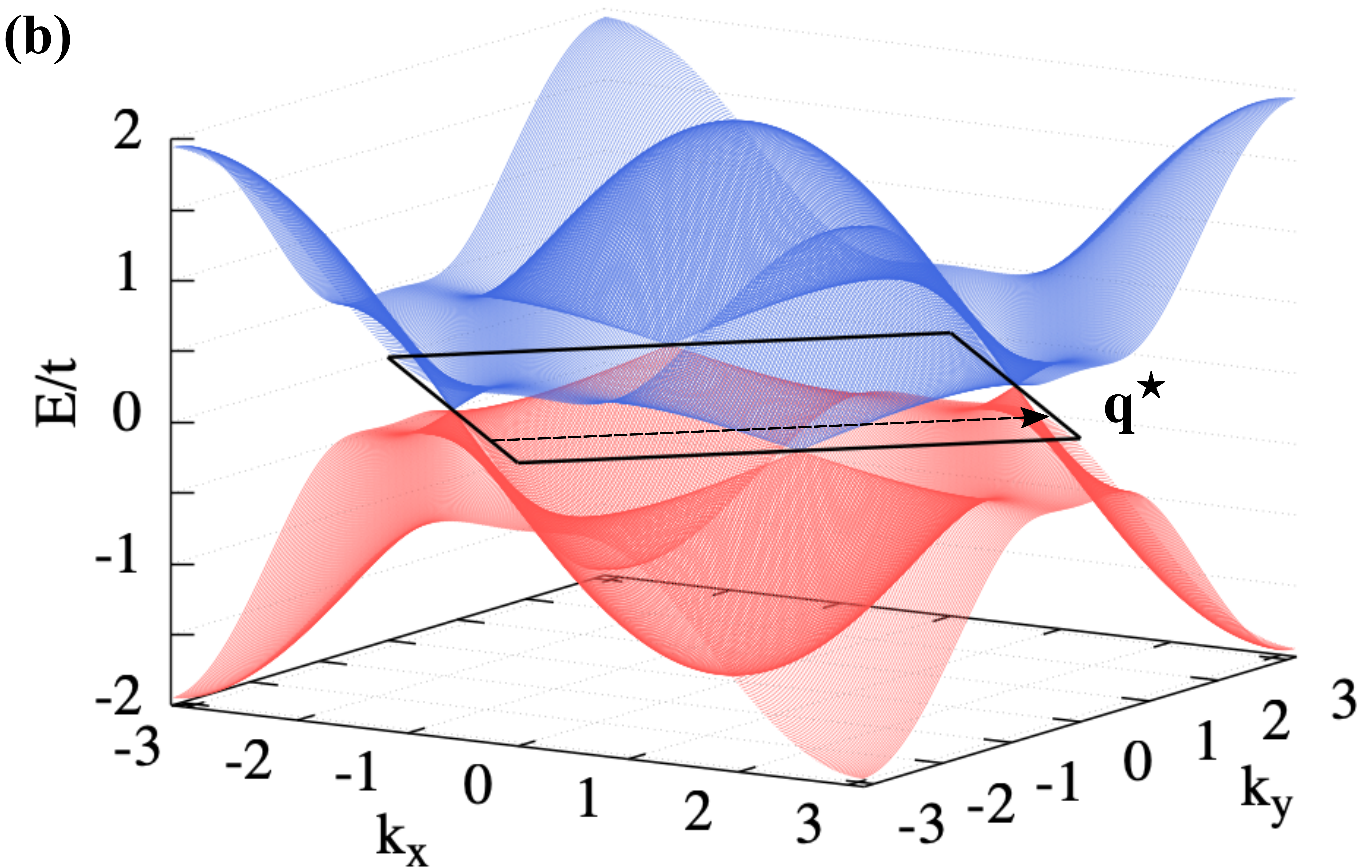}
\caption{(a) Current susceptibilities for $\bm{q}$ along high-symmetry lines. The longitudinal (black line) statisfies Eq.~\eqref{ql}, the transverse develops a  peak as the temperature is lowered. (inset) Alternating cavity field configuration in the superradiant phase. (b) Representative spectrum in the superradiant phase with the four Dirac points and the nesting vector  $\bm{q}^* = (\pi,\pi)$. The original Fermi surface is indicated by a solid black line.
}
\label{fig1}
\end{figure}
Following the above steps starting with the Peierls substitution, one arrives at Eq.~\eqref{paramagnetic} for the transverse and longitudinal responses, with the dipole elements
\begin{equation}\label{dipole}
  g_{ \bm{k},\bm{q}^*}^{T/L} =  -i \frac{\sqrt{2} t}{\pi} \left( \cos k_x \mp \cos k_y \right),
\end{equation}
coupling the valence and conduction bands. The nesting by $\bq^*$, represented in Fig.\ref{fig1} (b), implies a divergence of Eq.~\eqref{paramagnetic} when $\bk$ approaches one side of the Fermi surface while the transverse dipole~\eqref{dipole} remains finite. On the contrary, the longitudinal dipole vanishes ensuring a finite paramagnetic response.
This is illustrated in Fig.~\eqref{fig1} (a) where the divergence in the transverse response develops at low temperature 
at the nesting vector $\bq^*$ (M point). Extracting the divergence, we obtain the critical temperature $T_c \sim t e^{- \frac{\pi^2}{4} \sqrt{\hbar \omega_q/t} \frac{\hbar}{a \gamma}} $ ($a$ is the lattice spacing) below which the superradiant phase is energetically favorable. The occupied superradiant bosonic mode is not uniform in space but spatially modulated at the nesting wavector $\bq^*$.

Next we investigate the superradiant phase for the square lattice model. The ground state acquires a finite bosonic occupation $\langle a_{\bm q^*} \rangle/V \ne 0$. There is an absence of light-matter entanglement in the thermodynamic limit (see SM) and we assume a photon coherent state. Viewed from the electrons, we obtain an effective Hamiltonian~\eqref{Matter_Hamiltonian} with classical phases dressing the hoppings, similar to piercing a non-uniform magnetic flux through the lattice. For $\bm{q}^* = (\pi,\pi)$, the corresponding flux configuration alternates between plaquettes as $\pm \phi$, see Fig.~\ref{fig1}, thus breaking time-reversal symmetry (TRS). 
With this flux, the new unit cell has length $\sqrt{2}$, along the diagonals of the original square lattice, and contains two inequivalent sites $A$ and $B$~\cite{affleck1988}. The new Bloch Hamiltonian takes the form
\begin{equation}\label{bloch-form}
 h(\bk,\bm{A}) = - t \begin{pmatrix} 0 & d_{AB} \\ d_{AB}^* & 0 
  \end{pmatrix},
\end{equation}
with the matrix element $d_{AB} = e^{i \phi'} \cos k_x + e^{-i \phi'} \cos k_y$, $\phi'=\phi/4$. The alternating flux $\pm \phi$ hence opens a gap almost everywhere on the Fermi surface except at four $C_{4 z}$-related 
points, $k_x = \pm \pi/2$, $k_y = \pm \pi/2$, from which four Dirac cones emerge. The band spectrum is represented in Fig.~\ref{fig1}. 
Like in graphene, the $C_{2 z} T$ symmetry, $\sigma_x  h^*(\bk) \sigma_x = h(\bk)$, imposes a vanishing Berry curvature and protects the Dirac points~\cite{kim2015,ahn2019,mora2019} characterized by the Berry phases $\pm \pi$ ($C_{4 z}$ reversing the Berry phase). The similarities with graphene extend to zero-energy boundary modes~\cite{ruy2002,delplace2011}, which develop in graphene for zig-zag edges while they are absent at armchair termination~\cite{akhmerov2008}. Here, we find a collection of zero-energy states when the square lattice has a termination along the diagonals of the original lattice, but not for edges parallel to  the $x$ or $y$ direction.

The second model we discuss consists of electrons moving on a honeycomb with a quadratic band touching dispersion.
The Bloch Hamiltonian incorporates nearest- and third-nearest-neighbors hoppings~\cite{bena2011,montambaux2012}. It takes the form of Eq.~\eqref{bloch-form} with  $d_{AB} = \sum_{j=1}^3 e^{i \bm{k} \cdot \bm{\Delta}_{j}} + r  \sum_{j=1}^3 e^{-2 i \bm{k} \cdot \bm{\Delta}_{j}}$, where the three vectors $\bm{\Delta}_{j}$ connect nearest neighbors on the lattice. $r=0$ is the standard model describing electronic bands in graphene. It possesses two inequivalent Dirac cones centered at the K and K' points with Berry phases $\pm \pi$. Additional Dirac cones enter the Brillouin zone for non-zero $r$ and fuse with the original ones at $r=1/2$ resulting in $d_{AB} = - \frac{9}{8} ( \delta k_x \pm i \delta k_y)^2$ in the vicinity of the K (K') point. They give rise to two parabolic band contacts with Berry phases $\mp 2 \pi$ at K and K'. At half filling, setting $\bq^*$ to be the vector connecting K and K', we find a finite transverse dipole element
\begin{equation}
  | g_{ \bm{K},\bm{q}^*}^{T} | =  \frac{9 \sqrt{3}}{8 \pi} \, t,
\end{equation}
in Eq.~\eqref{paramagnetic}, resulting in a divergence in the zero-temperature current susceptibility $Q_T$, for details we refer to the SM, and therefore to a superradiant phase at arbitrary weak light-matter coupling.
\begin{figure}[t]
  \includegraphics[width=0.95\columnwidth]{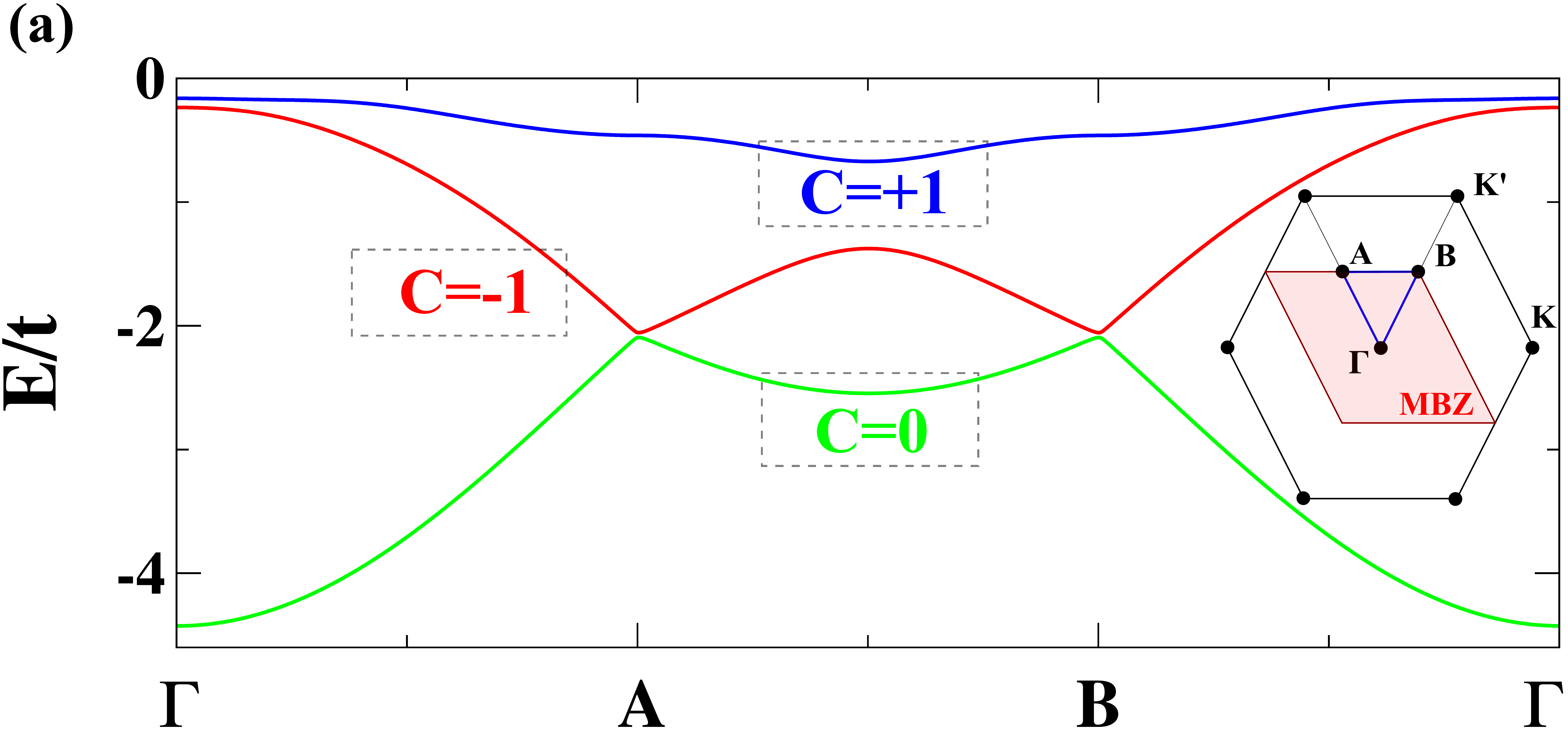}
  \includegraphics[width=0.95\columnwidth]{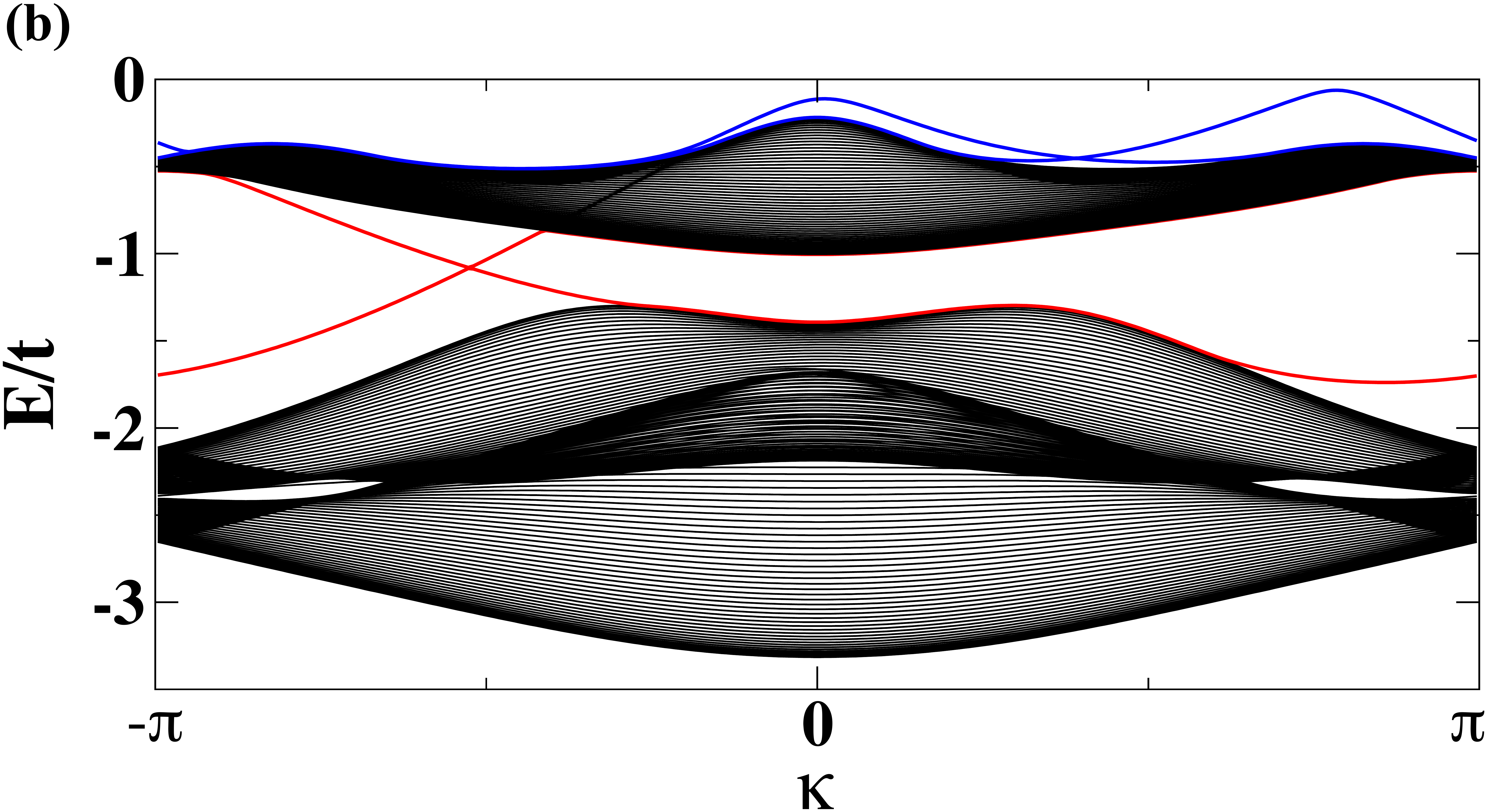}
  \caption{(a) Band spectrum (three lowest) of the superradiant state with non-zero Chern numbers obtained for $\bar{A}=0.3$, inset: reciprocal space. (b) Spectrum obtained in a ribbon geometry for $\bar{A}=1.5$. Black lines correspond to bulk states while the (non-)topologically protected edge states are shown in (blue) red.}
\label{fig2}
\end{figure}
In this case, the divergence is produced by the parabolic form of the energy difference in Eq.~\eqref{paramagnetic} (and not by line nesting) while the Fermi surface reduces to the two points K and K'. As expected, the longitudinal dipole element vanishes at $(\bm{K},\bm{q}^*)$, protected by the $f$-sum rule~\eqref{ql}.

The superradiant state is again described by a classical photon field modulated at $\bm{q}^*$, with $6$ sites per unit cell.
As shown in Fig.~\ref{fig2}(a), the photonic condensate opens a gap at the K and K' points and TRS breaking results in topological bands with non-zero Chern numbers and therefore in a topological superradiant phase. The model also displays a chiral symmetry anti-commuting with the Bloch Hamiltonian~\footnote{The chiral symmetry is diagonal and takes +1(-1) values on the $A$ ($B$) sublattice of the hexagonal lattice.} which imposes bands of opposite energies to have the same Chern number. 
For ribbon boundary conditions (periodic boundary conditions along $\bm{y}$, while open along the other principal direction) the superradiant phase presents topologically protected 1D edge states, displayed in Fig.~\ref{fig2}(b), crossing the band gap between bands $2$ and $3$ characterized by opposite Chern  numbers. 
Such ``superradiant edge state'' could be directly probed by light emission.

\textit{Conclusions and outlook ---} We established a framework for finding superradiant phase transitions in electronic systems. The divergence of the transverse current susceptibility is not necessary for obtaining superradiance and a sufficiently strong light-matter coupling works if it is simply negative. We also envision a superradiant phase transition close to magic angles~\cite{MacDonald_M-Model,cao_TBG1,cao_TBG2} in twisted bilayer graphene, where the scenario of parabolic band touching is very similar to one discussed here~\cite{song2019,hejazi2019}.

\textit{Acknowledgments---}  We would like to acknowledge fruitful discussions and correspondences with Marcello Andolina, Denis Basko, Cristiano Ciuti, Mark Goerbig and Marco Polini. This work was supported by the French National Research Agency (project  SIMCIRCUIT, ANR-18-CE47-0014-01).

\textit{Note added---}  As we were finalizing the writing of this manuscript, we learned about the theoretical work of Andolina  et al. \cite{Andolina_preprint_2020} which overlaps with the first part of our work and reaches a similar conclusion concerning the occurrence of a  superradiant phase transition.


\bibliography{mybiblio}

\begin{thebibliography}{58}%
\makeatletter
\providecommand \@ifxundefined [1]{%
 \@ifx{#1\undefined}
}%
\providecommand \@ifnum [1]{%
 \ifnum #1\expandafter \@firstoftwo
 \else \expandafter \@secondoftwo
 \fi
}%
\providecommand \@ifx [1]{%
 \ifx #1\expandafter \@firstoftwo
 \else \expandafter \@secondoftwo
 \fi
}%
\providecommand \natexlab [1]{#1}%
\providecommand \enquote  [1]{``#1''}%
\providecommand \bibnamefont  [1]{#1}%
\providecommand \bibfnamefont [1]{#1}%
\providecommand \citenamefont [1]{#1}%
\providecommand \href@noop [0]{\@secondoftwo}%
\providecommand \href [0]{\begingroup \@sanitize@url \@href}%
\providecommand \@href[1]{\@@startlink{#1}\@@href}%
\providecommand \@@href[1]{\endgroup#1\@@endlink}%
\providecommand \@sanitize@url [0]{\catcode `\\12\catcode `\$12\catcode
  `\&12\catcode `\#12\catcode `\^12\catcode `\_12\catcode `\%12\relax}%
\providecommand \@@startlink[1]{}%
\providecommand \@@endlink[0]{}%
\providecommand \url  [0]{\begingroup\@sanitize@url \@url }%
\providecommand \@url [1]{\endgroup\@href {#1}{\urlprefix }}%
\providecommand \urlprefix  [0]{URL }%
\providecommand \Eprint [0]{\href }%
\providecommand \doibase [0]{http://dx.doi.org/}%
\providecommand \selectlanguage [0]{\@gobble}%
\providecommand \bibinfo  [0]{\@secondoftwo}%
\providecommand \bibfield  [0]{\@secondoftwo}%
\providecommand \translation [1]{[#1]}%
\providecommand \BibitemOpen [0]{}%
\providecommand \bibitemStop [0]{}%
\providecommand \bibitemNoStop [0]{.\EOS\space}%
\providecommand \EOS [0]{\spacefactor3000\relax}%
\providecommand \BibitemShut  [1]{\csname bibitem#1\endcsname}%
\let\auto@bib@innerbib\@empty
\bibitem [{\citenamefont {Aspect}\ \emph {et~al.}(1988)\citenamefont {Aspect},
  \citenamefont {Arimondo}, \citenamefont {Kaiser}, \citenamefont
  {Vansteenkiste},\ and\ \citenamefont {Cohen-Tannoudji}}]{Aspect_1988}%
  \BibitemOpen
  \bibfield  {author} {\bibinfo {author} {\bibfnamefont {A.}~\bibnamefont
  {Aspect}}, \bibinfo {author} {\bibfnamefont {E.}~\bibnamefont {Arimondo}},
  \bibinfo {author} {\bibfnamefont {R.}~\bibnamefont {Kaiser}}, \bibinfo
  {author} {\bibfnamefont {N.}~\bibnamefont {Vansteenkiste}}, \ and\ \bibinfo
  {author} {\bibfnamefont {C.}~\bibnamefont {Cohen-Tannoudji}},\ }\href
  {\doibase 10.1103/PhysRevLett.61.826} {\bibfield  {journal} {\bibinfo
  {journal} {Phys. Rev. Lett.}\ }\textbf {\bibinfo {volume} {61}},\ \bibinfo
  {pages} {826} (\bibinfo {year} {1988})}\BibitemShut {NoStop}%
\bibitem [{\citenamefont {Phillips}(1998)}]{Philipps_1998}%
  \BibitemOpen
  \bibfield  {author} {\bibinfo {author} {\bibfnamefont {W.~D.}\ \bibnamefont
  {Phillips}},\ }\href {\doibase 10.1103/RevModPhys.70.721} {\bibfield
  {journal} {\bibinfo  {journal} {Rev. Mod. Phys.}\ }\textbf {\bibinfo {volume}
  {70}},\ \bibinfo {pages} {721} (\bibinfo {year} {1998})}\BibitemShut
  {NoStop}%
\bibitem [{\citenamefont {Monroe}\ \emph {et~al.}(1995)\citenamefont {Monroe},
  \citenamefont {Meekhof}, \citenamefont {King}, \citenamefont {Itano},\ and\
  \citenamefont {Wineland}}]{Monroe_1995}%
  \BibitemOpen
  \bibfield  {author} {\bibinfo {author} {\bibfnamefont {C.}~\bibnamefont
  {Monroe}}, \bibinfo {author} {\bibfnamefont {D.~M.}\ \bibnamefont {Meekhof}},
  \bibinfo {author} {\bibfnamefont {B.~E.}\ \bibnamefont {King}}, \bibinfo
  {author} {\bibfnamefont {W.~M.}\ \bibnamefont {Itano}}, \ and\ \bibinfo
  {author} {\bibfnamefont {D.~J.}\ \bibnamefont {Wineland}},\ }\href {\doibase
  10.1103/PhysRevLett.75.4714} {\bibfield  {journal} {\bibinfo  {journal}
  {Phys. Rev. Lett.}\ }\textbf {\bibinfo {volume} {75}},\ \bibinfo {pages}
  {4714} (\bibinfo {year} {1995})}\BibitemShut {NoStop}%
\bibitem [{\citenamefont {Cirac}\ \emph {et~al.}(1997)\citenamefont {Cirac},
  \citenamefont {Zoller}, \citenamefont {Kimble},\ and\ \citenamefont
  {Mabuchi}}]{Cirac_1997}%
  \BibitemOpen
  \bibfield  {author} {\bibinfo {author} {\bibfnamefont {J.~I.}\ \bibnamefont
  {Cirac}}, \bibinfo {author} {\bibfnamefont {P.}~\bibnamefont {Zoller}},
  \bibinfo {author} {\bibfnamefont {H.~J.}\ \bibnamefont {Kimble}}, \ and\
  \bibinfo {author} {\bibfnamefont {H.}~\bibnamefont {Mabuchi}},\ }\href
  {\doibase 10.1103/PhysRevLett.78.3221} {\bibfield  {journal} {\bibinfo
  {journal} {Phys. Rev. Lett.}\ }\textbf {\bibinfo {volume} {78}},\ \bibinfo
  {pages} {3221} (\bibinfo {year} {1997})}\BibitemShut {NoStop}%
\bibitem [{\citenamefont {Duan}\ and\ \citenamefont
  {Kimble}(2004)}]{Duan_2004}%
  \BibitemOpen
  \bibfield  {author} {\bibinfo {author} {\bibfnamefont {L.-M.}\ \bibnamefont
  {Duan}}\ and\ \bibinfo {author} {\bibfnamefont {H.~J.}\ \bibnamefont
  {Kimble}},\ }\href {\doibase 10.1103/PhysRevLett.92.127902} {\bibfield
  {journal} {\bibinfo  {journal} {Phys. Rev. Lett.}\ }\textbf {\bibinfo
  {volume} {92}},\ \bibinfo {pages} {127902} (\bibinfo {year}
  {2004})}\BibitemShut {NoStop}%
\bibitem [{\citenamefont {Raimond}\ \emph {et~al.}(2001)\citenamefont
  {Raimond}, \citenamefont {Brune},\ and\ \citenamefont
  {Haroche}}]{Haroche_2001}%
  \BibitemOpen
  \bibfield  {author} {\bibinfo {author} {\bibfnamefont {J.~M.}\ \bibnamefont
  {Raimond}}, \bibinfo {author} {\bibfnamefont {M.}~\bibnamefont {Brune}}, \
  and\ \bibinfo {author} {\bibfnamefont {S.}~\bibnamefont {Haroche}},\ }\href
  {\doibase 10.1103/RevModPhys.73.565} {\bibfield  {journal} {\bibinfo
  {journal} {Rev. Mod. Phys.}\ }\textbf {\bibinfo {volume} {73}},\ \bibinfo
  {pages} {565} (\bibinfo {year} {2001})}\BibitemShut {NoStop}%
\bibitem [{\citenamefont {Leibfried}\ \emph {et~al.}(2003)\citenamefont
  {Leibfried}, \citenamefont {Blatt}, \citenamefont {Monroe},\ and\
  \citenamefont {Wineland}}]{Wineland_2003}%
  \BibitemOpen
  \bibfield  {author} {\bibinfo {author} {\bibfnamefont {D.}~\bibnamefont
  {Leibfried}}, \bibinfo {author} {\bibfnamefont {R.}~\bibnamefont {Blatt}},
  \bibinfo {author} {\bibfnamefont {C.}~\bibnamefont {Monroe}}, \ and\ \bibinfo
  {author} {\bibfnamefont {D.}~\bibnamefont {Wineland}},\ }\href {\doibase
  10.1103/RevModPhys.75.281} {\bibfield  {journal} {\bibinfo  {journal} {Rev.
  Mod. Phys.}\ }\textbf {\bibinfo {volume} {75}},\ \bibinfo {pages} {281}
  (\bibinfo {year} {2003})}\BibitemShut {NoStop}%
\bibitem [{\citenamefont {Scalari}\ \emph {et~al.}(2012)\citenamefont
  {Scalari}, \citenamefont {Maissen}, \citenamefont {Tur{\v c}inkov{\'a}},
  \citenamefont {Hagenm{\"u}ller}, \citenamefont {De~Liberato}, \citenamefont
  {Ciuti}, \citenamefont {Reichl}, \citenamefont {Schuh}, \citenamefont
  {Wegscheider}, \citenamefont {Beck},\ and\ \citenamefont
  {Faist}}]{Faist_2012}%
  \BibitemOpen
  \bibfield  {author} {\bibinfo {author} {\bibfnamefont {G.}~\bibnamefont
  {Scalari}}, \bibinfo {author} {\bibfnamefont {C.}~\bibnamefont {Maissen}},
  \bibinfo {author} {\bibfnamefont {D.}~\bibnamefont {Tur{\v c}inkov{\'a}}},
  \bibinfo {author} {\bibfnamefont {D.}~\bibnamefont {Hagenm{\"u}ller}},
  \bibinfo {author} {\bibfnamefont {S.}~\bibnamefont {De~Liberato}}, \bibinfo
  {author} {\bibfnamefont {C.}~\bibnamefont {Ciuti}}, \bibinfo {author}
  {\bibfnamefont {C.}~\bibnamefont {Reichl}}, \bibinfo {author} {\bibfnamefont
  {D.}~\bibnamefont {Schuh}}, \bibinfo {author} {\bibfnamefont
  {W.}~\bibnamefont {Wegscheider}}, \bibinfo {author} {\bibfnamefont
  {M.}~\bibnamefont {Beck}}, \ and\ \bibinfo {author} {\bibfnamefont
  {J.}~\bibnamefont {Faist}},\ }\href {\doibase 10.1126/science.1216022}
  {\bibfield  {journal} {\bibinfo  {journal} {Science}\ }\textbf {\bibinfo
  {volume} {335}},\ \bibinfo {pages} {1323} (\bibinfo {year} {2012})},\ \Eprint
  {http://arxiv.org/abs/https://science.sciencemag.org/content/335/6074/1323.full.pdf}
  {https://science.sciencemag.org/content/335/6074/1323.full.pdf} \BibitemShut
  {NoStop}%
\bibitem [{\citenamefont {Maissen}\ \emph {et~al.}(2014)\citenamefont
  {Maissen}, \citenamefont {Scalari}, \citenamefont {Valmorra}, \citenamefont
  {Beck}, \citenamefont {Faist}, \citenamefont {Cibella}, \citenamefont
  {Leoni}, \citenamefont {Reichl}, \citenamefont {Charpentier},\ and\
  \citenamefont {Wegscheider}}]{Maissen_2014}%
  \BibitemOpen
  \bibfield  {author} {\bibinfo {author} {\bibfnamefont {C.}~\bibnamefont
  {Maissen}}, \bibinfo {author} {\bibfnamefont {G.}~\bibnamefont {Scalari}},
  \bibinfo {author} {\bibfnamefont {F.}~\bibnamefont {Valmorra}}, \bibinfo
  {author} {\bibfnamefont {M.}~\bibnamefont {Beck}}, \bibinfo {author}
  {\bibfnamefont {J.}~\bibnamefont {Faist}}, \bibinfo {author} {\bibfnamefont
  {S.}~\bibnamefont {Cibella}}, \bibinfo {author} {\bibfnamefont
  {R.}~\bibnamefont {Leoni}}, \bibinfo {author} {\bibfnamefont
  {C.}~\bibnamefont {Reichl}}, \bibinfo {author} {\bibfnamefont
  {C.}~\bibnamefont {Charpentier}}, \ and\ \bibinfo {author} {\bibfnamefont
  {W.}~\bibnamefont {Wegscheider}},\ }\href {\doibase
  10.1103/PhysRevB.90.205309} {\bibfield  {journal} {\bibinfo  {journal} {Phys.
  Rev. B}\ }\textbf {\bibinfo {volume} {90}},\ \bibinfo {pages} {205309}
  (\bibinfo {year} {2014})}\BibitemShut {NoStop}%
\bibitem [{\citenamefont {Smolka}\ \emph {et~al.}(2014)\citenamefont {Smolka},
  \citenamefont {Wuester}, \citenamefont {Haupt}, \citenamefont {Faelt},
  \citenamefont {Wegscheider},\ and\ \citenamefont {Imamoglu}}]{Smolka_2014}%
  \BibitemOpen
  \bibfield  {author} {\bibinfo {author} {\bibfnamefont {S.}~\bibnamefont
  {Smolka}}, \bibinfo {author} {\bibfnamefont {W.}~\bibnamefont {Wuester}},
  \bibinfo {author} {\bibfnamefont {F.}~\bibnamefont {Haupt}}, \bibinfo
  {author} {\bibfnamefont {S.}~\bibnamefont {Faelt}}, \bibinfo {author}
  {\bibfnamefont {W.}~\bibnamefont {Wegscheider}}, \ and\ \bibinfo {author}
  {\bibfnamefont {A.}~\bibnamefont {Imamoglu}},\ }\href {\doibase
  10.1126/science.1258595} {\bibfield  {journal} {\bibinfo  {journal}
  {Science}\ }\textbf {\bibinfo {volume} {346}},\ \bibinfo {pages} {332}
  (\bibinfo {year} {2014})},\ \Eprint
  {http://arxiv.org/abs/https://science.sciencemag.org/content/346/6207/332.full.pdf}
  {https://science.sciencemag.org/content/346/6207/332.full.pdf} \BibitemShut
  {NoStop}%
\bibitem [{\citenamefont {Liu}\ \emph {et~al.}(2015)\citenamefont {Liu},
  \citenamefont {Galfsky}, \citenamefont {Sun}, \citenamefont {Xia},
  \citenamefont {Lin}, \citenamefont {Lee}, \citenamefont {Kéna-Cohen},\ and\
  \citenamefont {Menon}}]{Liu_2015}%
  \BibitemOpen
  \bibfield  {author} {\bibinfo {author} {\bibfnamefont {X.}~\bibnamefont
  {Liu}}, \bibinfo {author} {\bibfnamefont {T.}~\bibnamefont {Galfsky}},
  \bibinfo {author} {\bibfnamefont {Z.}~\bibnamefont {Sun}}, \bibinfo {author}
  {\bibfnamefont {F.}~\bibnamefont {Xia}}, \bibinfo {author} {\bibfnamefont
  {E.-c.}\ \bibnamefont {Lin}}, \bibinfo {author} {\bibfnamefont {Y.-H.}\
  \bibnamefont {Lee}}, \bibinfo {author} {\bibfnamefont {S.}~\bibnamefont
  {Kéna-Cohen}}, \ and\ \bibinfo {author} {\bibfnamefont {V.~M.}\ \bibnamefont
  {Menon}},\ }\href {\doibase 10.1038/nphoton.2014.304} {\bibfield  {journal}
  {\bibinfo  {journal} {Nature Photonics}\ }\textbf {\bibinfo {volume} {9}},\
  \bibinfo {pages} {30} (\bibinfo {year} {2015})}\BibitemShut {NoStop}%
\bibitem [{\citenamefont {Basov}\ \emph {et~al.}(2016)\citenamefont {Basov},
  \citenamefont {Fogler},\ and\ \citenamefont {Garc{\'\i}a~de
  Abajo}}]{Basov_2016}%
  \BibitemOpen
  \bibfield  {author} {\bibinfo {author} {\bibfnamefont {D.~N.}\ \bibnamefont
  {Basov}}, \bibinfo {author} {\bibfnamefont {M.~M.}\ \bibnamefont {Fogler}}, \
  and\ \bibinfo {author} {\bibfnamefont {F.~J.}\ \bibnamefont {Garc{\'\i}a~de
  Abajo}},\ }\href {\doibase 10.1126/science.aag1992} {\bibfield  {journal}
  {\bibinfo  {journal} {Science}\ }\textbf {\bibinfo {volume} {354}} (\bibinfo
  {year} {2016}),\ 10.1126/science.aag1992},\ \Eprint
  {http://arxiv.org/abs/https://science.sciencemag.org/content/354/6309/aag1992.full.pdf}
  {https://science.sciencemag.org/content/354/6309/aag1992.full.pdf}
  \BibitemShut {NoStop}%
\bibitem [{\citenamefont {Foerst}\ \emph {et~al.}(2011)\citenamefont {Foerst},
  \citenamefont {Manzoni}, \citenamefont {Kaiser}, \citenamefont {Tomioka},
  \citenamefont {Tokura}, \citenamefont {Merlin},\ and\ \citenamefont
  {Cavalleri}}]{Forst_2011}%
  \BibitemOpen
  \bibfield  {author} {\bibinfo {author} {\bibfnamefont {M.}~\bibnamefont
  {Foerst}}, \bibinfo {author} {\bibfnamefont {C.}~\bibnamefont {Manzoni}},
  \bibinfo {author} {\bibfnamefont {S.}~\bibnamefont {Kaiser}}, \bibinfo
  {author} {\bibfnamefont {Y.}~\bibnamefont {Tomioka}}, \bibinfo {author}
  {\bibfnamefont {Y.}~\bibnamefont {Tokura}}, \bibinfo {author} {\bibfnamefont
  {R.}~\bibnamefont {Merlin}}, \ and\ \bibinfo {author} {\bibfnamefont
  {A.}~\bibnamefont {Cavalleri}},\ }\href {\doibase 10.1038/nphys2055}
  {\bibfield  {journal} {\bibinfo  {journal} {Nature Physics}\ }\textbf
  {\bibinfo {volume} {7}},\ \bibinfo {pages} {854} (\bibinfo {year}
  {2011})}\BibitemShut {NoStop}%
\bibitem [{\citenamefont {Subedi}\ \emph {et~al.}(2014)\citenamefont {Subedi},
  \citenamefont {Cavalleri},\ and\ \citenamefont {Georges}}]{Subedi_2014}%
  \BibitemOpen
  \bibfield  {author} {\bibinfo {author} {\bibfnamefont {A.}~\bibnamefont
  {Subedi}}, \bibinfo {author} {\bibfnamefont {A.}~\bibnamefont {Cavalleri}}, \
  and\ \bibinfo {author} {\bibfnamefont {A.}~\bibnamefont {Georges}},\ }\href
  {\doibase 10.1103/PhysRevB.89.220301} {\bibfield  {journal} {\bibinfo
  {journal} {Phys. Rev. B}\ }\textbf {\bibinfo {volume} {89}},\ \bibinfo
  {pages} {220301} (\bibinfo {year} {2014})}\BibitemShut {NoStop}%
\bibitem [{\citenamefont {Schlawin}\ \emph {et~al.}(2019)\citenamefont
  {Schlawin}, \citenamefont {Cavalleri},\ and\ \citenamefont
  {Jaksch}}]{Jaksch_2019}%
  \BibitemOpen
  \bibfield  {author} {\bibinfo {author} {\bibfnamefont {F.}~\bibnamefont
  {Schlawin}}, \bibinfo {author} {\bibfnamefont {A.}~\bibnamefont {Cavalleri}},
  \ and\ \bibinfo {author} {\bibfnamefont {D.}~\bibnamefont {Jaksch}},\ }\href
  {\doibase 10.1103/PhysRevLett.122.133602} {\bibfield  {journal} {\bibinfo
  {journal} {Phys. Rev. Lett.}\ }\textbf {\bibinfo {volume} {122}},\ \bibinfo
  {pages} {133602} (\bibinfo {year} {2019})}\BibitemShut {NoStop}%
\bibitem [{\citenamefont {Mazza}\ and\ \citenamefont
  {Georges}(2019)}]{Mazza_2019}%
  \BibitemOpen
  \bibfield  {author} {\bibinfo {author} {\bibfnamefont {G.}~\bibnamefont
  {Mazza}}\ and\ \bibinfo {author} {\bibfnamefont {A.}~\bibnamefont
  {Georges}},\ }\href {\doibase 10.1103/PhysRevLett.122.017401} {\bibfield
  {journal} {\bibinfo  {journal} {Phys. Rev. Lett.}\ }\textbf {\bibinfo
  {volume} {122}},\ \bibinfo {pages} {017401} (\bibinfo {year}
  {2019})}\BibitemShut {NoStop}%
\bibitem [{\citenamefont {Claassen}\ \emph {et~al.}(2019)\citenamefont
  {Claassen}, \citenamefont {Kennes}, \citenamefont {Zingl}, \citenamefont
  {Sentef},\ and\ \citenamefont {Rubio}}]{Claassen_2019}%
  \BibitemOpen
  \bibfield  {author} {\bibinfo {author} {\bibfnamefont {M.}~\bibnamefont
  {Claassen}}, \bibinfo {author} {\bibfnamefont {D.~M.}\ \bibnamefont
  {Kennes}}, \bibinfo {author} {\bibfnamefont {M.}~\bibnamefont {Zingl}},
  \bibinfo {author} {\bibfnamefont {M.~A.}\ \bibnamefont {Sentef}}, \ and\
  \bibinfo {author} {\bibfnamefont {A.}~\bibnamefont {Rubio}},\ }\href
  {\doibase 10.1038/s41567-019-0532-6} {\bibfield  {journal} {\bibinfo
  {journal} {Nature Physics}\ }\textbf {\bibinfo {volume} {15}},\ \bibinfo
  {pages} {766} (\bibinfo {year} {2019})}\BibitemShut {NoStop}%
\bibitem [{\citenamefont {Gross}\ and\ \citenamefont
  {Haroche}(1982)}]{Haroche_1982}%
  \BibitemOpen
  \bibfield  {author} {\bibinfo {author} {\bibfnamefont {M.}~\bibnamefont
  {Gross}}\ and\ \bibinfo {author} {\bibfnamefont {S.}~\bibnamefont
  {Haroche}},\ }\href {\doibase https://doi.org/10.1016/0370-1573(82)90102-8}
  {\bibfield  {journal} {\bibinfo  {journal} {Physics Reports}\ }\textbf
  {\bibinfo {volume} {93}},\ \bibinfo {pages} {301 } (\bibinfo {year}
  {1982})}\BibitemShut {NoStop}%
\bibitem [{\citenamefont {Dicke}(1954)}]{Dicke}%
  \BibitemOpen
  \bibfield  {author} {\bibinfo {author} {\bibfnamefont {R.~H.}\ \bibnamefont
  {Dicke}},\ }\href {\doibase 10.1103/PhysRev.93.99} {\bibfield  {journal}
  {\bibinfo  {journal} {Phys. Rev.}\ }\textbf {\bibinfo {volume} {93}},\
  \bibinfo {pages} {99} (\bibinfo {year} {1954})}\BibitemShut {NoStop}%
\bibitem [{\citenamefont {Hepp}\ and\ \citenamefont {Lieb}(1973)}]{Lieb_1973}%
  \BibitemOpen
  \bibfield  {author} {\bibinfo {author} {\bibfnamefont {K.}~\bibnamefont
  {Hepp}}\ and\ \bibinfo {author} {\bibfnamefont {E.~H.}\ \bibnamefont
  {Lieb}},\ }\href {\doibase https://doi.org/10.1016/0003-4916(73)90039-0}
  {\bibfield  {journal} {\bibinfo  {journal} {Annals of Physics}\ }\textbf
  {\bibinfo {volume} {76}},\ \bibinfo {pages} {360 } (\bibinfo {year}
  {1973})}\BibitemShut {NoStop}%
\bibitem [{\citenamefont {Wang}\ and\ \citenamefont {Hioe}(1973)}]{Wang_1973}%
  \BibitemOpen
  \bibfield  {author} {\bibinfo {author} {\bibfnamefont {Y.~K.}\ \bibnamefont
  {Wang}}\ and\ \bibinfo {author} {\bibfnamefont {F.~T.}\ \bibnamefont
  {Hioe}},\ }\href {\doibase 10.1103/PhysRevA.7.831} {\bibfield  {journal}
  {\bibinfo  {journal} {Phys. Rev. A}\ }\textbf {\bibinfo {volume} {7}},\
  \bibinfo {pages} {831} (\bibinfo {year} {1973})}\BibitemShut {NoStop}%
\bibitem [{\citenamefont {Skribanowitz}\ \emph {et~al.}(1973)\citenamefont
  {Skribanowitz}, \citenamefont {Herman}, \citenamefont {MacGillivray},\ and\
  \citenamefont {Feld}}]{Feld_1973}%
  \BibitemOpen
  \bibfield  {author} {\bibinfo {author} {\bibfnamefont {N.}~\bibnamefont
  {Skribanowitz}}, \bibinfo {author} {\bibfnamefont {I.~P.}\ \bibnamefont
  {Herman}}, \bibinfo {author} {\bibfnamefont {J.~C.}\ \bibnamefont
  {MacGillivray}}, \ and\ \bibinfo {author} {\bibfnamefont {M.~S.}\
  \bibnamefont {Feld}},\ }\href {\doibase 10.1103/PhysRevLett.30.309}
  {\bibfield  {journal} {\bibinfo  {journal} {Phys. Rev. Lett.}\ }\textbf
  {\bibinfo {volume} {30}},\ \bibinfo {pages} {309} (\bibinfo {year}
  {1973})}\BibitemShut {NoStop}%
\bibitem [{\citenamefont {Scheibner}\ \emph {et~al.}(2007)\citenamefont
  {Scheibner}, \citenamefont {Schmidt}, \citenamefont {Worschech},
  \citenamefont {Forchel}, \citenamefont {B.}, \citenamefont {Passow},\ and\
  \citenamefont {Hommel}}]{Scheibner_2007}%
  \BibitemOpen
  \bibfield  {author} {\bibinfo {author} {\bibfnamefont {M.}~\bibnamefont
  {Scheibner}}, \bibinfo {author} {\bibfnamefont {T.}~\bibnamefont {Schmidt}},
  \bibinfo {author} {\bibfnamefont {L.}~\bibnamefont {Worschech}}, \bibinfo
  {author} {\bibfnamefont {A.}~\bibnamefont {Forchel}}, \bibinfo {author}
  {\bibfnamefont {G.}~\bibnamefont {B.}}, \bibinfo {author} {\bibfnamefont
  {T.}~\bibnamefont {Passow}}, \ and\ \bibinfo {author} {\bibfnamefont
  {D.}~\bibnamefont {Hommel}},\ }\href {\doibase
  https://doi.org/10.1038/nphys494} {\bibfield  {journal} {\bibinfo  {journal}
  {Nature Physics}\ }\textbf {\bibinfo {volume} {3}},\ \bibinfo {pages} {106}
  (\bibinfo {year} {2007})}\BibitemShut {NoStop}%
\bibitem [{\citenamefont {Timothy Noe~II}\ \emph {et~al.}(2012)\citenamefont
  {Timothy Noe~II}, \citenamefont {Kim}, \citenamefont {Lee}, \citenamefont
  {Wang}, \citenamefont {Wójcik}, \citenamefont {McGill}, \citenamefont
  {Reitze}, \citenamefont {Belyanin},\ and\ \citenamefont
  {Kono}}]{Timothy_2012}%
  \BibitemOpen
  \bibfield  {author} {\bibinfo {author} {\bibfnamefont {G.}~\bibnamefont
  {Timothy Noe~II}}, \bibinfo {author} {\bibfnamefont {J.-H.}\ \bibnamefont
  {Kim}}, \bibinfo {author} {\bibfnamefont {J.}~\bibnamefont {Lee}}, \bibinfo
  {author} {\bibfnamefont {Y.}~\bibnamefont {Wang}}, \bibinfo {author}
  {\bibfnamefont {A.~K.}\ \bibnamefont {Wójcik}}, \bibinfo {author}
  {\bibfnamefont {S.~A.}\ \bibnamefont {McGill}}, \bibinfo {author}
  {\bibfnamefont {D.~H.}\ \bibnamefont {Reitze}}, \bibinfo {author}
  {\bibfnamefont {A.~A.}\ \bibnamefont {Belyanin}}, \ and\ \bibinfo {author}
  {\bibfnamefont {J.}~\bibnamefont {Kono}},\ }\href {\doibase
  10.1038/nphys2207} {\bibfield  {journal} {\bibinfo  {journal} {Nature
  Physics}\ }\textbf {\bibinfo {volume} {8}},\ \bibinfo {pages} {219} (\bibinfo
  {year} {2012})}\BibitemShut {NoStop}%
\bibitem [{\citenamefont {Baumann}\ \emph {et~al.}(2010)\citenamefont
  {Baumann}, \citenamefont {Guerlin}, \citenamefont {Brennecke},\ and\
  \citenamefont {Esslinger}}]{Baumann_2010}%
  \BibitemOpen
  \bibfield  {author} {\bibinfo {author} {\bibfnamefont {K.}~\bibnamefont
  {Baumann}}, \bibinfo {author} {\bibfnamefont {C.}~\bibnamefont {Guerlin}},
  \bibinfo {author} {\bibfnamefont {F.}~\bibnamefont {Brennecke}}, \ and\
  \bibinfo {author} {\bibfnamefont {T.}~\bibnamefont {Esslinger}},\ }\href
  {\doibase 10.1038/nature09009} {\bibfield  {journal} {\bibinfo  {journal}
  {Nature}\ }\textbf {\bibinfo {volume} {464}},\ \bibinfo {pages} {1301}
  (\bibinfo {year} {2010})}\BibitemShut {NoStop}%
\bibitem [{\citenamefont {Rza\ifmmode~\dot{z}\else \.{z}\fi{}ewski}\ \emph
  {et~al.}(1975)\citenamefont {Rza\ifmmode~\dot{z}\else \.{z}\fi{}ewski},
  \citenamefont {W\'odkiewicz},\ and\ \citenamefont {\ifmmode~\dot{Z}\else
  \.{Z}\fi{}akowicz}}]{Rzazewski_1975}%
  \BibitemOpen
  \bibfield  {author} {\bibinfo {author} {\bibfnamefont {K.}~\bibnamefont
  {Rza\ifmmode~\dot{z}\else \.{z}\fi{}ewski}}, \bibinfo {author} {\bibfnamefont
  {K.}~\bibnamefont {W\'odkiewicz}}, \ and\ \bibinfo {author} {\bibfnamefont
  {W.}~\bibnamefont {\ifmmode~\dot{Z}\else \.{Z}\fi{}akowicz}},\ }\href
  {\doibase 10.1103/PhysRevLett.35.432} {\bibfield  {journal} {\bibinfo
  {journal} {Phys. Rev. Lett.}\ }\textbf {\bibinfo {volume} {35}},\ \bibinfo
  {pages} {432} (\bibinfo {year} {1975})}\BibitemShut {NoStop}%
\bibitem [{\citenamefont {Bialynicki-Birula}\ and\ \citenamefont {Rza\ifmmode
  \mbox{\c{}}\else \c{}\fi{}\ifmmode~\dot{z}\else
  \.{z}\fi{}ewski}(1979)}]{Rzazewski_1979}%
  \BibitemOpen
  \bibfield  {author} {\bibinfo {author} {\bibfnamefont {I.}~\bibnamefont
  {Bialynicki-Birula}}\ and\ \bibinfo {author} {\bibfnamefont {K.}~\bibnamefont
  {Rza\ifmmode \mbox{\c{}}\else \c{}\fi{}\ifmmode~\dot{z}\else
  \.{z}\fi{}ewski}},\ }\href {\doibase 10.1103/PhysRevA.19.301} {\bibfield
  {journal} {\bibinfo  {journal} {Phys. Rev. A}\ }\textbf {\bibinfo {volume}
  {19}},\ \bibinfo {pages} {301} (\bibinfo {year} {1979})}\BibitemShut
  {NoStop}%
\bibitem [{\citenamefont {Gaw\ifmmode~\mbox{\c{e}}\else \c{e}\fi{}dzki}\ and\
  \citenamefont {Rza\ifmmode \mbox{\c{}}\else \c{}\fi{}\ifmmode~\acute{z}\else
  \'{z}\fi{}ewski}(1981)}]{Rzazewski_1981}%
  \BibitemOpen
  \bibfield  {author} {\bibinfo {author} {\bibfnamefont {K.}~\bibnamefont
  {Gaw\ifmmode~\mbox{\c{e}}\else \c{e}\fi{}dzki}}\ and\ \bibinfo {author}
  {\bibfnamefont {K.}~\bibnamefont {Rza\ifmmode \mbox{\c{}}\else
  \c{}\fi{}\ifmmode~\acute{z}\else \'{z}\fi{}ewski}},\ }\href {\doibase
  10.1103/PhysRevA.23.2134} {\bibfield  {journal} {\bibinfo  {journal} {Phys.
  Rev. A}\ }\textbf {\bibinfo {volume} {23}},\ \bibinfo {pages} {2134}
  (\bibinfo {year} {1981})}\BibitemShut {NoStop}%
\bibitem [{\citenamefont {Emeljanov}\ and\ \citenamefont
  {Klimontovich}(1976)}]{EMELJANOV1976366}%
  \BibitemOpen
  \bibfield  {author} {\bibinfo {author} {\bibfnamefont {V.}~\bibnamefont
  {Emeljanov}}\ and\ \bibinfo {author} {\bibfnamefont {Y.}~\bibnamefont
  {Klimontovich}},\ }\href {\doibase
  https://doi.org/10.1016/0375-9601(76)90411-4} {\bibfield  {journal} {\bibinfo
   {journal} {Physics Letters A}\ }\textbf {\bibinfo {volume} {59}},\ \bibinfo
  {pages} {366 } (\bibinfo {year} {1976})}\BibitemShut {NoStop}%
\bibitem [{\citenamefont {Keeling}(2007)}]{Keeling_2007}%
  \BibitemOpen
  \bibfield  {author} {\bibinfo {author} {\bibfnamefont {J.}~\bibnamefont
  {Keeling}},\ }\href {\doibase 10.1088/0953-8984/19/29/295213} {\bibfield
  {journal} {\bibinfo  {journal} {Journal of Physics: Condensed Matter}\
  }\textbf {\bibinfo {volume} {19}},\ \bibinfo {pages} {295213} (\bibinfo
  {year} {2007})}\BibitemShut {NoStop}%
\bibitem [{\citenamefont {De~Bernardis}\ \emph {et~al.}(2018)\citenamefont
  {De~Bernardis}, \citenamefont {Jaako},\ and\ \citenamefont
  {Rabl}}]{Rabl_2018}%
  \BibitemOpen
  \bibfield  {author} {\bibinfo {author} {\bibfnamefont {D.}~\bibnamefont
  {De~Bernardis}}, \bibinfo {author} {\bibfnamefont {T.}~\bibnamefont {Jaako}},
  \ and\ \bibinfo {author} {\bibfnamefont {P.}~\bibnamefont {Rabl}},\ }\href
  {\doibase 10.1103/PhysRevA.97.043820} {\bibfield  {journal} {\bibinfo
  {journal} {Phys. Rev. A}\ }\textbf {\bibinfo {volume} {97}},\ \bibinfo
  {pages} {043820} (\bibinfo {year} {2018})}\BibitemShut {NoStop}%
\bibitem [{\citenamefont {Pellegrino}\ \emph {et~al.}(2016)\citenamefont
  {Pellegrino}, \citenamefont {Giovannetti}, \citenamefont {MacDonald},\ and\
  \citenamefont {Polini}}]{Pellegrino_2016}%
  \BibitemOpen
  \bibfield  {author} {\bibinfo {author} {\bibfnamefont {F.~M.~D.}\
  \bibnamefont {Pellegrino}}, \bibinfo {author} {\bibfnamefont
  {V.}~\bibnamefont {Giovannetti}}, \bibinfo {author} {\bibfnamefont {A.~H.}\
  \bibnamefont {MacDonald}}, \ and\ \bibinfo {author} {\bibfnamefont
  {M.}~\bibnamefont {Polini}},\ }\href {\doibase 10.1038/ncomms13355}
  {\bibfield  {journal} {\bibinfo  {journal} {Nature Communications}\ }\textbf
  {\bibinfo {volume} {7}},\ \bibinfo {pages} {13355} (\bibinfo {year}
  {2016})}\BibitemShut {NoStop}%
\bibitem [{\citenamefont {Nataf}\ and\ \citenamefont
  {Ciuti}(2010)}]{Nataf_2010}%
  \BibitemOpen
  \bibfield  {author} {\bibinfo {author} {\bibfnamefont {P.}~\bibnamefont
  {Nataf}}\ and\ \bibinfo {author} {\bibfnamefont {C.}~\bibnamefont {Ciuti}},\
  }\href {\doibase 10.1038/ncomms1069} {\bibfield  {journal} {\bibinfo
  {journal} {Nature Communications}\ }\textbf {\bibinfo {volume} {1}},\
  \bibinfo {pages} {72} (\bibinfo {year} {2010})}\BibitemShut {NoStop}%
\bibitem [{\citenamefont {Viehmann}\ \emph {et~al.}(2011)\citenamefont
  {Viehmann}, \citenamefont {von Delft},\ and\ \citenamefont
  {Marquardt}}]{Marquardt_2011}%
  \BibitemOpen
  \bibfield  {author} {\bibinfo {author} {\bibfnamefont {O.}~\bibnamefont
  {Viehmann}}, \bibinfo {author} {\bibfnamefont {J.}~\bibnamefont {von Delft}},
  \ and\ \bibinfo {author} {\bibfnamefont {F.}~\bibnamefont {Marquardt}},\
  }\href {\doibase 10.1103/PhysRevLett.107.113602} {\bibfield  {journal}
  {\bibinfo  {journal} {Phys. Rev. Lett.}\ }\textbf {\bibinfo {volume} {107}},\
  \bibinfo {pages} {113602} (\bibinfo {year} {2011})}\BibitemShut {NoStop}%
\bibitem [{\citenamefont {Ciuti}\ and\ \citenamefont
  {Nataf}(2012)}]{Ciuti_2012}%
  \BibitemOpen
  \bibfield  {author} {\bibinfo {author} {\bibfnamefont {C.}~\bibnamefont
  {Ciuti}}\ and\ \bibinfo {author} {\bibfnamefont {P.}~\bibnamefont {Nataf}},\
  }\href {\doibase 10.1103/PhysRevLett.109.179301} {\bibfield  {journal}
  {\bibinfo  {journal} {Phys. Rev. Lett.}\ }\textbf {\bibinfo {volume} {109}},\
  \bibinfo {pages} {179301} (\bibinfo {year} {2012})}\BibitemShut {NoStop}%
\bibitem [{\citenamefont {Hagenm\"uller}\ and\ \citenamefont
  {Ciuti}(2012)}]{Hagenmuller_2012}%
  \BibitemOpen
  \bibfield  {author} {\bibinfo {author} {\bibfnamefont {D.}~\bibnamefont
  {Hagenm\"uller}}\ and\ \bibinfo {author} {\bibfnamefont {C.}~\bibnamefont
  {Ciuti}},\ }\href {\doibase 10.1103/PhysRevLett.109.267403} {\bibfield
  {journal} {\bibinfo  {journal} {Phys. Rev. Lett.}\ }\textbf {\bibinfo
  {volume} {109}},\ \bibinfo {pages} {267403} (\bibinfo {year}
  {2012})}\BibitemShut {NoStop}%
\bibitem [{\citenamefont {Hayn}\ \emph {et~al.}(2012)\citenamefont {Hayn},
  \citenamefont {Emary},\ and\ \citenamefont {Brandes}}]{Brandes_2012}%
  \BibitemOpen
  \bibfield  {author} {\bibinfo {author} {\bibfnamefont {M.}~\bibnamefont
  {Hayn}}, \bibinfo {author} {\bibfnamefont {C.}~\bibnamefont {Emary}}, \ and\
  \bibinfo {author} {\bibfnamefont {T.}~\bibnamefont {Brandes}},\ }\href
  {\doibase 10.1103/PhysRevA.86.063822} {\bibfield  {journal} {\bibinfo
  {journal} {Phys. Rev. A}\ }\textbf {\bibinfo {volume} {86}},\ \bibinfo
  {pages} {063822} (\bibinfo {year} {2012})}\BibitemShut {NoStop}%
\bibitem [{\citenamefont {Chirolli}\ \emph {et~al.}(2012)\citenamefont
  {Chirolli}, \citenamefont {Polini}, \citenamefont {Giovannetti},\ and\
  \citenamefont {MacDonald}}]{Chirolli_2012}%
  \BibitemOpen
  \bibfield  {author} {\bibinfo {author} {\bibfnamefont {L.}~\bibnamefont
  {Chirolli}}, \bibinfo {author} {\bibfnamefont {M.}~\bibnamefont {Polini}},
  \bibinfo {author} {\bibfnamefont {V.}~\bibnamefont {Giovannetti}}, \ and\
  \bibinfo {author} {\bibfnamefont {A.~H.}\ \bibnamefont {MacDonald}},\ }\href
  {\doibase 10.1103/PhysRevLett.109.267404} {\bibfield  {journal} {\bibinfo
  {journal} {Phys. Rev. Lett.}\ }\textbf {\bibinfo {volume} {109}},\ \bibinfo
  {pages} {267404} (\bibinfo {year} {2012})}\BibitemShut {NoStop}%
\bibitem [{\citenamefont {Bamba}\ and\ \citenamefont
  {Ogawa}(2014)}]{Bamba_2014}%
  \BibitemOpen
  \bibfield  {author} {\bibinfo {author} {\bibfnamefont {M.}~\bibnamefont
  {Bamba}}\ and\ \bibinfo {author} {\bibfnamefont {T.}~\bibnamefont {Ogawa}},\
  }\href {\doibase 10.1103/PhysRevA.90.063825} {\bibfield  {journal} {\bibinfo
  {journal} {Phys. Rev. A}\ }\textbf {\bibinfo {volume} {90}},\ \bibinfo
  {pages} {063825} (\bibinfo {year} {2014})}\BibitemShut {NoStop}%
\bibitem [{\citenamefont {Andolina}\ \emph {et~al.}(2019)\citenamefont
  {Andolina}, \citenamefont {Pellegrino}, \citenamefont {Giovannetti},
  \citenamefont {MacDonald},\ and\ \citenamefont {Polini}}]{Andolina_2019}%
  \BibitemOpen
  \bibfield  {author} {\bibinfo {author} {\bibfnamefont {G.~M.}\ \bibnamefont
  {Andolina}}, \bibinfo {author} {\bibfnamefont {F.~M.~D.}\ \bibnamefont
  {Pellegrino}}, \bibinfo {author} {\bibfnamefont {V.}~\bibnamefont
  {Giovannetti}}, \bibinfo {author} {\bibfnamefont {A.~H.}\ \bibnamefont
  {MacDonald}}, \ and\ \bibinfo {author} {\bibfnamefont {M.}~\bibnamefont
  {Polini}},\ }\href {\doibase 10.1103/PhysRevB.100.121109} {\bibfield
  {journal} {\bibinfo  {journal} {Phys. Rev. B}\ }\textbf {\bibinfo {volume}
  {100}},\ \bibinfo {pages} {121109} (\bibinfo {year} {2019})}\BibitemShut
  {NoStop}%
\bibitem [{\citenamefont {Nataf}\ \emph {et~al.}(2019)\citenamefont {Nataf},
  \citenamefont {Champel}, \citenamefont {Blatter},\ and\ \citenamefont
  {Basko}}]{Basko_PRL2019}%
  \BibitemOpen
  \bibfield  {author} {\bibinfo {author} {\bibfnamefont {P.}~\bibnamefont
  {Nataf}}, \bibinfo {author} {\bibfnamefont {T.}~\bibnamefont {Champel}},
  \bibinfo {author} {\bibfnamefont {G.}~\bibnamefont {Blatter}}, \ and\
  \bibinfo {author} {\bibfnamefont {D.~M.}\ \bibnamefont {Basko}},\ }\href
  {\doibase 10.1103/PhysRevLett.123.207402} {\bibfield  {journal} {\bibinfo
  {journal} {Phys. Rev. Lett.}\ }\textbf {\bibinfo {volume} {123}},\ \bibinfo
  {pages} {207402} (\bibinfo {year} {2019})}\BibitemShut {NoStop}%
\bibitem [{Note1()}]{Note1}%
  \BibitemOpen
  \bibinfo {note} {Even in the Coulomb gauge considered in this work, the
  projection of the 3D cavity field to a 2D material may have a longitudinal
  component.}\BibitemShut {Stop}%
\bibitem [{\citenamefont {Affleck}\ and\ \citenamefont
  {Marston}(1988)}]{affleck1988}%
  \BibitemOpen
  \bibfield  {author} {\bibinfo {author} {\bibfnamefont {I.}~\bibnamefont
  {Affleck}}\ and\ \bibinfo {author} {\bibfnamefont {J.~B.}\ \bibnamefont
  {Marston}},\ }\href {\doibase 10.1103/PhysRevB.37.3774} {\bibfield  {journal}
  {\bibinfo  {journal} {Phys. Rev. B}\ }\textbf {\bibinfo {volume} {37}},\
  \bibinfo {pages} {3774} (\bibinfo {year} {1988})}\BibitemShut {NoStop}%
\bibitem [{\citenamefont {Kim}\ \emph {et~al.}(2015)\citenamefont {Kim},
  \citenamefont {Wieder}, \citenamefont {Kane},\ and\ \citenamefont
  {Rappe}}]{kim2015}%
  \BibitemOpen
  \bibfield  {author} {\bibinfo {author} {\bibfnamefont {Y.}~\bibnamefont
  {Kim}}, \bibinfo {author} {\bibfnamefont {B.~J.}\ \bibnamefont {Wieder}},
  \bibinfo {author} {\bibfnamefont {C.~L.}\ \bibnamefont {Kane}}, \ and\
  \bibinfo {author} {\bibfnamefont {A.~M.}\ \bibnamefont {Rappe}},\ }\href
  {\doibase 10.1103/PhysRevLett.115.036806} {\bibfield  {journal} {\bibinfo
  {journal} {Phys. Rev. Lett.}\ }\textbf {\bibinfo {volume} {115}},\ \bibinfo
  {pages} {036806} (\bibinfo {year} {2015})}\BibitemShut {NoStop}%
\bibitem [{\citenamefont {Ahn}\ \emph {et~al.}(2019)\citenamefont {Ahn},
  \citenamefont {Park},\ and\ \citenamefont {Yang}}]{ahn2019}%
  \BibitemOpen
  \bibfield  {author} {\bibinfo {author} {\bibfnamefont {J.}~\bibnamefont
  {Ahn}}, \bibinfo {author} {\bibfnamefont {S.}~\bibnamefont {Park}}, \ and\
  \bibinfo {author} {\bibfnamefont {B.-J.}\ \bibnamefont {Yang}},\ }\href
  {\doibase 10.1103/PhysRevX.9.021013} {\bibfield  {journal} {\bibinfo
  {journal} {Phys. Rev. X}\ }\textbf {\bibinfo {volume} {9}},\ \bibinfo {pages}
  {021013} (\bibinfo {year} {2019})}\BibitemShut {NoStop}%
\bibitem [{\citenamefont {Mora}\ \emph {et~al.}(2019)\citenamefont {Mora},
  \citenamefont {Regnault},\ and\ \citenamefont {Bernevig}}]{mora2019}%
  \BibitemOpen
  \bibfield  {author} {\bibinfo {author} {\bibfnamefont {C.}~\bibnamefont
  {Mora}}, \bibinfo {author} {\bibfnamefont {N.}~\bibnamefont {Regnault}}, \
  and\ \bibinfo {author} {\bibfnamefont {B.~A.}\ \bibnamefont {Bernevig}},\
  }\href {\doibase 10.1103/PhysRevLett.123.026402} {\bibfield  {journal}
  {\bibinfo  {journal} {Phys. Rev. Lett.}\ }\textbf {\bibinfo {volume} {123}},\
  \bibinfo {pages} {026402} (\bibinfo {year} {2019})}\BibitemShut {NoStop}%
\bibitem [{\citenamefont {Ryu}\ and\ \citenamefont {Hatsugai}(2002)}]{ruy2002}%
  \BibitemOpen
  \bibfield  {author} {\bibinfo {author} {\bibfnamefont {S.}~\bibnamefont
  {Ryu}}\ and\ \bibinfo {author} {\bibfnamefont {Y.}~\bibnamefont {Hatsugai}},\
  }\href {\doibase 10.1103/PhysRevLett.89.077002} {\bibfield  {journal}
  {\bibinfo  {journal} {Phys. Rev. Lett.}\ }\textbf {\bibinfo {volume} {89}},\
  \bibinfo {pages} {077002} (\bibinfo {year} {2002})}\BibitemShut {NoStop}%
\bibitem [{\citenamefont {Delplace}\ \emph {et~al.}(2011)\citenamefont
  {Delplace}, \citenamefont {Ullmo},\ and\ \citenamefont
  {Montambaux}}]{delplace2011}%
  \BibitemOpen
  \bibfield  {author} {\bibinfo {author} {\bibfnamefont {P.}~\bibnamefont
  {Delplace}}, \bibinfo {author} {\bibfnamefont {D.}~\bibnamefont {Ullmo}}, \
  and\ \bibinfo {author} {\bibfnamefont {G.}~\bibnamefont {Montambaux}},\
  }\href {\doibase 10.1103/PhysRevB.84.195452} {\bibfield  {journal} {\bibinfo
  {journal} {Phys. Rev. B}\ }\textbf {\bibinfo {volume} {84}},\ \bibinfo
  {pages} {195452} (\bibinfo {year} {2011})}\BibitemShut {NoStop}%
\bibitem [{\citenamefont {Akhmerov}\ and\ \citenamefont
  {Beenakker}(2008)}]{akhmerov2008}%
  \BibitemOpen
  \bibfield  {author} {\bibinfo {author} {\bibfnamefont {A.~R.}\ \bibnamefont
  {Akhmerov}}\ and\ \bibinfo {author} {\bibfnamefont {C.~W.~J.}\ \bibnamefont
  {Beenakker}},\ }\href {\doibase 10.1103/PhysRevB.77.085423} {\bibfield
  {journal} {\bibinfo  {journal} {Phys. Rev. B}\ }\textbf {\bibinfo {volume}
  {77}},\ \bibinfo {pages} {085423} (\bibinfo {year} {2008})}\BibitemShut
  {NoStop}%
\bibitem [{\citenamefont {Bena}\ and\ \citenamefont {Simon}(2011)}]{bena2011}%
  \BibitemOpen
  \bibfield  {author} {\bibinfo {author} {\bibfnamefont {C.}~\bibnamefont
  {Bena}}\ and\ \bibinfo {author} {\bibfnamefont {L.}~\bibnamefont {Simon}},\
  }\href {\doibase 10.1103/PhysRevB.83.115404} {\bibfield  {journal} {\bibinfo
  {journal} {Phys. Rev. B}\ }\textbf {\bibinfo {volume} {83}},\ \bibinfo
  {pages} {115404} (\bibinfo {year} {2011})}\BibitemShut {NoStop}%
\bibitem [{\citenamefont {Montambaux}(2012)}]{montambaux2012}%
  \BibitemOpen
  \bibfield  {author} {\bibinfo {author} {\bibfnamefont {G.}~\bibnamefont
  {Montambaux}},\ }\href@noop {} {\bibfield  {journal} {\bibinfo  {journal}
  {Eur. Phys. J. B.}\ }\textbf {\bibinfo {volume} {85}},\ \bibinfo {pages}
  {375} (\bibinfo {year} {2012})}\BibitemShut {NoStop}%
\bibitem [{Note2()}]{Note2}%
  \BibitemOpen
  \bibinfo {note} {The chiral symmetry is diagonal and takes +1(-1) values on
  the $A$ ($B$) sublattice of the hexagonal lattice.}\BibitemShut {Stop}%
\bibitem [{\citenamefont {Bistritzer}\ and\ \citenamefont
  {MacDonald}(2011)}]{MacDonald_M-Model}%
  \BibitemOpen
  \bibfield  {author} {\bibinfo {author} {\bibfnamefont {R.}~\bibnamefont
  {Bistritzer}}\ and\ \bibinfo {author} {\bibfnamefont {A.~H.}\ \bibnamefont
  {MacDonald}},\ }\href@noop {} {\bibfield  {journal} {\bibinfo  {journal}
  {Proceedings of the National Academy of Sciences}\ }\textbf {\bibinfo
  {volume} {108}},\ \bibinfo {pages} {12233} (\bibinfo {year}
  {2011})}\BibitemShut {NoStop}%
\bibitem [{\citenamefont {Cao}\ \emph {et~al.}(2018{\natexlab{a}})\citenamefont
  {Cao}, \citenamefont {Fatemi}, \citenamefont {Demir}, \citenamefont {Fang},
  \citenamefont {Tomarken}, \citenamefont {Luo}, \citenamefont
  {Sanchez-Yamagishi}, \citenamefont {Watanabe}, \citenamefont {Taniguchi},
  \citenamefont {Kaxiras} \emph {et~al.}}]{cao_TBG1}%
  \BibitemOpen
  \bibfield  {author} {\bibinfo {author} {\bibfnamefont {Y.}~\bibnamefont
  {Cao}}, \bibinfo {author} {\bibfnamefont {V.}~\bibnamefont {Fatemi}},
  \bibinfo {author} {\bibfnamefont {A.}~\bibnamefont {Demir}}, \bibinfo
  {author} {\bibfnamefont {S.}~\bibnamefont {Fang}}, \bibinfo {author}
  {\bibfnamefont {S.~L.}\ \bibnamefont {Tomarken}}, \bibinfo {author}
  {\bibfnamefont {J.~Y.}\ \bibnamefont {Luo}}, \bibinfo {author} {\bibfnamefont
  {J.~D.}\ \bibnamefont {Sanchez-Yamagishi}}, \bibinfo {author} {\bibfnamefont
  {K.}~\bibnamefont {Watanabe}}, \bibinfo {author} {\bibfnamefont
  {T.}~\bibnamefont {Taniguchi}}, \bibinfo {author} {\bibfnamefont
  {E.}~\bibnamefont {Kaxiras}},  \emph {et~al.},\ }\href@noop {} {\bibfield
  {journal} {\bibinfo  {journal} {Nature}\ }\textbf {\bibinfo {volume} {556}},\
  \bibinfo {pages} {80} (\bibinfo {year} {2018}{\natexlab{a}})}\BibitemShut
  {NoStop}%
\bibitem [{\citenamefont {Cao}\ \emph {et~al.}(2018{\natexlab{b}})\citenamefont
  {Cao}, \citenamefont {Fatemi}, \citenamefont {Fang}, \citenamefont
  {Watanabe}, \citenamefont {Taniguchi}, \citenamefont {Kaxiras},\ and\
  \citenamefont {Jarillo-Herrero}}]{cao_TBG2}%
  \BibitemOpen
  \bibfield  {author} {\bibinfo {author} {\bibfnamefont {Y.}~\bibnamefont
  {Cao}}, \bibinfo {author} {\bibfnamefont {V.}~\bibnamefont {Fatemi}},
  \bibinfo {author} {\bibfnamefont {S.}~\bibnamefont {Fang}}, \bibinfo {author}
  {\bibfnamefont {K.}~\bibnamefont {Watanabe}}, \bibinfo {author}
  {\bibfnamefont {T.}~\bibnamefont {Taniguchi}}, \bibinfo {author}
  {\bibfnamefont {E.}~\bibnamefont {Kaxiras}}, \ and\ \bibinfo {author}
  {\bibfnamefont {P.}~\bibnamefont {Jarillo-Herrero}},\ }\href@noop {}
  {\bibfield  {journal} {\bibinfo  {journal} {Nature}\ }\textbf {\bibinfo
  {volume} {556}},\ \bibinfo {pages} {43} (\bibinfo {year}
  {2018}{\natexlab{b}})}\BibitemShut {NoStop}%
\bibitem [{\citenamefont {Song}\ \emph {et~al.}(2019)\citenamefont {Song},
  \citenamefont {Wang}, \citenamefont {Shi}, \citenamefont {Li}, \citenamefont
  {Fang},\ and\ \citenamefont {Bernevig}}]{song2019}%
  \BibitemOpen
  \bibfield  {author} {\bibinfo {author} {\bibfnamefont {Z.}~\bibnamefont
  {Song}}, \bibinfo {author} {\bibfnamefont {Z.}~\bibnamefont {Wang}}, \bibinfo
  {author} {\bibfnamefont {W.}~\bibnamefont {Shi}}, \bibinfo {author}
  {\bibfnamefont {G.}~\bibnamefont {Li}}, \bibinfo {author} {\bibfnamefont
  {C.}~\bibnamefont {Fang}}, \ and\ \bibinfo {author} {\bibfnamefont {B.~A.}\
  \bibnamefont {Bernevig}},\ }\href {\doibase 10.1103/PhysRevLett.123.036401}
  {\bibfield  {journal} {\bibinfo  {journal} {Phys. Rev. Lett.}\ }\textbf
  {\bibinfo {volume} {123}},\ \bibinfo {pages} {036401} (\bibinfo {year}
  {2019})}\BibitemShut {NoStop}%
\bibitem [{\citenamefont {Hejazi}\ \emph {et~al.}(2019)\citenamefont {Hejazi},
  \citenamefont {Liu}, \citenamefont {Shapourian}, \citenamefont {Chen},\ and\
  \citenamefont {Balents}}]{hejazi2019}%
  \BibitemOpen
  \bibfield  {author} {\bibinfo {author} {\bibfnamefont {K.}~\bibnamefont
  {Hejazi}}, \bibinfo {author} {\bibfnamefont {C.}~\bibnamefont {Liu}},
  \bibinfo {author} {\bibfnamefont {H.}~\bibnamefont {Shapourian}}, \bibinfo
  {author} {\bibfnamefont {X.}~\bibnamefont {Chen}}, \ and\ \bibinfo {author}
  {\bibfnamefont {L.}~\bibnamefont {Balents}},\ }\href {\doibase
  10.1103/PhysRevB.99.035111} {\bibfield  {journal} {\bibinfo  {journal} {Phys.
  Rev. B}\ }\textbf {\bibinfo {volume} {99}},\ \bibinfo {pages} {035111}
  (\bibinfo {year} {2019})}\BibitemShut {NoStop}%
\bibitem [{\citenamefont {Andolina}\ \emph {et~al.}(2020)\citenamefont
  {Andolina}, \citenamefont {Pellegrino}, \citenamefont {Giovannetti},
  \citenamefont {MacDonald},\ and\ \citenamefont
  {Polini}}]{Andolina_preprint_2020}%
  \BibitemOpen
  \bibfield  {author} {\bibinfo {author} {\bibfnamefont {G.~M.}\ \bibnamefont
  {Andolina}}, \bibinfo {author} {\bibfnamefont {F.~M.~D.}\ \bibnamefont
  {Pellegrino}}, \bibinfo {author} {\bibfnamefont {V.}~\bibnamefont
  {Giovannetti}}, \bibinfo {author} {\bibfnamefont {A.~H.}\ \bibnamefont
  {MacDonald}}, \ and\ \bibinfo {author} {\bibfnamefont {M.}~\bibnamefont
  {Polini}},\ }\href@noop {} {\enquote {\bibinfo {title} {Theory of photon
  condensation in a spatially-varying electromagnetic field},}\ } (\bibinfo
  {year} {2020}),\ \bibinfo {note} {arXiv:2005:xxxxx}\BibitemShut {NoStop}%
\end{thebibliography}%


\begin{thebibliography}{12}
\expandafter\ifx\csname natexlab\endcsname\relax\def\natexlab#1{#1}\fi
\expandafter\ifx\csname bibnamefont\endcsname\relax
  \def\bibnamefont#1{#1}\fi
\expandafter\ifx\csname bibfnamefont\endcsname\relax
  \def\bibfnamefont#1{#1}\fi
\expandafter\ifx\csname citenamefont\endcsname\relax
  \def\citenamefont#1{#1}\fi
\expandafter\ifx\csname url\endcsname\relax
  \def\url#1{\texttt{#1}}\fi
\expandafter\ifx\csname urlprefix\endcsname\relax\def\urlprefix{URL }\fi
\providecommand{\bibinfo}[2]{#2}
\providecommand{\eprint}[2][]{\url{#2}}

\bibitem[{\citenamefont{Cohen-Tannoudji
  et~al.}(1989)\citenamefont{Cohen-Tannoudji, Dupont-Roc, and
  Grynberg}}]{Cohen_1989}
\bibinfo{author}{\bibfnamefont{C.}~\bibnamefont{Cohen-Tannoudji}},
  \bibinfo{author}{\bibfnamefont{J.}~\bibnamefont{Dupont-Roc}},
  \bibnamefont{and} \bibinfo{author}{\bibfnamefont{G.}~\bibnamefont{Grynberg}},
  \emph{\bibinfo{title}{{Photons and atoms: introduction to quantum
  electrodynamics}}} (\bibinfo{publisher}{Wiley}, \bibinfo{address}{New York,
  NY}, \bibinfo{year}{1989}), \bibinfo{note}{trans. of : Photons et atomes.
  Paris, InterEditions, 1987},
  \urlprefix\url{https://cds.cern.ch/record/113864}.

\bibitem[{\citenamefont{Jackson}(1975)}]{Jackson_1975}
\bibinfo{author}{\bibfnamefont{J.~D.} \bibnamefont{Jackson}},
  \emph{\bibinfo{title}{{Classical electrodynamics; 2nd ed.}}}
  (\bibinfo{publisher}{Wiley}, \bibinfo{address}{New York, NY},
  \bibinfo{year}{1975}), \urlprefix\url{https://cds.cern.ch/record/100964}.

\bibitem[{\citenamefont{Kakazu and Kim}(1994)}]{Kakazu_1994}
\bibinfo{author}{\bibfnamefont{K.}~\bibnamefont{Kakazu}} \bibnamefont{and}
  \bibinfo{author}{\bibfnamefont{Y.~S.} \bibnamefont{Kim}},
  \bibinfo{journal}{Phys. Rev. A} \textbf{\bibinfo{volume}{50}},
  \bibinfo{pages}{1830} (\bibinfo{year}{1994}),
  \urlprefix\url{https://link.aps.org/doi/10.1103/PhysRevA.50.1830}.

\bibitem[{\citenamefont{Peierls}(1933)}]{Peierls_1938}
\bibinfo{author}{\bibfnamefont{R.}~\bibnamefont{Peierls}},
  \bibinfo{journal}{Zeitschrift f{\"u}r Physik} \textbf{\bibinfo{volume}{80}},
  \bibinfo{pages}{763} (\bibinfo{year}{1933}),
  \urlprefix\url{https://doi.org/10.1007/BF01342591}.

\bibitem[{\citenamefont{Mahan}(2000)}]{Mahan_2000}
\bibinfo{author}{\bibfnamefont{G.~D.} \bibnamefont{Mahan}},
  \emph{\bibinfo{title}{Many Particle Physics, Third Edition}}
  (\bibinfo{publisher}{Plenum}, \bibinfo{address}{New York},
  \bibinfo{year}{2000}).

\bibitem[{\citenamefont{Giuliani and Vignale}(2005)}]{Giuliani_book}
\bibinfo{author}{\bibfnamefont{G.~F.} \bibnamefont{Giuliani}} \bibnamefont{and}
  \bibinfo{author}{\bibfnamefont{G.}~\bibnamefont{Vignale}},
  \emph{\bibinfo{title}{{Quantum theory of the electron liquid}}}
  (\bibinfo{publisher}{Cambridge Univ. Press}, \bibinfo{address}{Cambridge},
  \bibinfo{year}{2005}), \urlprefix\url{https://cds.cern.ch/record/826125}.

\bibitem[{\citenamefont{Pines and Nozi{\`e}res}(1966)}]{Pines_1966}
\bibinfo{author}{\bibfnamefont{D.}~\bibnamefont{Pines}} \bibnamefont{and}
  \bibinfo{author}{\bibfnamefont{P.}~\bibnamefont{Nozi{\`e}res}},
  \emph{\bibinfo{title}{The Theory of Quantum Liquids: Normal Fermi liquids}},
  Advanced book classics (\bibinfo{publisher}{W.A. Benjamin},
  \bibinfo{year}{1966}),
  \urlprefix\url{https://books.google.it/books?id=GP1QAAAAMAAJ}.

\bibitem[{\citenamefont{Andolina et~al.}(2019)\citenamefont{Andolina,
  Pellegrino, Giovannetti, MacDonald, and Polini}}]{Andolina_2019}
\bibinfo{author}{\bibfnamefont{G.~M.} \bibnamefont{Andolina}},
  \bibinfo{author}{\bibfnamefont{F.~M.~D.} \bibnamefont{Pellegrino}},
  \bibinfo{author}{\bibfnamefont{V.}~\bibnamefont{Giovannetti}},
  \bibinfo{author}{\bibfnamefont{A.~H.} \bibnamefont{MacDonald}},
  \bibnamefont{and} \bibinfo{author}{\bibfnamefont{M.}~\bibnamefont{Polini}},
  \bibinfo{journal}{Phys. Rev. B} \textbf{\bibinfo{volume}{100}},
  \bibinfo{pages}{121109} (\bibinfo{year}{2019}),
  \urlprefix\url{https://link.aps.org/doi/10.1103/PhysRevB.100.121109}.

\bibitem[{\citenamefont{Nataf et~al.}(2019)\citenamefont{Nataf, Champel,
  Blatter, and Basko}}]{Basko_PRL2019}
\bibinfo{author}{\bibfnamefont{P.}~\bibnamefont{Nataf}},
  \bibinfo{author}{\bibfnamefont{T.}~\bibnamefont{Champel}},
  \bibinfo{author}{\bibfnamefont{G.}~\bibnamefont{Blatter}}, \bibnamefont{and}
  \bibinfo{author}{\bibfnamefont{D.~M.} \bibnamefont{Basko}},
  \bibinfo{journal}{Phys. Rev. Lett.} \textbf{\bibinfo{volume}{123}},
  \bibinfo{pages}{207402} (\bibinfo{year}{2019}),
  \urlprefix\url{https://link.aps.org/doi/10.1103/PhysRevLett.123.207402}.

\bibitem[{\citenamefont{Bernevig and Hughes}(2013)}]{Bernevig_book}
\bibinfo{author}{\bibfnamefont{B.~A.} \bibnamefont{Bernevig}} \bibnamefont{and}
  \bibinfo{author}{\bibfnamefont{T.~L.} \bibnamefont{Hughes}},
  \emph{\bibinfo{title}{Topological Insulators and Topological
  Superconductors}} (\bibinfo{publisher}{Princeton University Press},
  \bibinfo{year}{2013}), ISBN \bibinfo{isbn}{9780691151755},
  \urlprefix\url{http://www.jstor.org/stable/j.ctt19cc2gc}.

\bibitem[{\citenamefont{Bena and Simon}(2011)}]{bena2011}
\bibinfo{author}{\bibfnamefont{C.}~\bibnamefont{Bena}} \bibnamefont{and}
  \bibinfo{author}{\bibfnamefont{L.}~\bibnamefont{Simon}},
  \bibinfo{journal}{Phys. Rev. B} \textbf{\bibinfo{volume}{83}},
  \bibinfo{pages}{115404} (\bibinfo{year}{2011}),
  \urlprefix\url{https://link.aps.org/doi/10.1103/PhysRevB.83.115404}.

\bibitem[{\citenamefont{Montambaux}(2012)}]{montambaux2012}
\bibinfo{author}{\bibfnamefont{G.}~\bibnamefont{Montambaux}},
  \bibinfo{journal}{Eur. Phys. J. B.} \textbf{\bibinfo{volume}{85}},
  \bibinfo{pages}{375} (\bibinfo{year}{2012}).

\end{thebibliography}

\end{document}


\title{Supplementary Material: Superradiant phase transition in electronic systems and emergent topological phases}
\author{Daniele Guerci}
\affiliation{Universit{\'e} de Paris, Laboratoire Mat{\'e}riaux et Ph{\'e}nom{\`e}nes Quantiques, CNRS, F-75013 Paris, France} 
\affiliation{Universit{\'e} Paris-Saclay, CNRS, Laboratoire de Physique des Solides, 91405, Orsay, France.} 
\author{Pascal Simon}
\affiliation{Universit{\'e} Paris-Saclay, CNRS, Laboratoire de Physique des Solides, 91405, Orsay, France.} 
\author{Christophe Mora}
\affiliation{Universit{\'e} de Paris, Laboratoire Mat{\'e}riaux et Ph{\'e}nom{\`e}nes Quantiques, CNRS, F-75013 Paris, France}

\date{\today} 
\pacs{}

\maketitle

\section{The light-matter interaction for tight-binding electron models}
\label{sec1}

In this section we derive the light-matter interaction for tight-binding electron models. 
The analysis introduces a general expression for the paramagnetic current and diamagnetic tensor operators. 
We begin by recalling the kinetic part of the electronic Hamiltonian: 
\be
\label{model_tb_paper}
H_0=-\sum_{\bm{j}}\sum_{\alpha\beta}\sum_{\bm{\delta}}c^\dagger_{\bm{R_{j,\alpha}}}\,t^{\bm{\delta}}_{\alpha\beta}\,c^\dagga_{\bm{R_{j,\beta}+\delta}},
\ee
where $\bm{R_{j,\alpha}}=\bm{R_j}+\bm{v}_{\alpha},\,\,\bm{R_{j,\beta}}+\bm{\delta}=\bm{R_j}+\bm{v}_{\beta}+\bm{\delta}$, $\bm{v}_{\alpha}$ and $\bm{v}_{\beta}$ are basis vectors, while $\bm{\delta}$ connects different unit cells.
In Fourier space $H_0$ \eqn{model_tb_paper} becomes:
\be
H_0=\sum_{\bk}\sum_{\alpha\beta}\,c^\dagger_{\bk,\alpha}\, h_{\alpha\beta}(\bk)\,\,c^\dagga_{\bk,\beta}
\ee
where
\be
\label{model_h_paper}
h_{\alpha\beta}(\bk) = -\sum_{\bm{\delta}}\,t^{\bm{\delta}}_{\alpha\beta}\,e^{i\bk\cdot(\bm{R_{j,\beta}}+\bm{\delta}-\bm{R_{j,\alpha}})},
\ee
the sum is extended over the lattice vectors $\bm{\delta}$ connecting different unit cells. 
Before going on it is convenient to define the vector connecting sites $\bm{R_{j,\alpha}}$ and $\bm{R_{j,\beta}}+\bm{\delta}$, $\bm{\Delta R^{\bm{\delta}}_{\alpha,\beta}}=\bm{v}_{\beta}-\bm{v}_{\alpha}+\bm{\delta}$.
We shall work in the Coulomb gauge $\nabla\cdot\hat{\bm{A}}(\br)=0$, so that the three-dimensional cavity field is quantized in terms of transverse modes \cite{Cohen_1989,Jackson_1975,Kakazu_1994}. 
For lattice models the equivalent of the minimal coupling prescription is the so-called $Peierls$ substitution \cite{Peierls_1938,Mahan_2000}:
\be
\label{Peierls_ansatz}
t^{\bm{\delta}}_{\alpha\beta}\to t^{\bm{\delta}}_{\alpha\beta}\,e^{ie\lambda/c},
\ee
where
\bal
\label{path_phase}
\lambda&=\int_{\bm{R_{j,\alpha}}}^{\bm{R_{j,\beta}+\delta}}\,d\bm{l}\cdot\hat{\bm{A}}(\bm{l})=2\sum_{\bq}\sin\left(\frac{\bq\cdot\bm{\Delta R^{\bm{\delta}}_{\alpha,\beta}}}{2}\right)\frac{\bm{\Delta R^{\bm{\delta}}_{\alpha,\beta}}\cdot\hat{\bm{A}}(\bq)}{\bm{\Delta R^{\bm{\delta}}_{\alpha,\beta}}\cdot\bq}\, e^{i\bq\cdot(\bm{R_{j,\alpha}}+\bm{R_{j,\beta}+\delta})/2}.
\eal
By introducing the Fourier series in the right hand side of the previous equation 
\be
\hat{\bm{A}}(\bm{l})=\sum_{\bq}\,e^{i\bq\cdot\bm{l}}\hat{\bm{A}}(\bq),
\ee 
we readily find
\be
\lambda = 2\sum_{\bq}\sin\left(\frac{\bq\cdot\bm{\Delta R^{\bm{\delta}}_{\alpha,\beta}}}{2}\right)\frac{\bm{\Delta R^{\bm{\delta}}_{\alpha,\beta}}\cdot\hat{\bm{A}}(\bq)}{\bm{\Delta R^{\bm{\delta}}_{\alpha,\beta}}\cdot\bq}\, e^{i\bq\cdot(\bm{R_{j,\alpha}}+\bm{R_{j,\beta}+\delta})/2}.
\ee
Expanding the phase factor \eqn{Peierls_ansatz} up to terms of order $\hat{\bm{A}}^2$ one obtains
\be
\label{light_matter}
H_0(\hat{\bm{A}})=H_0+H_A+H_{A^2},
\ee  
where
\be
\label{HA_linear}
H_A = \frac{e}{c}\sum_{\bq}\,\hat{\bm{A}}(\bq)\cdot\bm{J}_p(-\bq),\quad
H_{A^2} = -\frac{e^2}{2c^2}\sum_{\bq_1,i}\sum_{\bq_2,j}\,\hat{A}_{i}(\bq_1)\,\mathcal{T}^{i,j}(-\bq_1,-\bq_2)\,\hat{A}_{j}(\bq_2).
\ee    
In Eq. \eqn{HA_linear} we have introduced the paramagnetic current operator:
\be
\bm{J}_p(\bq)=\sum_{\bp}\,\sum_{\alpha\beta}c^\dagger_{\bp-\bq/2,\,\alpha}\,\bm{j}_{\alpha\beta}(\bp,\bq)\,c_{\bp+\bq/2,\,\beta},
\ee
where:
\bal
\label{general_current_vertex}
\bm{j}_{\alpha\beta}(\bk,\bq) &= \sum_{\bm{\delta}}\frac{\bm{\Delta R^{\bm{\delta}}_{\alpha,\beta}}}{\bm{\Delta R^{\bm{\delta}}_{\alpha,\beta}}\cdot\bq}t^{\bm{\delta}}_{\alpha\beta}
\frac{\delta}{\delta t^{\bm{\delta}}_{\alpha\beta}}\left[h_{\alpha\beta}(\bk+\bq/2)-h_{\alpha\beta}(\bk-\bq/2)\right]\\
&=-2i\sum_{\bm{\delta}}\sin\left(\frac{\bq\cdot\bm{\Delta R^{\bm{\delta}}_{\alpha,\beta}}}{2}\right)\frac{\bm{\Delta R^{\bm{\delta}}_{\alpha,\beta}}}{\bm{\Delta R^{\bm{\delta}}_{\alpha,\beta}}\cdot\bq}
\,t^{\bm{\delta}}_{\alpha\beta}\,e^{i\bk\cdot\bm{\Delta R^{\bm{\delta}}_{\alpha,\beta}}}.
\eal
One can easily check that from Eq. \eqn{general_current_vertex} follows the continuity equation:
\be
\label{continuity_eq}
i\partial_t\rho(\bq)=[\rho(\bq),H_0]=\sum_{\bk}\sum_{\alpha\beta} c^\dagger_{\bk-\bq/2,\,\alpha}\,\left[h_{\alpha\beta}\left(\bk+\bq/2\right)-h_{\alpha\beta}\left(\bk-\bq/2\right)\right]\,c^\dagga_{\bk+\bq/2,\,\beta}=\bq\cdot\bm{J}_{p}(\bq),
\ee
where $\rho(\bq)=\sum_{\bk}\sum_{\alpha}\,c^\dagger_{\bk,\alpha}c^\dagga_{\bk+\bq,\alpha}$.
Moreover, we notice that $\bm{J}^\dagger_{p}(\bq)=\bm{J}_{p}(-\bq)$ which implies $\bm{j}_{\alpha\beta}(\bk,-\bq) = \bm{j}^*_{\beta\alpha}(\bk,\bq)$. 
The diamagnetic tensor operator $\bm{\mathcal{T}}(-\bq_1,-\bq_2)$ in $H_{A^2}$ reads:
\be
\label{diamagnetic_tensor_general}
\mathcal{T}^{i,j} (-\bq_1,-\bq_2)=\sum_{\bk}\sum_{\alpha\beta}\, c^\dagger_{\bk+(\bq_1+\bq_2)/2,\,\alpha}\,\tau^{i,j}_{\alpha\beta}(\bk,-\bq_1,-\bq_2)\,c^\dagga_{\bk-(\bq_1+\bq_2)/2,\,\beta},
\ee
where $i,j=x,y,z$ $(x,y)$ in three (two) dimensions, and
\bal
\label{diamagnetic_papero}
\tau^{i,j}_{\alpha\beta}(\bk,-\bq_1,-\bq_2)&=\sum_{\bm{\delta}}\,\frac{\left(\bm{\Delta R^{\bm{\delta}}_{\alpha,\beta}}\right)_i}{\bm{\Delta R^{\bm{\delta}}_{\alpha,\beta}}\cdot\bq_1}\,\frac{\left(\bm{\Delta R^{\bm{\delta}}_{\alpha,\beta}}\right)_j}{\bm{\Delta R^{\bm{\delta}}_{\alpha,\beta}}\cdot\bq_2}
t^{\bm{\delta}}_{\alpha\beta}\frac{\delta}{\delta t^{\bm{\delta}}_{\alpha\beta}}\,\Bigg[\left(h_{\alpha\beta}\left(\bk+\frac{\bq_1}{2}-\frac{\bq_2}{2}\right)-h_{\alpha\beta}\left(\bk-\frac{\bq_1}{2}-\frac{\bq_2}{2}\right)\right)\\
&-\left(h_{\alpha\beta}\left(\bk+\frac{\bq_1}{2}+\frac{\bq_2}{2}\right)-h_{\alpha\beta}\left(\bk-\frac{\bq_1}{2}+\frac{\bq_2}{2}\right)\right)\Bigg]\\
&=-4\sum_{\bm{\delta}}\,\frac{\left(\bm{\Delta R^{\bm{\delta}}_{\alpha,\beta}}\right)_i}{\bm{\Delta R^{\bm{\delta}}_{\alpha,\beta}}\cdot\bq_1}\,\sin\left(\frac{\bq_1\cdot\bm{\Delta R^{\bm{\delta}}_{\alpha,\beta}}}{2}\right)
\,t^{\bm{\delta}}_{\alpha\beta}\,e^{i\bk\cdot\bm{\Delta R^{\bm{\delta}}_{\alpha,\beta}}}\,\sin\left(\frac{\bq_2\cdot\bm{\Delta R^{\bm{\delta}}_{\alpha,\beta}}}{2}\right)\,\frac{\left(\bm{\Delta R^{\bm{\delta}}_{\alpha,\beta}}\right)_j}{\bm{\Delta R^{\bm{\delta}}_{\alpha,\beta}}\cdot\bq_2}.
\eal
Finally, we observe that the current operator is obtained as the functional derivative of the electronic Hamiltonian respect the vector potential: 
\be
J_i(\bq)=\frac{c}{e}\frac{\delta\,H_{el}}{\delta \hat{A}_i(-\bq)}=J_{p,i}(\bq)-\frac{e}{c}\sum_{\bq^\prime,j}\,\mathcal{T}^{i,j}\left(\bq,-\bq^\prime\right)\,\hat{A}_{j}\left(\bq^\prime\right).
\ee
We underline that the light-matter Hamiltonian \eqn{light_matter} obtained by performing the Peierls substitution and expanding to second order in $\hat{\bm{A}}$ is by construction gauge invariant. 
This property plays a key role in the context of strong light-matter interaction where models or approximations that break the gauge invariance lead to incorrect results. 
Before discussing the ground state obtained by a quantum description of the cavity field $\hat{\bm{A}}(\bq)$ it is crucial to remind some basics consequences of the charge conservation law and of the gauge invariance.

\section{Gauge invariance and charge conservation}
\label{sec2}

In this section we are going to recall some important constraints on the electron response to the cavity field, that are direct consequences of fundamental laws, i.e. gauge invariance of the electromagnetic field and the charge continuity equation. For the sake of clarity, since quantization of the cavity field is not crucial, we are going to treat the cavity field classically $\hat{\bm{A}}\to\bm{A}(t,\bx)$. 

Within linear response theory the current response to the cavity electromagnetic field is described by:
\be
j_{i}(\omega,\bq)=\frac{e}{c}\sum_{j}\,Q^{i,j}(\omega,\bq)\,A_j(\omega,\bq),
\ee
where $\bm{j}(\omega,\bq)$ is the current density $\bm{j}(\omega,\bq)=\langle \bm{J}(\omega,\bq)\rangle/V$, and $\bm{Q}(\omega,\bq)$ is the current-current response tensor:
\be
\label{current_current_response}
Q^{i,j}(\omega,\bq)=K^{i,j}(\omega,\bq)-\,\left\langle\mathcal{T}^{i,j}(\bq,-\bq)\right\rangle/V.
\ee
Here $K^{i,j}(\omega,\bq)$ is the paramagnetic current response function that in the Lehmann representation reads:
\be
\label{paramagnetic_response}
 K^{i,j}(\omega,\bq)=\sum_n\,\frac{P_n}{V}\,\sum_m\,\left(\frac{\bra{\Phi_n}J_{p,i}(\bq)\ket{\Phi_m}\,\bra{\Phi_m}J_{p,j}(-\bq)\ket{\Phi_n}}{\omega-\left(E_m-E_n\right)+i0^+}-\frac{\bra{\Phi_n}J_{p,j}(-\bq)\ket{\Phi_m}\,\bra{\Phi_m}J_{p,i}(\bq)\ket{\Phi_n}}{\omega+\left(E_m-E_n\right)+i0^+}\right),
 \ee
 where $P_n=\exp(-\beta\,E_n)/Z$, $\ket{\Phi_n}$ and $\ket{\Phi_m}$ are eigenstates of the electronic Hamiltonian $H_{el}$ and $E_n$ and $E_m$ the corresponding eigenvalues. 

The value of the response function \eqn{current_current_response} in some particular limits of physical significance can be obtained without making any complicated calculation. 
In particular, the gauge invariance of the electromagnetic field \cite{Cohen_1989,Jackson_1975} imposes that a physical system cannot respond to a static and spatially homogeneous vector potential, i.e. in the \textit{static} limit $\left(\omega=0,\bq\to0\right)$\footnote{The identity \eqn{diamagnetic_s_r} does not hold for superconductors, where the ground state breaks spontaneously the gauge symmetry.}
\be
\label{diamagnetic_s_r}
\lim_{q_i\to0}\bm{Q}(\omega=0,q_i,q_j=0)=\bm{0},
\ee
or
\be
\lim_{q_i\to0}\bm{K}(\omega=0,q_i,q_j=0)=\left\langle\bm{\mathcal{T}}(\bm{0},\bm{0})\right\rangle/V.
\ee
The latter constraint, also know as \textit{diamagnetic} or TRK sum-rule \cite{Giuliani_book,Pines_1966}, is explicitly satisfied by the tight-binding model in Eq. \eqn{light_matter} where the minimal coupling corresponds to dress the hopping amplitude by the gauge field $\bm{A}$ \eqn{Peierls_ansatz}. Indeed, for an uniform and static vector potential $\bm{A}_0$ the phase factor \eqn{path_phase} reads $\lambda=\bm{A}_0\cdot(\bm{R_{j,\beta}}+\bm{\delta}-\bm{R_{j,\alpha}})$ and the Bloch Hamiltonian \eqn{model_h_paper} is modified as $h_{\alpha\beta}(\bk-e\bm{A}_0/c)$, i.e. a simple momentum shift. The momentum shift is harmless as the Brillouin zone is a compact space so that the matter energy variation due to $\bm{A}_0$ is zero, $\Delta E_{el}=E_{el}(\bm{A}_0)-E_{el}(\bm{0})=0$. 
Making use of the Hellmann-Feynman theorem we can write:
\be
\frac{\partial E_{el}(\bm{A}_0)}{\partial\bm{A}_0}=\left\langle\Phi\right|\frac{\partial H_{el}}{\partial \bm{A}_0}\left|\Phi\right\rangle=\frac{e}{c}\lim_{\bq\to0}\left\langle\Phi\right|\bm{J}(\bq)\left|\Phi\right\rangle=0.
\ee
Within linear response theory we have: 
\be
\lim_{\bq\to0}\left\langle\Phi\right|J_i(\bq)\left|\Phi\right\rangle=\lim_{\bq\to0}\sum_{j}\,Q^{i,j}(0,\bq)\,{A}_{0,j}=0,
\ee
the TRK sum-rule \eqn{diamagnetic_s_r} follows as a result of arbitrariness of $\bm{A}_0$.

An additional condition on the current-current response is provided by the charge continuity equation \eqn{continuity_eq}. We observe that Eq. \eqn{continuity_eq} implies:
\be
\label{A_1_0}
\left[\rho(\bq),H_{el}\right]=\bq\cdot\bm{J}_{p}(\bq) 
\ee
which connects the charge density operator to the longitudinal paramagnetic current:
\be
\label{longitudinal_para_current}
J^{L}_{p}(\bq)=-i\frac{\bq}{|\bq|}\cdot\bm{J}_{p}(\bq).
\ee
We notice that:
\be
\label{A_1_1}
\left\langle\left[\left[\rho(\bq),H_{el}\right],\rho(-\bq)\right]\right\rangle=\sum_n\,P_n\sum_m\left(E_m-E_n\right)\left[\left|\bra{\Phi_n}\rho(\bq)\ket{\Phi_m}\right|^2+\left|\bra{\Phi_n}\rho(-\bq)\ket{\Phi_m}\right|^2\right].
\ee
On the other hand, the left hand side of Eq. \eqn{A_1_1} is also equal to:
\be
\label{A_1_2}
\left\langle\left[\bq\cdot\bm{J}_{p}(\bq),\rho(-\bq)\right]\right\rangle=-\sum_{i,j}\,q_{i}\,\left\langle\mathcal{T}^{i,j}\left(\bq,-\bq\right)\right\rangle\,q_{j}.
\ee
Finally, by inserting Eq. \eqn{A_1_0} in \eqn{A_1_1} we obtain the $f$-sum rule \cite{Giuliani_book,Pines_1966}:
\be
\sum_n\,P_n\sum_m\frac{\left|\bra{\Phi_n}J^{L}_{p}(\bq)\ket{\Phi_m}\right|^2+\left|\bra{\Phi_n}J^{L}_{p}(-\bq)\ket{\Phi_m}\right|^2}{E_m-E_n}=-\sum_{i,j}\,\frac{q_i}{|\bq|}\left\langle\mathcal{T}^{i,j}\left(\bq,-\bq\right)\right\rangle\frac{q_j}{|\bq|}.
\ee 
The latter equation can be also written as:
\be
\label{f_sum_rule}
Q_{L}(\omega=0,\bq) = 0,
\ee
where $Q_{L}(\omega,\bq)$ is the longitudinal component of the current-current response tensor that reads:
\be
Q_{L}(\omega,\bq)=\sum_{i,j}\frac{q_i}{|\bq|}\,Q^{i,j}(\omega,\bq)\,\frac{q_j}{|\bq|}.
\ee 
We anticipate that Eq. \eqn{f_sum_rule}, also known as $f$-sum rule \cite{Giuliani_book,Pines_1966}, forbids the condensation of longitudinal photons, i.e. photons with polarization parallel to the wave vector $\bq$ of the electronic excitations. 
In addition to \eqn{longitudinal_para_current} we define the transverse current operator as: 
\be
\label{transverse_para_current}
J^{T}_{p}(\bq)=\bm{u}^*_{\bq\,T}\cdot\bm{J}_{p}(\bq),
\ee  
where $\bm{u}_{\bq\,T}$ is the transverse polarization vector, that in two dimensions reads:
\be
\bm{u}_{\bq\,T}=i\frac{\bm{z}\times\bq}{|\bq|}.
\ee
The component of the current response tensor along the transverse direction is:
\be
Q_{T}(\omega,\bq)=\sum_{i,j}\frac{(\bm{z}\times\bq)_i}{|\bq|}\,Q^{i,j}(\omega,\bq)\,\frac{(\bm{z}\times\bq)_j}{|\bq|}.
\ee
Finally, we report the response function to a static and spatially-modulated transverse field:
\be
\label{transverse}
Q_{T}(0,\bq)=-\frac{1}{V}\sum_n\,P_n\sum_m\frac{\left|\bra{\Phi_n}J^{T}_{p}(\bq)\ket{\Phi_m}\right|^2+\left|\bra{\Phi_n}J^{T}_{p}(-\bq)\ket{\Phi_m}\right|^2}{E_m-E_n}-\sum_{i,j}\,(\bm{u^*}_{\bq\,T})_i\,\frac{\left\langle\mathcal{T}^{i,j}\left(\bq,-\bq\right)\right\rangle}{V}\,(\bm{u}_{\bq\,T})_j.
\ee
Differently from the longitudinal component, the transverse one is not constrained by any sum-rule at $\omega=0$.  

\section{The criterion for the superradiant quantum phase transition}
\label{sec3}

In the following we consider the instability of the normal ground state towards the superradiant phase. 
We characterize the superradiant quantum phase transition (SQPT) by two different methods.  
Firstly, in the proximity of the superradiant critical point, we perform a ground state calculation that determines the condition for spontaneous condensation of the polariton mode. Then, we apply a field-theoretical approach to compute the self-consistent cavity mode Green's function. Remarkably, the condition for photon condensation coincides with the vanishing of the lowest polariton frequency. 
From general arguments we obtain a no-go theorem that forbids superradiant phase transition for longitudinal modes, but not for the transverse ones.

\subsection{A single-mode ground state calculation}
\label{subsec3_1}

The minimal model that manifests the SQPT can be obtained by picking a single space-dependent photon out from the collection of cavity modes. It is important to remark that the three-dimensional field, once projected in the low-dimensional material embedded in the cavity, exhibits transverse and longitudinal components respect the wave vector $\bq$ exchanged with the electronic excitations \cite{Kakazu_1994}.
We treat the spatially-modulated cavity mode in a quantum fashion. 
Following the main text we drop the polarisation index $\sigma$ as well as the wave vector $\bq$ so that $\bm{u}_{\bq\,\sigma}$ is simply $\bm{u}$, and
\be
\label{single_mode}
\hat{\bm{A}}(\bq) = \bar{A}\,\bm{u}\,\left(a^\dagga_{\bq}+a^\dagger_{-\bq}\right)/\sqrt{V},
\ee 
where $a^\dagga_\bq$ is an annihilation operator for a cavity photon with wave vector $\bq$ and polarisation $\bm{u}$, and $V$ the volume of the cavity. The linear and quadratic couplings are
\be
\label{linear_pick}
H_{A}= \frac{e}{c} \left[ \hat{\bm{A}}(\bq)\cdot\bm{J}_p(-\bq) + \hat{\bm{A}}(-\bq)\cdot\bm{J}_p(\bq)\right],
\ee 
\bal
\label{quad_pick}
H_{A^2}=&-\frac{e^2}{2c^2} \Big[ \hat{A}_{i}(\bq)\,\mathcal{T}^{i,j}(-\bq,\bq)\,\hat{A}_{j}(-\bq) + \hat{A}_{i}(-\bq)\,\mathcal{T}^{i,j}(\bq,-\bq)\,\hat{A}_{j}(\bq) \\
&+ \hat{A}_{i}(\bq)\,\mathcal{T}^{i,j}(-\bq,-\bq)\,\hat{A}_{j}(\bq)+ \hat{A}_i(-\bq)\,\mathcal{T}^{i,j}(\bq,\bq)\,\hat{A}_{j}(-\bq)\Big],
\eal
where the summation over $i$ and $j$ is assumed. For the rest of the Supplementary Material whenever an index is repeated, we will assume that it is summed over, unless otherwise noted.
In the thermodynamic limit, $V\to\infty$, the ground state wave function factorises \cite{Andolina_2019}:  
\be
\ket{\Psi}=\ket{\Phi}\otimes\ket{\chi}
\ee
where $\ket{\Phi}$ is the matter wave function, while $\ket{\chi}$ the photon one. 
The latter is the ground state of the bosonic Hamiltonian: 
\bal
\label{bosonic_H}
H_a\equiv\left\langle\Phi\left| H \right|\Phi\right\rangle=&\hbar \omega_{\bq}\left(a^\dagger_{\bq} a^\dagga_{\bq}+a^\dagger_{-\bq} a^\dagga_{-\bq}+1\right)+
\gamma\,\sqrt{V}\left[\left(a^\dagga_{\bq}+a^\dagger_{-\bq}\right)\mathcal{J}^{*}_\bq+\left(a^\dagga_{-\bq}+a^\dagger_{\bq}\right)\mathcal{J}^\dagga_\bq\right]\\
&+2\kappa^\dagga_\bq\,\left(a^\dagga_{\bq}+a^\dagger_{-\bq}\right)\left(a^\dagga_{-\bq}+a^\dagger_{\bq}\right)+\Delta\kappa^\dagga_\bq\left(a^\dagga_{\bq}+a^\dagger_{-\bq}\right)^2+\Delta\kappa^*_\bq\left(a^\dagga_{-\bq}+a^\dagger_{\bq}\right)^2,
\eal
where we have defined the coupling
\be
\label{coupling}
\gamma=e\bar{A}/c.
\ee
Moreover, in Eq. \eqn{bosonic_H} we have introduced the fermionic averages:
\be
\label{fermionic_average_current}
\mathcal{J}^\dagga_\bq=\frac{\left\langle\Phi\left| \bm{u}^*\cdot\bm{J}_p(\bq)\right|\Phi\right\rangle}{V},
\ee
and
\be
\label{fermionic_average_diam}
\kappa^\dagga_\bq=-\frac{\gamma^2}{2}u^*_i\,\frac{\left\langle\Phi\left|\mathcal{T}^{i,j}(\bq,-\bq)\right|\Phi\right\rangle}{V}\,u_j,\quad\Delta\kappa^\dagga_\bq=-\frac{\gamma^2}{2}\,u_i\,\frac{\left\langle\Phi\left|\mathcal{T}^{i,j}(-\bq,-\bq)\right|\Phi\right\rangle}{V}\,u_j.
\ee
On the other hand, the fermionic Hamiltonian reads:
\bal
\label{fermionic_H}
H_f\equiv\left\langle\chi\left|H\right|\chi\right\rangle=&H_{el}+\sqrt{2}\gamma\,\left[\frac{\left\langle X_{\bq} \right\rangle}{\sqrt{V}}\,  \bm{u}\cdot\bm{J}_p(-\bq)+\frac{\left\langle X_{-\bq} \right\rangle}{\sqrt{V}}\,  \bm{u}^*\cdot\bm{J}_p(\bq)\right]\\
&-\gamma^2\,\Bigg[\frac{\left\langle X_{\bq}X_{-\bq} \right\rangle}{V}\,u_i\,\mathcal{T}^{i,j}(-\bq,\bq)\,u^*_j+\frac{\left\langle X_{-\bq}X_{\bq} \right\rangle}{V}\,u^*_i\,\mathcal{T}^{i,j}(\bq,-\bq)\,u_{j}\\
&+\frac{\left\langle X_{\bq}X_{\bq} \right\rangle}{V}\,u_i\,\mathcal{T}^{i,j}(-\bq,-\bq)\,u_j+\frac{\left\langle X_{-\bq}X_{-\bq} \right\rangle}{V}\,u^*_i\,\mathcal{T}^{i,j}(\bq,\bq)\,u^*_j\Bigg],
\eal
where we have introduced $X_{\bq}=(a^\dagga_{\bq}+a^\dagger_{-\bq})/\sqrt{2}$. 
 
 In proximity to the SQPT, where $\left|\left\langle X_{\bq} \right\rangle\right|/\sqrt{V}\ll1$, we perform an expansion in the photon condensate density and we keep term up to second order in $\left|\left\langle X_{\bq} \right\rangle\right|/\sqrt{V}$. 
 We assume that the ground state in the absence of light-matter interaction $\ket{\Phi_0}$ is homogeneous: $\bra{\Phi_0}\mathcal{T}^{i,j}(\bq,\bq^\prime)\ket{\Phi_0}\propto\delta_{\bq^\prime,-\bq}$.
 Under these assumptions and as a consequence of the form of the Hamiltonian \eqn{fermionic_H} we realize:
\be
\mathcal{J}^\dagga_\bq\sim\mathcal{O}\left(\frac{\left\langle X_{\bq} \right\rangle}{\sqrt{V}}\right),\quad\kappa^\dagga_\bq\sim\mathcal{O}\left(1\right),\quad\Delta\kappa^\dagga_\bq\sim\mathcal{O}\left(\frac{\left\langle X_{-\bq}X_{-\bq} \right\rangle}{V}\right).
\ee
Therefore, the bosonic Hamiltonian \eqn{bosonic_H} becomes:
\bal
\label{bosonic_H_1}
H_a=&\hbar \omega_{\bq}\left(a^\dagger_{\bq} a^\dagga_{\bq}+a^\dagger_{-\bq} a^\dagga_{-\bq}+1\right)+\gamma\,\sqrt{V}\left[\left(a^\dagga_{\bq}+a^\dagger_{-\bq}\right)\mathcal{J}^{*}_\bq+\left(a^\dagga_{-\bq}+a^\dagger_{\bq}\right)\mathcal{J}^\dagga_\bq\right]+2\kappa^\dagga_\bq\left(a^\dagga_{\bq}+a^\dagger_{-\bq}\right)\left(a^\dagga_{-\bq}+a^\dagger_{\bq}\right).
\eal
Eq. \eqn{bosonic_H_1} can be diagonalized via the following Bogoliubov transformation:
\bal
&\phi^\dagga_{\bq}=\frac{\lambda_\bq+1}{2\sqrt{\lambda_\bq}}\,a^\dagga_{\bq}+\frac{\lambda_\bq-1}{2\sqrt{\lambda_\bq}}\,a^\dagger_{-\bq},\\
&\phi^\dagga_{-\bq}=\frac{\lambda_\bq+1}{2\sqrt{\lambda_\bq}}\,a^\dagga_{-\bq}+\frac{\lambda_\bq-1}{2\sqrt{\lambda_\bq}}\,a^\dagger_{\bq}.
\eal
We notice that the previous transformation preserves the canonical commutation rules $\left[\phi_i,\phi^\dagger_j\right]=\delta_{ij}$ and $\left[\phi_i,\phi_j\right]=0$. By choosing $\lambda_\bq=\sqrt{1+4\kappa_\bq/\hbar\omega_{\bq}}$ we obtain:
\be
\label{bosonic_H_2}
H_a=\hbar\omega_{\bq}\lambda_\bq\sum_{\alpha=\pm}\left(\phi^\dagger_{\alpha\bq}\phi^\dagga_{\alpha\bq}+\frac{1}{2}\right)+\gamma\,\sqrt{\frac{V}{\lambda_\bq}}\left[\mathcal{J}^\dagga_\bq\left(\phi^\dagga_{-\bq}+\phi^\dagger_\bq\right)+\mathcal{J}^*_\bq\left(\phi^\dagga_\bq+\phi^\dagger_{-\bq}\right)\right].
\ee
The model \eqn{bosonic_H_2} corresponds to the sum of two displaced harmonic oscillators $\phi^\dagga_\bq$ and $\phi^\dagga_{-\bq}$, so that the bosonic ground state wave function is given by: $\ket{\chi}=\ket{\alpha^\dagga_\bq}\otimes\ket{\alpha^\dagga_{-\bq}}$ where $\ket{\alpha^\dagga_\bq}$ and $\ket{\alpha^\dagga_{-\bq}}$ are coherent states, i.e. $\phi^\dagga_\bq\ket{\alpha^\dagga_\bq}=\alpha_\bq\ket{\alpha^\dagga_\bq}$.
The minimisation of the energy functional:
\be
\left\langle\chi\left| H_a \right|\chi\right\rangle=\hbar\omega_{\bq}\lambda_\bq\left(\left|\alpha^\dagga_{\bq}\right|^2+\left|\alpha^\dagga_{-\bq}\right|^2+1\right)+\sqrt{\frac{\gamma}{\lambda_\bq}}\left[\mathcal{J}_\bq\left(\alpha^\dagga_{-\bq}+\alpha^*_{\bq}\right)+\mathcal{J}^*_\bq\left(\alpha^\dagga_\bq+\alpha^*_{-\bq}\right)\right]\ee
gives the saddle point values $\bar{\alpha}^\dagga_\bq$ and $\bar{\alpha}^\dagga_{-\bq}$:
\be
\label{saddle_point}
\bar{\alpha}^\dagga_\bq=-\frac{\gamma}{\hbar\omega_{\bq}\lambda_\bq}\sqrt{\frac{V}{\lambda_\bq}}\,\mathcal{J}^\dagga_\bq,\quad\bar{\alpha}^\dagga_{-\bq}=-\frac{\gamma}{\hbar\omega_{\bq}\lambda_\bq}\sqrt{\frac{V}{\lambda_\bq}}\,\mathcal{J}^*_\bq,
\ee
and $\bar{\alpha}^*_\bq=\bar{\alpha}^\dagga_{-\bq}$. The ground state problem is reduced to the optimisation of the energy 
\be
\label{light_matter_E}
E(\bar{\alpha}_{\bq})=\bra{\Phi}\,H_{el}\ket{\Phi}-\hbar\omega_\bq\lambda_\bq\left(|\bar{\alpha}_\bq|^2+|\bar{\alpha}_{-\bq}|\right),
\ee
where the photon configuration is self-consistently determined by the electronic configuration $\ket{\Phi}$ through Eq. \eqn{saddle_point}. 
In the following, we expand the ground state energy for fixed but small photon density which, due to Eq. \eqn{saddle_point}, corresponds to a non-vanishing spatially-modulated electronic current density. 
Under these circumstances we employ the stiffness theorem \cite{Giuliani_book,Pines_1966} to compute exactly the ground state energy for a given average value of the current operator $\bm{u}^*\cdot\bm{J}_p(\bq)$ up to order $|\mathcal{J}_\bq|^2$:
\be
\label{stiffness0}
E(\bar{\alpha}_{\bq})\simeq E(0)-V\,\frac{\mathcal{J}^\dagga_\bq\,\mathcal{J}^*_\bq+\mathcal{J}^*_\bq\,\mathcal{J}^\dagga_\bq}{2\,K(0,\bq)}-\hbar\omega_{\bq}\,\lambda\,\left(|\bar{\alpha}_\bq|^2+|\bar{\alpha}_{-\bq}|\right),
\ee
where in the Lehmann representation
\be
K(0,\bq)=-\frac{1}{V}\sum_{m>0}\frac{\left|\left\langle\Phi_0\left|\bm{u}^*\cdot\bm{J}_p(\bq)\right|\Phi_m\right\rangle \right|^{2}+\left|\left\langle \Phi_0\left|\bm{u}\cdot\bm{J}_p(-\bq)\right|\Phi_m\right\rangle \right|^{2}}{E_{m}-E_{0}},
\ee
with $\ket{\Phi_0}$ the ground state and $\ket{\Phi_m}$ the excited state of the electron-only Hamiltonian $H_{el}$.
By using the self-consistency equation \eqn{saddle_point} we obtain: 
\be
E(\bar{\alpha}_{\bq})- E(0)=\mathcal{N}_\bq\,\left[\hbar\omega_\bq+2\gamma^2\,Q(0,\bq)\right]\,|\bar{\alpha}_\bq|^2<0
\ee
where the positive constant $\mathcal{N}_\bq$ is: 
\be
\mathcal{N}_\bq=-\frac{\hbar\omega_\bq\,\lambda_\bq}{2\gamma^2\,K(0,\bq)}.
\ee
Thus, the normal ground state is unstable towards photon condensation if:
\be
\label{inequality_A}
-2\gamma^2\,Q(0,\bq)>\hbar\omega_{\bq},
\ee
with $Q(0,\bq)=u^*_i\,Q^{i,j}(0,\bq)\,u_j$.
As a consequence of the TRK sum-rule \eqn{diamagnetic_s_r}, a static and homogeneous cavity field cannot satisfy the previous inequality. 
Therefore, in agreement with Ref. \cite{Andolina_2019} we conclude that the ground state cannot lower its energy by spontaneously generating a constant vector potential. 
At finite wave vector $\bq\neq\bm{0}$ of the cavity field, the $f$-sum rule \eqn{inequality_A} forbids SQPT for photons with longitudinal polarization, i.e. $\bm{{u}}\parallel\bq$ where $\bq$ is the wave vector exchanged with the low-dimensional electronic excitations. Indeed, Eq. \eqn{f_sum_rule} implies the vanishing of the left hand side of Eq. \eqn{inequality_A} and the inequality cannot be satisfied. 
Before going to specific electronic models presenting SQPT, we show that the criterion \eqn{inequality_A} coincides with the softening of a polariton mode in the spectrum of the cavity excitations.

\subsection{Soft polariton mode at the superradiance instability}
\label{subsec3_2}

In this section, we compute the polariton Green's function to show that the criterion for the superradiance instability \eqn{inequality_A} coincides with the softening of the lowest polariton mode. 
We introduce the conjugate operators $X_{\bq}$ and $P_{\bq}$:
\be
\label{normal_variables}
X_{\bq}=(a^\dagga_{\bq}+a^\dagger_{-\bq})/\sqrt{2},\quad P_{\bq}=-i(a^\dagga_{\bq}-a^\dagger_{-\bq})/\sqrt{2}, 
\ee
and the corresponding Green's functions:
\be
\Pi^{XX}(\tau,\bq)=-\left\langle T_\tau\left(X_{\bq}(\tau)\,X_{-\bq}\right) \right\rangle,\quad\Pi^{PX}(\tau,\bq)=-\left\langle T_\tau\left(P_{\bq}(\tau)\,X_{-\bq}\right) \right\rangle.
\ee 
To second order in the light-matter interaction we obtain: 
\bal
\label{Dyson_Pi}
\omega\,\Pi^{XX}(\omega,\bq)&=i\omega_{\bq}\,\Pi^{PX}(\omega,\bq),\\
\omega\,\Pi^{PX}(\omega,\bq)&=-i-i\omega_{\bq}\,\Pi^{XX}(\omega,\bq)-i2\gamma^2\,u^*_i\,Q^{i,j}(\omega,\bq)\,u_j\,\Pi^{XX}(\omega,\bq),
\eal
where $u^*_i\,Q^{i,j}(\omega,\bq)\,u_j$ is the current correlation function projected along the polarization direction of the cavity mode. 
By solving Eq. \eqn{Dyson_Pi} for $\Pi^{XX}(\omega,\bq)$ we obtain: 
\be
\Pi^{XX}(\omega,\bq)=\left[(\omega^2-\omega^2_{\bq})/\omega_{\bq}-2\gamma^2\,u^*_i\,Q^{i,j}(\omega,\bq)\,u_j\right]^{-1}.
\ee 
 Finally, the polariton frequencies $\omega^{pol}_{\bq}$ are the real and positive roots of the characteristic equation:
 \be
 \label{polariton_spectrum}
\omega^2-\omega^2_{\bq}-2\gamma^2\,\omega_\bq\,u^*_i\,Q^{i,j}(\omega,\bq)\,u_j=0.
 \ee  
 The instability of the normal ground state with respect to the spontaneous generation of a cavity field with wave vector $\bq$ occurs in the presence of a zero energy $\omega=0$ solution of Eq. \eqn{polariton_spectrum}. 
 The latter condition is nothing but the saturation of the inequality in Eq. \eqn{inequality_A}. 
  
We conclude by observing that the Green's function approach presented in this Section can be conveniently applied to more realistic situations where a set of cavity photons is considered \cite{Basko_PRL2019}.   
It is important to remark that, by including additional cavity modes, one improves the quantitative characterisation of the superradiant critical point without changing the qualitative picture.

 \section{The current-current response function for non-interacting electrons}
\label{sec4}

In this section we give technical details on the evaluation of the current-current response function \eqn{current_current_response} valid for non-interacting electron systems. The analysis introduces the dipole matrix elements $g^{n,m}_{\bk,\bq}$. 
By performing straightforward, calculations we find that the average value of the diamagnetic tensor reads: 
\be
\left\langle\,\mathcal{T}^{i,j}(\bq,-\bq)\right\rangle=\int_{BZ}\frac{d^d\bk}{(2\pi)^d}\,\sum_n\,f(\epsilon_{\bk,n})\,\left[v^n_\bk\right]^*_{\alpha}\,\tau^{i,j}_{\alpha\beta}(\bk,\bq,-\bq)\,\left[v^n_\bk\right]_{\beta},
\ee
where $\left[v^n_\bk\right]_{\alpha}$ is the $\alpha$ component of eigenstate of the Hamiltonian $h(\bk)$ \eqn{model_h_paper}, $\epsilon_{\bk,n}$ the corresponding eigenvalue and $f(\epsilon)$ the Fermi-Dirac distribution function. 
On the other hand the paramagnetic contribution \eqn{paramagnetic_response} can be easily computed by observing that the excited states $\ket{\Phi_m}$ are electron-hole pairs:
\be
\label{paramagnetic_response_A_0}
K^{i,j}(\omega,\bq)=\int_{BZ}\frac{d^d\bk}{(2\pi)^d}\sum_{nm}\,\left[g^{n,m}_{\bk,\bq}\right]_i\,\left[g^{n,m}_{\bk,\bq}\right]^*_j\frac{f(\epsilon_{\bk,n})-f(\epsilon_{\bk+\bq,m})}{\omega-(\epsilon_{\bk+\bq,m}-\epsilon_{\bk,n})+i0^+},
\ee
where we have introduced the dipole matrix element: 
\be
\label{dipole_m_elements}
\left[g^{n,m}_{\bk,\bq}\right]_i=\bra{0}J_{p,i}(\bq)\ket{\bk,\bq}_{n,m}=\sum_{\alpha\beta}\,\left[v^n_\bk\right]^*_\alpha\,j^i_{\alpha\beta}(\bk+\bq/2,\bq)\,\left[v^m_{\bk+\bq}\right]_\beta.
\ee
In the previous expression, $\ket{0}$ is the electronic ground state and $\ket{\bk,\bq}_{n,m}=\Gamma^\dagger_{\bk+\bq,m}\Gamma^\dagga_{\bk,n}\ket{0}$, 
where $\Gamma_{\bk,n}$ and $\Gamma_{\bk+\bq,m}$ are the Bloch orbital annihilation operators.
We observe that the previous matrix element satisfies the following property $\left[g^{n,m}_{\bk-\bq,\bq}\right]_i=\left[g^{m,n}_{\bk,-\bq}\right]^*_i$.
 By taking the zero temperature limit of Eq. \eqn{paramagnetic_response_A_0} we obtain:
\be
\label{paramagnetic_response_A_1}
 K^{i,j}(\omega,\bq)=\int_{BZ}\frac{d^d\bk}{(2\pi)^d}\sum_{n<0<m}\left(\frac{\left[g^{n,m}_{\bk,\bq}\right]_i\,\left[g^{n,m}_{\bk,\bq}\right]^*_j}{\omega-(\epsilon_{\bk+\bq,m}-\epsilon_{\bk,n})+i0^+}-\frac{\left[g^{n,m}_{\bk,-\bq}\right]_j\,\left[g^{n,m}_{\bk,-\bq}\right]^*_i}{\omega+(\epsilon_{\bk-\bq,m}-\epsilon_{\bk,n})+i0^+}\right),
 \ee 
 where we label the states with $n<0\,(>0)$ for negative (positive) energies.
 Particularly relevant for the analysis of the superradiance instability is the \textit{static} limit of the previous expression 
 \be
\label{paramagnetic_response_A_2}
 K^{i,j}(\omega=0,\bq)=\int_{BZ}\frac{d^d\bk}{(2\pi)^d}\sum_{n<0<m}\left(\frac{\left[g^{n,m}_{\bk,\bq}\right]_i\,\left[g^{n,m}_{\bk,\bq}\right]^*_j}{\epsilon_{\bk,n}-\epsilon_{\bk+\bq,m}}+\frac{\left[g^{n,m}_{\bk,-\bq}\right]_j\,\left[g^{n,m}_{\bk,-\bq}\right]^*_i}{\epsilon_{\bk,n}-\epsilon_{\bk-\bq,m}}\right).
 \ee 
Finally, by projecting the paramagnetic response tensor along the polarisation direction of the cavity field $\bm{u}_\bq$ we obtain Eq. (11) of the main text:  
\be
\label{paramagnetic_response_A_3}
\left(\bm{u}^*_\bq\right)_i\,K^{i,j}(\omega=0,\bq)\left(\bm{u}_\bq\right)_j=\int_{BZ}\frac{d^d\bk}{(2\pi)^d}\sum_{n<0<m}\sum_{\pm}\frac{\left|g^{n,m}_{\bk,\pm\bq}\right|^2}{\epsilon_{\bk,n}-\epsilon_{\bk\pm\bq,m}},
\ee
where we have used the property $\bm{u}^*_\bq=\bm{u}_{-\bq}$ and $g^{n,m}_{\bk,\bq}=\bm{u}^*_\bq\cdot\bm{g}^{n,m}_{\bk,\bq}$.

\section{The superradiant instability in the square lattice model}
\label{sec5}

In the first example we characterize the superradiance instability for the textbook two-dimensional square lattice with nearest-neighbor hopping.  Moreover, we present a detailed characterisation of the superradiant phase. 
The Hamiltonian in Fourier space reads $h(\bk)=-t\left[\cos(k_x\,a) +\cos(k_y\,a) \right]$, where $a$ is the lattice parameter.
By applying Eqs. \eqn{general_current_vertex} and \eqn{diamagnetic_papero} we readily find that the paramagnetic current and diamagnetic tensor operators are: 
\bal
&\bm{j}(\bk,\bq)=2t\left[\frac{\bm{x}}{q_x}\sin\left(\frac{q_x\,a}{2}\right)\sin\left(k_x\,a\right)+\frac{\bm{y}}{q_y}\sin\left(\frac{q_y\,a}{2}\right)\sin\left(k_y\,a\right)\right],\\
&\tau^{i,j}(\bk,\bq_1,\bq_2)=-4t\,\frac{\delta_{i,j}}{q^i_1\,q^i_2}\sin\left(\frac{q^i_1\,a}{2}\right)\sin\left(\frac{q^i_2\,a}{2}\right)\cos(k_i\,a).
\eal
From Eq. \eqn{dipole_m_elements} we readily obtain that the dipole matrix element is $g^i_{\bk,\bq}=2t\sin\left(q_i\,a/2\right)\sin\left(k_i\,a+q_i\,a/2\right)/q_i$. 

 \begin{figure}
\begin{center}
\includegraphics[width=0.35\textwidth]{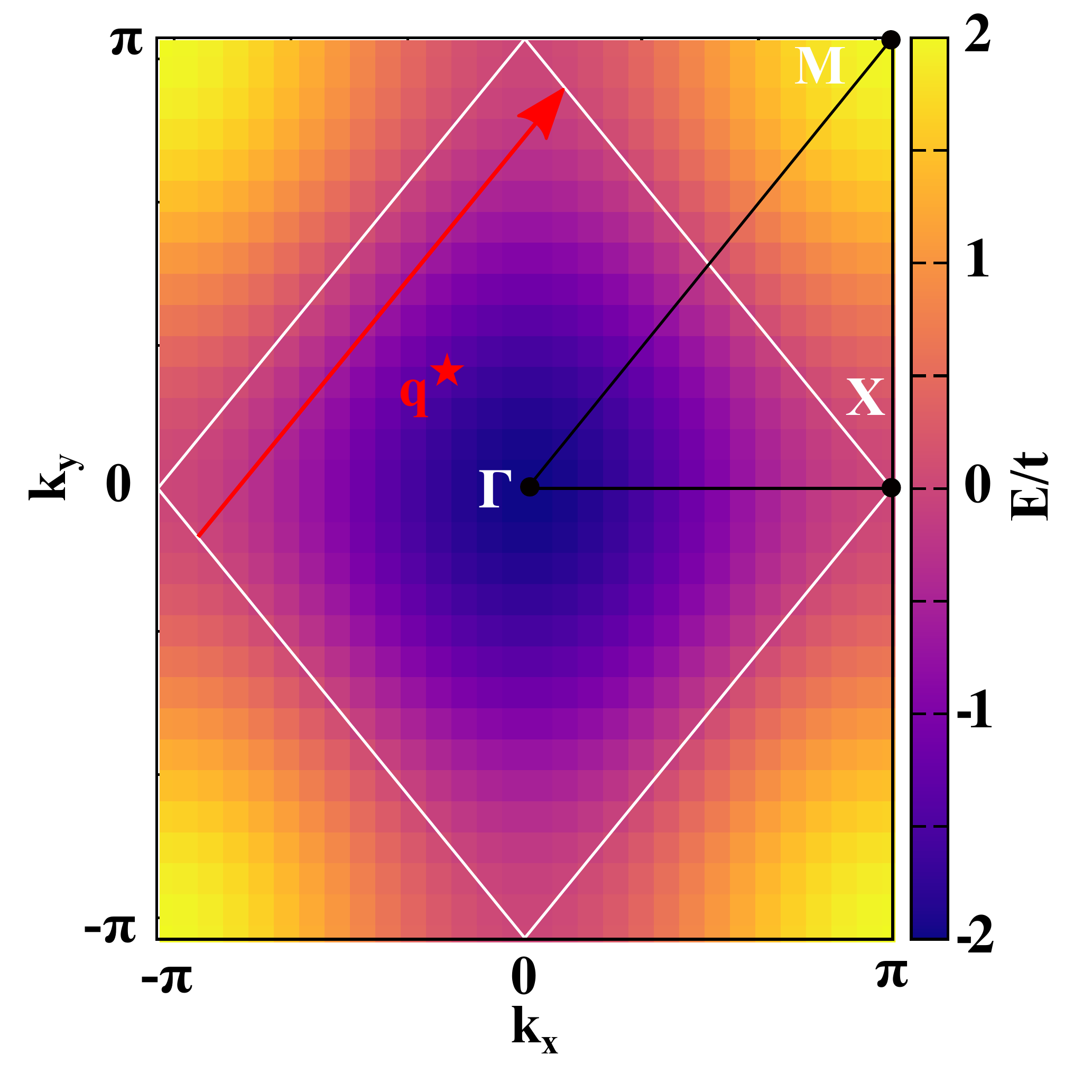}
\vspace{-0.15cm}
\caption{The color map shows the band structure of the square lattice model with $t=1$. The solid white line denotes the Fermi surface at half-filling. We also report the high-symmetry points and the nesting wave vector $\bq^\star=(\pi,\pi)/a$.}
\label{normal_GS_bands}
\end{center}
\end{figure}
\begin{figure}
\begin{center}
\includegraphics[width=0.7\textwidth]{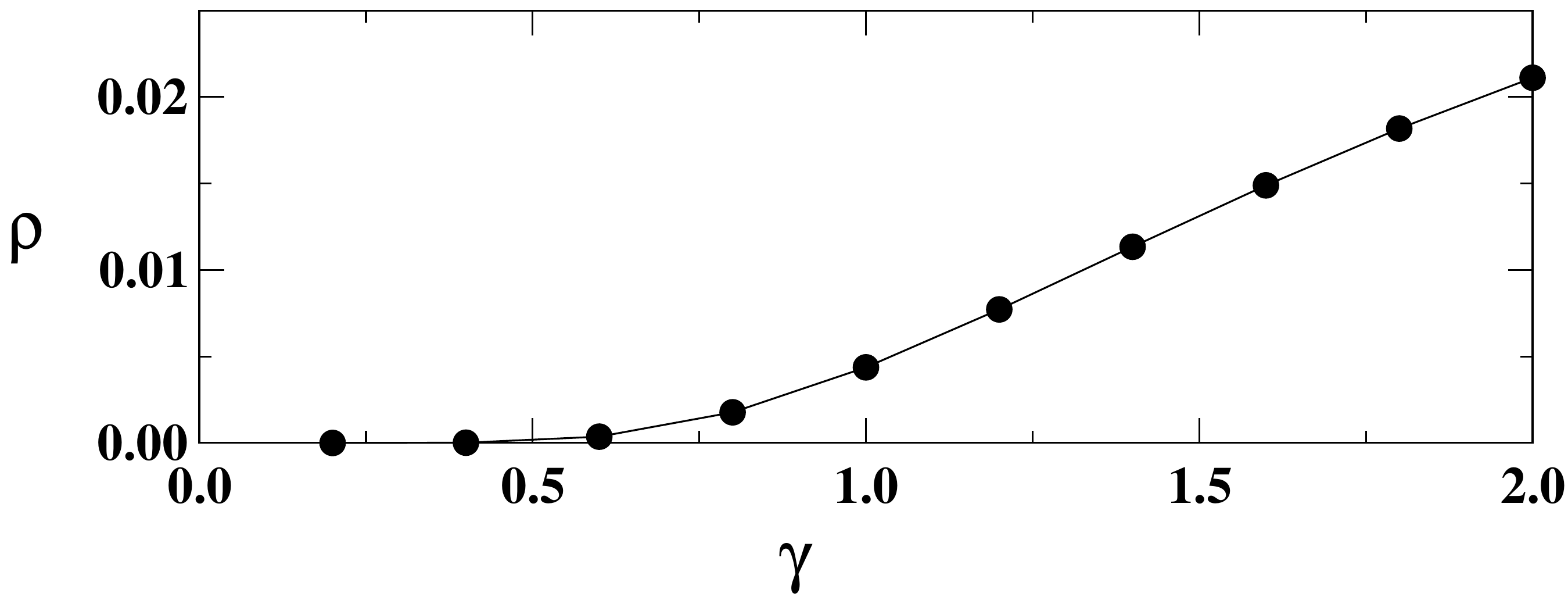}
\vspace{-0.15cm}
\caption{The cavity photon density at $\bq^\star=(\pi,\pi)/a$ as a function of $\gamma$. The numerical calculation is performed with $t=1$, $\hbar\omega_{\bq^\star}/t=1$ and $T/t=0.002$.}
\label{rho_photons}
\end{center}
\end{figure}

At half-filling ($n=1$ one-electron per unit cell) the Fermi surface line has energy $\ep_\bk=0$, represented as the white solid line in Fig. \ref{normal_GS_bands}.   Also shown in Fig. \ref{normal_GS_bands} is the vector $\bq^\star=(\pi,\pi)/a$ which spans the Fermi surface. We notice that the wave vector $(\bq^\star)^\prime=(\pi,-\pi)/a$ is not shown, but it spans the Fermi surface in the direction perpendicular to $\bq^\star$. The vector $\bq^\star\pm(\bq^\star)^\prime$ equals to a reciprocal lattice vector, so that $\bq^\star$ and $(\bq^\star)^\prime$ are regarded as equivalent.
For $\bq=\bq^\star$, the longitudinal and transverse dipole matrix elements are: 
\be
\label{dipole_square_long_transv}
g^L_{\bk,\bq^\star}=-i\frac{\sqrt{2}t\,a}{\pi}\left[\cos(k_x\,a)+\cos(k_y\,a)\right],\quad
g^T_{\bk,\bq^\star}=-i\frac{\sqrt{2}t\,a}{\pi}\left[\cos(k_x\,a)-\cos(k_y\,a)\right].
\ee
It is easy to realise that for $\bk$ and $\bk+\bq^\star$ approaching the Fermi surface the longitudinal matrix element vanishes while the transverse one is finite apart from four specific points $\pm(\pi/2a,\pi/2a)$ and $\pm(\pi/2a,-\pi/2a)$, that corresponds to the Fermi points in the superradiant phase.
The remaining part of the section is devoted to the evaluation of the transverse and longitudinal components of the current-current response function. By inserting the first expression of \eqn{dipole_square_long_transv} in Eq. \eqn{paramagnetic_response_A_3} we find: 
\be
\label{paramagnetic_response_square_long}
K_L(0,\bq^\star)=-\frac{t\,a^2}{2\pi^4}\int_{MBZ}\,d^2\bk\,\left[\cos(k_x\,a)+\cos(k_y\,a)\right]=-\frac{8t}{\pi^4},
\ee
that exactly cancels the diamagnetic contribution $\mathcal{T}_L(\bq^\star,-\bq^\star)=-8t/\pi^4$, as expected from the $f$ sum-rule \eqn{f_sum_rule}.  In the previous integral \eqn{paramagnetic_response_square_long} we have introduced the magnetic Brillouin zone (MBZ) that is defined as the region of points with $\epsilon_{\bk}\le0$, see Fig. \ref{normal_GS_bands}.
Concerning the transverse response we have: 
\be
\label{paramagnetic_response_square}
K_T(0,\bq^\star)=-\frac{t\,a^2}{2\pi^4}\int_{MBZ}\,d^2\bk\,\frac{\left[\cos(k_y\,a)-\cos(k_x\,a)\right]^2}{\cos(k_y\,a)+\cos(k_x\,a)}=-\frac{8t}{\pi^4}\log^2\left(\frac{2}{\epsilon}\right),
\ee 
where $\epsilon$ in Eq. \eqn{paramagnetic_response_square} is an infrared cutoff introduced to regularise the integral. Thus, the current-current response to the transverse cavity field with $\bq^\star=(\pi,\pi)/a$ is: 
\be
Q_T(0,\bq^\star)=-8t\,\left[\log^2\left(2/\epsilon\right)-1\right]/\pi^4.
\ee
In Fig. 1(a) of the main text we show the behavior of $Q_T(0,\bq^\star)$ along the high-symmetry directions of the square lattice for various temperatures $T$. 
 At low temperature we start to observe a peak at $M=(\pi,\pi)/a=\bq^\star$, that eventually at $T/t=0$ becomes the logarithmic singularity \eqn{paramagnetic_response_square}.
From Eq. \eqn{paramagnetic_response_square} we can extract the critical temperature $T_c$ below which the superradiant phase sets in $T_c\sim\,t\,\exp\left(-\hbar\pi^2\,\sqrt{(\hbar\omega_\bq^\star/t\,\gamma^2)}/4a\right)$.  We observe that the same result can be obtained by minimising the light-matter energy \eqn{light_matter_E} constrained to the self-consistency condition \eqn{saddle_point}, see Fig. \ref{rho_photons} where the cavity photon density as a function of $\gamma$ is shown.

\subsection{The superradiant phase}
\label{subsec_5_1}
The analysis presented in the previous Section shows that, below a certain critical temperature, there is a macroscopic occupation of the transverse cavity mode with momentum $\bq^\star$, $\langle a_{\bq^\star}\rangle/V\neq0$. 
\begin{figure}
\begin{center}
\includegraphics[width=0.3\textwidth]{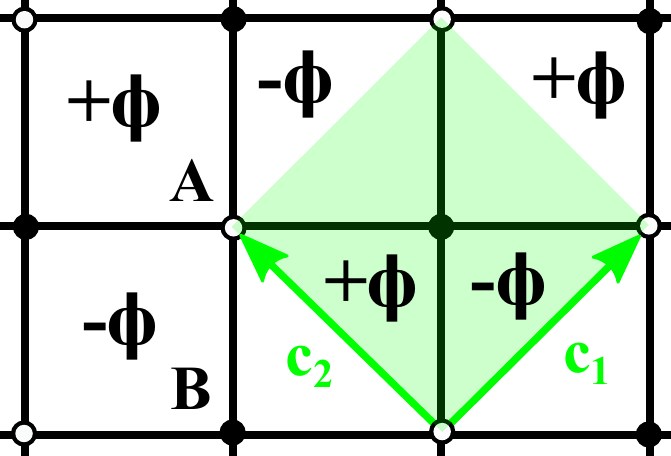}
\vspace{-0.15cm}
\caption{ The photon condensate gives rise to the staggered flux $\phi$, that breaks the translational invariance and the time reversal symmetry. Open and solid points mark the $A$ and $B$ sublattices, respectively. The unit cell, green shaded region, is generated by $\bm{c}_1$ and $\bm{c}_2$. The net flux per unit cell is zero.}
\label{mag_flux}
\end{center}
\end{figure}
\begin{figure}
\begin{center}
\includegraphics[width=0.6\textwidth]{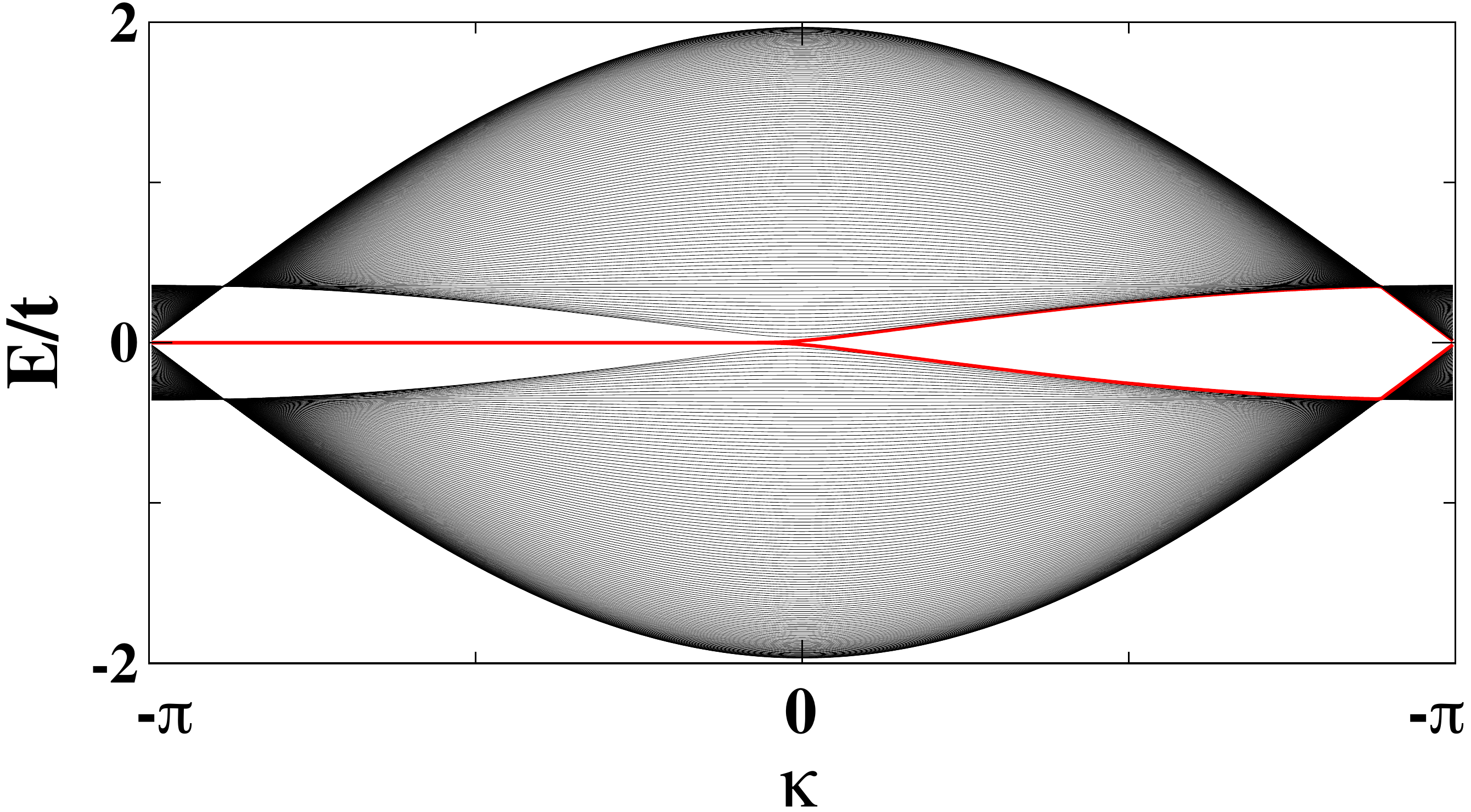}
\caption{Spectrum as a function of $\kappa=\bk\cdot\bm{c}_1$ obtained with ribbon boundary conditions. The black lines denotes the bulk bands, while solid red lines represent the zero-energy edge states. The numerical calculations are performed with $A_0=0.4$ and $t=1$.}
\label{edge_ribbon_square}
\end{center}
\end{figure}
In this regime the light-matter ground state spontaneously develops a cavity vector potential: 
\be
\label{vec_pot}
\langle \hat{\bm{A}}(\br)\rangle=-A_0\left[(\bm{y}-\bm{x})\sin\left(\pi(x+y)/a\right)+(\bm{y}+\bm{x})\,\sin\left(\pi(x-y)/a\right)\right]/\sqrt{2},
\ee
where $A_0\propto\sqrt{\rho_{ph}}$, $\rho_{ph}$ density of cavity photons, breaks the time-reversal symmetry (TRS). 
Viewed from the electrons, we obtain an effective tight-binding model with a staggered magnetic flux (Fig. \ref{mag_flux}), $\phi=\int_{\msquare}\,d\bm{S}\cdot\nabla\times\langle \hat{\bm{A}}(\br)\rangle=4\sqrt{2}a\,A_0/\pi$, and with zero total flux within the magnetic unit cell, that is the green shaded region in Fig. \ref{mag_flux} generated by the primitive vectors $\bm{c}_1=a(\bm{x}+\bm{y})$ and $\bm{c}_2=a(-\bm{x}+\bm{y})$.
Moreover, we observe that the model consists of two interpenetrating square lattices ($A$ and $B$ sublattices) as shown in Fig. \ref{mag_flux}. 
In the basis of two-component spinor $(c^\dagga_{\bk,A},\,c^\dagga_{\bk,B})$ the electronic Hamiltonian reads: 
\be
\label{electron_SX_graphene}
h(\bk,\bm{A})=-t\left[\cos\left(e\phi/4c\right)\left(\cos\,k_xa+\cos\,k_ya\right)\,\sigma^x+\sin\left(e\phi/4c\right)\left(\cos\,k_ya-\cos\,k_xa\right)\,\sigma^y\right]
\ee
where $\sigma^x$ and $\sigma^y$ are Pauli matrices. The bands of the Hamiltonian are given by $E_\pm(\bk)=\pm t\sqrt{\cos^2(k_x\,a)+\cos^2(k_x\,a)+2\cos^2(k_x\,a)\cos^2(k_x\,a)\cos(e\phi/2c)}$.
From the latter expression and from Fig. 1(b) of the manuscript we can clearly see that the flux opens a gap almost everywhere on the Fermi surface except at four points $\bm{K}_{1\pm}=\pm(\pi/2a,\pi/2a),\quad\bm{K}_{2\pm}=\pm(\pi/2a,-\pi/2a)$.
These exceptional points correspond to the $\bk$-points of the original Fermi surface where the transverse dipole matrix element $g^T_{\bk,\bq^\star}$ \eqn{dipole_square_long_transv} vanishes. 
In the vicinity of the Fermi point $\bm{K}_{1+}$, $\delta\bk=\bm{K}_{1+}-\bk$, the Hamiltonian \eqn{dipole_square_long_transv} takes the form: 
\be
\label{square_graphene_like}
h(\delta\bk,\bm{A})=-t\left[\cos(e\phi/4c)\left(\delta k_x+\delta k_y\right)\,\sigma^x+\sin(e\phi/4c)\left(\delta k_y-\delta k_x \right)\,\sigma^y\right],
\ee
that, for $e\phi/c=\pi$, is equivalent to the effective Hamiltonian close to the Dirac points of monolayer graphene, while for generic flux $h(\delta\bk)$ gives rise to deformed Dirac cones (we notice that the Hamiltonian in the vicinity of $\bm{K}_{1-}$ is obtained from \eqn{square_graphene_like} by the substitution $\delta\bk\to-\delta\bk$, while close to $\bm{K}_{2\pm}$ by sending in \eqn{square_graphene_like} $\delta k_x\to\pm\delta k_x$ and $\delta k_y\to\mp\delta k_y$). 
It is important to remark that the model breaks both TRS $T$, $h^*(\bk,\bm{A})\neq h(-\bk,\bm{A})$, and inversion symmetry $C_{2z}$, $\sigma^x\cdot h(\bk,\bm{A})\cdot\sigma^x\neq h(-\bk,\bm{A})$. Interestingly, the combinations of $T$ and $C_{2z}$, $C_{2z}\,T$, leaves the electronic Hamiltonian invariant, $\sigma^x\cdot h^*(\bk,\bm{A})\cdot\sigma^x=h(\bk,\bm{A})$. The latter symmetry imposes a vanishing Berry curvature and protects the Dirac points characterized by the Berry phases $\pm\pi$.

\subsection{Superradiant edge states and band structure for ribbon boundary conditions}
\label{subsec_5_2}
 In this section, we consider the electronic model \eqn{electron_SX_graphene} in the presence of ribbon boundary conditions, i.e. open boundary conditions along $\bm{c}_2$ while periodic along $\bm{c}_1$. By performing standard calculations, see \cite{Bernevig_book} for more details, we obtain the spectrum shown in Fig. \ref{edge_ribbon_square}. In addition to bulk states, black lines in Fig. \ref{edge_ribbon_square}, we observe the presence of zero-energy edge modes, red line in Fig. \ref{edge_ribbon_square}, as expected from the similarities of the superradiant model with graphene.   

\section{The superradiant instability in monolayer graphene}
\label{sec6} 
 
\begin{figure}
\begin{center}
\includegraphics[width=0.3\textwidth]{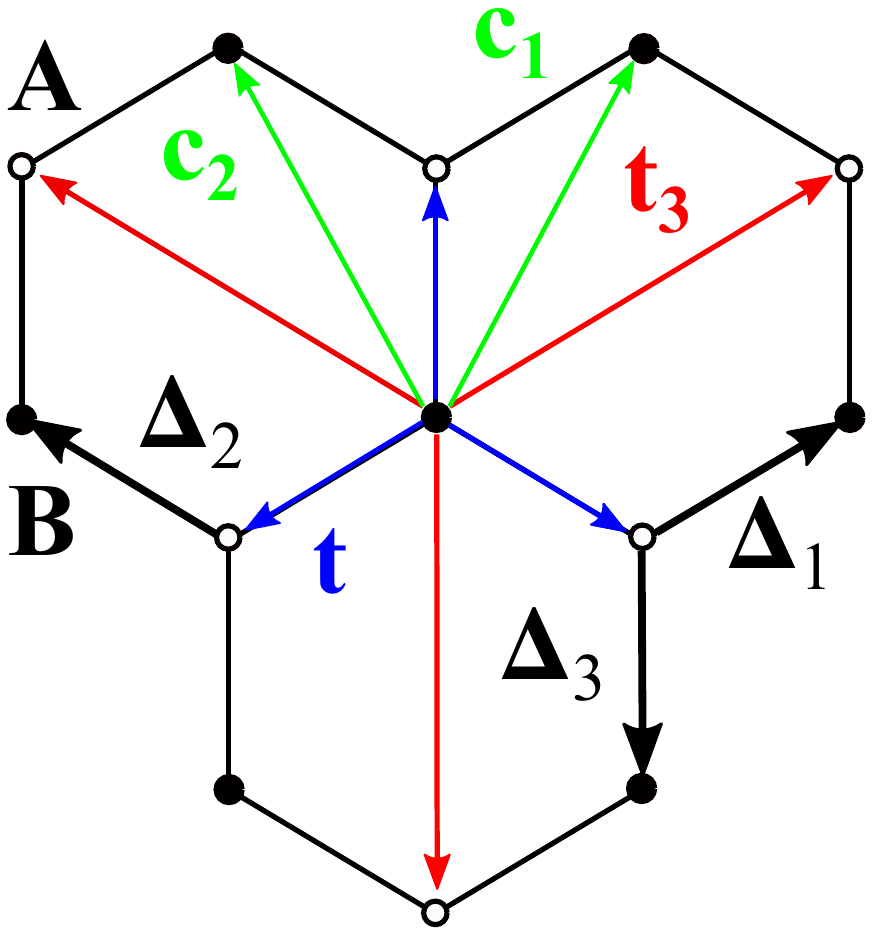}
\vspace{-0.15cm}
\caption{ Honeycomb lattice. Open and solid points mark the $A$ and $B$ sublattices, respectively.
The $\bm{\Delta}_j$ vectors connect one atom to its nearest-neighbors. Nearest- $(t)$ and third-neighbors-hoppings $(t_3)$ are denotes by blue and red arrows respectively. In green we report the primitive lattice vectors.}
\label{Lattice_graphene}
\end{center}
\end{figure}

\begin{figure}
\begin{center}
\includegraphics[width=0.75\textwidth]{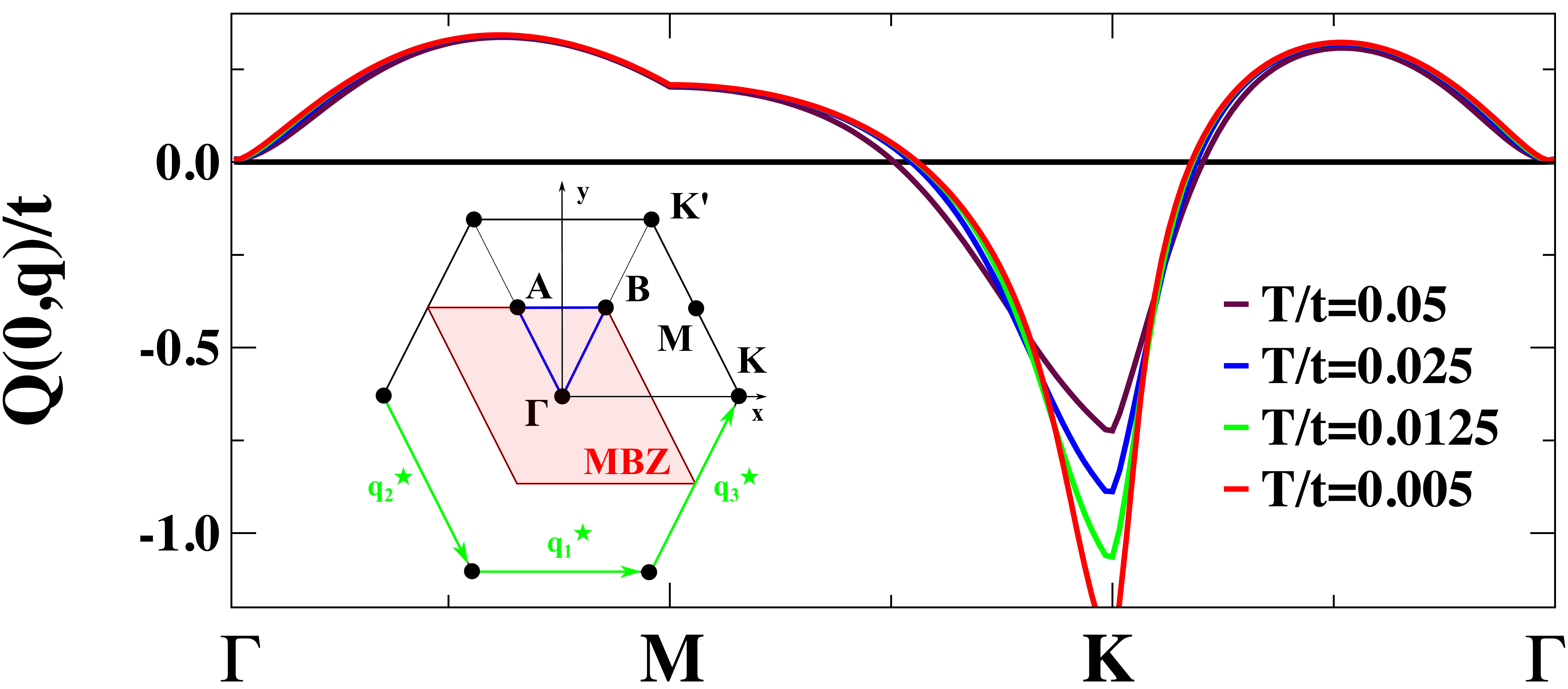}
\vspace{-0.15cm}
\caption{Longitudinal (black solid line) and transverse (maroon, blue, green and red lines) current-current correlation function as a function of the transferred momentum $\bq$. We perform calculations for different values of the temperature $T$. As you lower the temperature we start to observe a peak at $\bq=\bm{K}$ that eventually at $T/t=0$ becomes a logarithmic singularity. \textbf{Inset:} Brillouin zone of the Honeycomb lattice. 
We report the high-symmetry points $\Gamma$, $\bm{K}$, $\bm{K}^\prime$ and $\bm{M}$. The Fermi surface points are connected by the wave vector $\bq^\star_1$, $\bq^\star_2$ and $\bq^\star_3$ depicted as green arrows. In the presence of condensed cavity photons the Brillouin zone reduces to the magnetic brillouin zone (MBZ), red shaded region. In the MBZ we report the high-symmetry points $\bm{A}$ and $\bm{B}$.}
\label{QR_LT_graphene}
\end{center}
\end{figure}

 In this section we consider the superradiant instability in a tight-binding Hamiltonian characterised by quadratic band touching dispersion close to the Fermi points. 
 The lattice model consists of the monolayer honeycomb lattice with hopping between nearest-neighbors ($t$) and third-nearest-neighbors ($t_3$), see Fig. \ref{Lattice_graphene}. 
Following the manuscript we fix $t$ as the energy unit $t=1$ and we define the adimensional parameter $r=t_3/t$. In the basis of two-component spinor $(c_{\bk,A},c_{\bk,B})$, the model Hamiltonian reads: 
 \be
 \label{graphene_third}
 h(\bk)=-\left(\sum_{j=1}^{3}\,e^{i\bk\cdot\bm{\Delta}_{j}}+r\,\sum_{j=1}^{3}\,e^{-2i\bk\cdot\bm{\Delta}_{j}}\right)\,\sigma^+-\left(\sum_{j=1}^{3}\,e^{-i\bk\cdot\bm{\Delta}_{j}}+r\,\sum_{j=1}^{3}\,e^{2i\bk\cdot\bm{\Delta}_{j}}\right)\,\sigma^-,
 \ee
 where $\sigma^{\pm}=\sigma^x\pm\,i\,\sigma^y/2$, the vectors $\bm{\Delta}_j$ connect one atom to its nearest neighbors $\bm{\Delta}_{3}=a\,(0,-1),\quad\bm{\Delta}_{1(2)}=a\,\left(\pm\sqrt{3},1\right)/2$.
We observe that the third-nearest neighbor hopping term preserves the particle-hole symmetry so that at half-filling the Fermi energy is $\mu=0$. Once again, we apply Eq. \eqn{general_current_vertex} and we readily find 
\be
\bm{j}(\bk,\bq)=\bm{j}_{AB}(\bk,\bq)\,\sigma^++\bm{j}^*_{AB}(\bk,\bq)\,\sigma^-,
\ee
where $\bm{j}_{AB}(\bk,\bq)=-2i\sum_{j=1}^{3}\,\bm{\Delta}_j\,\left[e^{i\bk\cdot\bm{\Delta}_j}\,\sin\left(\bq\cdot\bm{\Delta}_j/2\right)-r\,e^{-2i\bk\cdot\bm{\Delta}_j}\,\sin\left(\bq\cdot\bm{\Delta}_j\right)\right]/(\bq\cdot\bm{\Delta}_j)$. 
On the other hand, Eq. \eqn{diamagnetic_papero} gives the following expression for the diamagnetic tensor:  
\be
\bm{\tau}^{i,j}(\bk,\bq_1,\bq_2)=\tau^{i,j}_{AB}(\bk,\bq_1,\bq_2)\,\sigma^++\left[\tau^{i,j}_{AB}(\bk,\bq_1,\bq_2)\right]^*\,\sigma^-,
\ee
where
\bal
\tau^{i,j}_{AB}(\bk,\bq_1,\bq_2)=&-4\sum_{a=1}^{3}\,\frac{\left(\bm{\Delta}_{a}\right)_i\,\left(\bm{\Delta}_{a}\right)_j}{(\bq_1\cdot\bm{\Delta}_a)(\bq_2\cdot\bm{\Delta}_a)}\Bigg[e^{i\bk\cdot\bm{\Delta}_a}\,\sin\left(\frac{\bq_1\cdot\bm{\Delta}_a}{2}\right)\,\sin\left(\frac{\bq_2\cdot\bm{\Delta}_a}{2}\right)\\
&+r\,\sin\left(\bq_1\cdot\bm{\Delta}_a\right)\sin\left(\bq_2\cdot\bm{\Delta}_a\right)e^{-2i\bk\cdot\bm{\Delta}_{a}}\Bigg].
\eal
The band structure of the Hamiltonian \eqn{graphene_third} evolves in a non-trivial way as a function of $r$, we refer to \cite{bena2011,montambaux2012} for a detailed review.
We are particularly interested to the collective response to cavity modes for the specific value $r=1/2$, where the model is characterized by quadratic band touching at $\bm{K}$ and $\bm{K}^\prime$ similarly to Bernal stacked bilayer graphene \cite{montambaux2012}. In particular, in the vicinity of $\bm{K}^\prime$, $\bq=\bk-\bm{K}^\prime$ and $|\bq|\ll1$, the Hamiltonian takes the form $h(\bq)=-9|\bq|^2\,\left(e^{2i\theta_{\bq}}\,\sigma^++e^{-2i\theta_{\bq}}\,\sigma^-\right)/8$, where $e^{i\theta_{\bq}}=(q_x+iq_y)/|\bq|$ and its eigenspectrum is given by: 
\be
\bm{v}^\pm_{\bm{K}^\prime+\bq}=\frac{1}{\sqrt{2}}\left(\begin{array}{c}1 \\\mp e^{-2i\theta_{\bq}}\end{array}\right),\quad\epsilon^\pm_{\bm{K}^\prime+\bq}=\pm\frac{9}{8}|\bq|^2.
\ee
The properties of $h(\bk)$ close to $\bm{K}$ can be deduced from $\bm{K}^\prime$ by TRS. 
At half-filling, the inequivalent Fermi points $\bm{K}$ and $\bm{K}^\prime$ points, see the inset in Fig. \ref{QR_LT_graphene}, are connected by the wave vectors $\bq^\star_1$, $\bq^\star_2$ and $\bq^\star_3$ that are depicted in the inset of Fig. \ref{QR_LT_graphene}.
These wave vectors are not independent, indeed $\bq^\star_1+\bq^\star_2=-\bm{g}_2$, and $\bq^\star_1+\bq^\star_3=\bm{g}_1$, where $\bm{g}_1=4\pi(\sqrt{3}/2,1/2)/3a$ and $\bm{g}_2=4\pi(-\sqrt{3}/2,1/2)/3a$ are reciprocal lattice vectors so that $\bq^\star_1$, $\bq^\star_2$ and $\bq^\star_3$ are regarded as equivalent. 
From now on, we will focus on the electronic response to a cavity field with wave vector $\bq^\star_1=4\pi\bm{x}/3\sqrt{3}a$. In particular we compute the dipole matrix element for an electron-hole excitation with relative wave vector $\bq^\star_1$ and center of mass momentum $\bm{K}+\bm{K}^\prime/2$, $|\bm{K},\bq^\star_1\rangle=\Gamma^\dagger_{\bm{K}+\bq^\star_1,+}\,\Gamma^\dagga_{\bm{K},-}\ket{0}$, where $\Gamma_{\bk,n}$ is the fermionic operator in the basis that diagonalises Hamiltonian \eqn{graphene_third}. By performing straightforward calculations we find: 
\be
\bm{j}\left(\bm{K}+\bm{K}^\prime/2,\bq^\star_1\right)=i\,\bm{y}\,a\,9\sqrt{3}\,\left[e^{-i2\pi/3}\sigma^+-e^{i2\pi/3}\sigma^-\right]/4\pi.
\ee 
Since the transferred momentum is along $\bm{x}$ the longitudinal component of the dipole matrix element is vanishing, $\left[g^{-,+}_{\bm{K},\bq^\star_1}\right]^L=0$, while the transverse one reads:
\bal
\label{dipole_g_T_graphene}
\left[g^{-,+}_{\bm{K},\bq^\star_1}\right]^T&=a\,t\,9\sqrt{3}\,\left[e^{-i2\pi/3}\,\left(\bm{v}^-_{\bm{K}+\delta\bk}\right)^*\cdot\sigma^+\cdot\bm{v}^+_{\bm{K}^\prime+\delta\bk}-e^{i2\pi/3}\,\left(\bm{v}^-_{\bm{K}+\delta\bk}\right)^*\cdot\sigma^-\cdot\bm{v}^+_{\bm{K}^\prime+\delta\bk}\right]/4\pi\\
&=a\,t\,9\sqrt{3}\,e^{-i2\theta_{\delta\bk}}/{8\pi},
\eal
where $\delta\bk$ is an infinitesimal displacement.
As a consequence of the parabolic form of the bands at $\bm{K}$ and $\bm{K}^\prime$, the finite dipole matrix element gives rise to a divergence in the zero-temperature current susceptibility $Q_T(0,\bq)$, as shown in Fig. \ref{QR_LT_graphene}. 
Therefore, the stable ground state is superradiant at arbitrary weak light-matter coupling. 
Finally, we observe that for $r\neq1/2$, the band touching is no more quadratic and the logarithmic singularity is softened. However, we still expect a large collective response to the cavity modes with wave vector $\bq^\star_1$  and a superradiant phase above a critical value of the light-matter interaction. The detailed analysis of the superradiant phase diagram as a function of $r$ is left to future investigations.

\subsection{The superradiant phase}
\label{subsec_6_1}

In the superradiant phase the average value of the vector potential is: 
\be
\label{graphene_vec_pot}
\langle \hat{\bm{A}}(\br)\rangle=-\bm{y}\,A_0\,\sin\left(4\pi\,x/3\sqrt{3}a+\varphi\right)
\ee
where $A_0$ is proportional to $\sqrt{\rho_{ph}}$ while $\varphi$ is the phase of the complex photon order parameter.
\begin{figure}
\begin{center}
\includegraphics[width=0.3\textwidth]{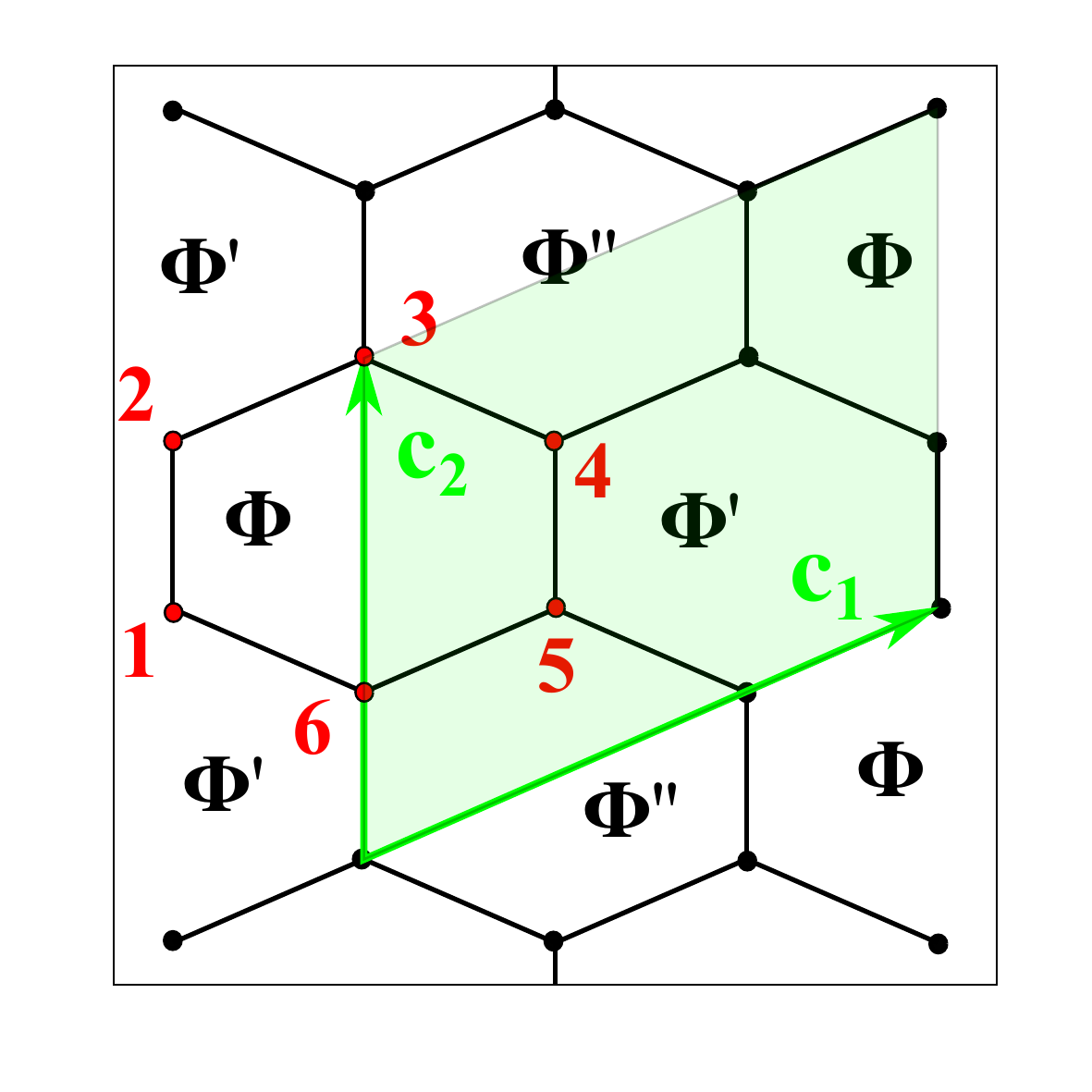}
\vspace{-0.15cm}
\caption{ Honeycomb lattice dressed by the cavity vector potential \eqn{graphene_vec_pot}. The red dots indicate the $6$ independent sites. The green arrows depict the primitive lattice vectors $\bm{c}_1$ and $\bm{c}_2$ that determine the unit cell.
The cavity vector potential gives rise to three independent fluxes $\Phi$, $\Phi^\prime$ and $\Phi^{\prime\prime}$.
The total flux per unit cell is zero, $\Phi+\Phi^\prime+\Phi^{\prime\prime}=0$.}
\label{Lattice_extended_graphene}
\end{center}
\end{figure}
In the presence of $\langle \hat{\bm{A}}(\br)\rangle$ the new unit cell is three times larger, green shaded region in Fig. \ref{Lattice_extended_graphene}, and correspondingly the Brillouin zone becomes one-third of the original one, red shaded region in the inset of Fig. \ref{QR_LT_graphene}. 
In the basis of the six-dimensional spinor $(c_{\bk,1},c_{\bk,2},c_{\bk,3},c_{\bk,4},c_{\bk,5},c_{\bk,6})$ the superradiant electronic Hamiltonian becomes $h(\bk,\bm{A})=h^0(\bk,\bm{A})+h^3(\bk,\bm{A})$ where the nearest-neighbor term reads: 
\be
h^0(\bk,\bm{A})=-\left(\begin{array}{cccccc}0 & e^{i\phi_{12}(\bm{A})} & 0 & e^{i\phi_{14}(\bm{A})-i\kappa_1} & 0 & e^{i\phi_{16}(\bm{A})} \\c.c. & 0 & e^{i\phi_{23}(\bm{A})} & 0 & e^{i\phi_{25}(\bm{A})+i\kappa_2-i\kappa_1} & 0 \\0 & c.c. & 0 & e^{i\phi_{34}(\bm{A})} & 0 & e^{i\phi_{36}(\bm{A})+i\kappa_2} \\c.c. & 0 & c.c. & 0 & e^{i\phi_{45}(\bm{A})} & 0 \\0 & c.c. & 0 & c.c. & 0 & e^{i\phi_{56}(\bm{A})} \\c.c. & 0 & c.c. & 0 & c.c. & 0\end{array}\right),
\ee
while the third nearest-neighbor contribution is: 
\be
h^3(\bk,\bm{A})=-\frac{1}{2}\left(\begin{array}{cccccc}0 & e^{i\phi^3_{12}(\bm{A})-i\kappa_2} & 0 & e^{i\phi^3_{14}(\bm{A})} & 0 & e^{i\phi^3_{16}(\bm{A})+i\kappa_2-i\kappa_1} \\c.c. & 0 & e^{i\phi^3_{23}(\bm{A})-i\kappa_1} & 0 & e^{i\phi^3_{25}(\bm{A})} & 0 \\0 & c.c. & 0 & e^{i\phi^3_{34}(\bm{A})+i\kappa_2-i\kappa_1} & 0 & e^{i\phi^3_{36}(\bm{A})} \\c.c. & 0 & c.c. & 0 & e^{i\phi^3_{45}(\bm{A})+i\kappa_2} & 0 \\0 & c.c. & 0 & c.c. & 0 & e^{i\phi^3_{56}(\bm{A})+i\kappa_1} \\c.c. & 0 & c.c. & 0 & c.c. & 0\end{array}\right).
\ee

In the previous expressions we have defined $\kappa_1$ and $\kappa_2$ projections of the wave vector $\bk$ along the primitive lattice vectors see Fig. \ref{Lattice_extended_graphene}, $\kappa_1=\bk\cdot\bm{c}_1$, $\kappa_2=\bk\cdot\bm{c}_2$, where
$\bm{c}_1=a(3\sqrt{3},\,3)/2$, $\bm{c}_2=a(0,\,3)$.
Moreover, we have introduced the phases $\phi_{ij}(\bm{A})$ and $\phi^3_{ij}(\bm{A})$ associated with nearest and third-nearest neighbors hopping processes. By performing straightforward calculations we find: 
\bal
&\phi_{12}(\bm{A})=-A_0a\,\sin(\varphi),\quad\phi_{36}(\bm{A})=-A_0a\,\sin\left(\varphi+2\pi/3\right),\quad\phi_{45}(\bm{A})=-A_0a\,\sin\left(\varphi+4\pi/3\right)\\
&\phi_{14}(\bm{A})=-\frac{3\,A_0a}{4\pi}\,\left[\cos(\varphi)+\sin\left(\frac{\pi}{6}-\varphi\right)\right],\quad\phi_{25}(\bm{A})=-\phi_{14}(\bm{A}),\\
&\phi_{16}(\bm{A})=\frac{3\,A_0a}{4\pi}\,\left[\cos(\varphi)+\sin\left(\frac{\pi}{6}+\varphi\right)\right],\quad\phi_{23}(\bm{A})=-\phi_{16}(\bm{A}),\\
&\phi_{34}(\bm{A})=-\frac{3\sqrt{3}\,A_0a}{4\pi}\sin(\varphi),\quad\phi_{56}(\bm{A})=\phi_{34}(\bm{A}).
\eal
Regarding $\phi^3_{ij}(\bm{A})$, we find that $\phi^3_{ij}(\bm{A})=\phi_{ij}(\bm{A})$ with the exception of the phases:  
\be
\phi^3_{12}(\bm{A})=2aA_0\,\sin(\varphi),\quad\phi^3_{36}(\bm{A})=2aA_0\,\sin(\varphi+\,2\pi/3),\quad\phi^3_{45}(\bm{A})=-2aA_0\,\sin(\varphi+\,4\pi/3).
\ee
We observe that the vector potential gives rise to a finite flux per hexagon, in particular we have three independent fluxes, see the pattern shown in Fig. \ref{Lattice_extended_graphene}. 
However, the total flux per unit cell is zero $\Phi(\varphi)+\Phi^\prime(\varphi)+\Phi^{\prime\prime}(\varphi)=0$. Despite its simplicity the model is characterised by non-trivial behavior as a function of the amplitude $A_0$ and the phase $\varphi$. In the next section we focus on the topological properties of the model in the regime of small $A_0$, leaving a detailed characterization of the topological properties as a function of $A_0$ and $\varphi$ to further studies. 

\subsection{Topological properties of the superradiant electronic insulator}
 
\begin{figure}
\begin{center}
\includegraphics[width=0.65\textwidth]{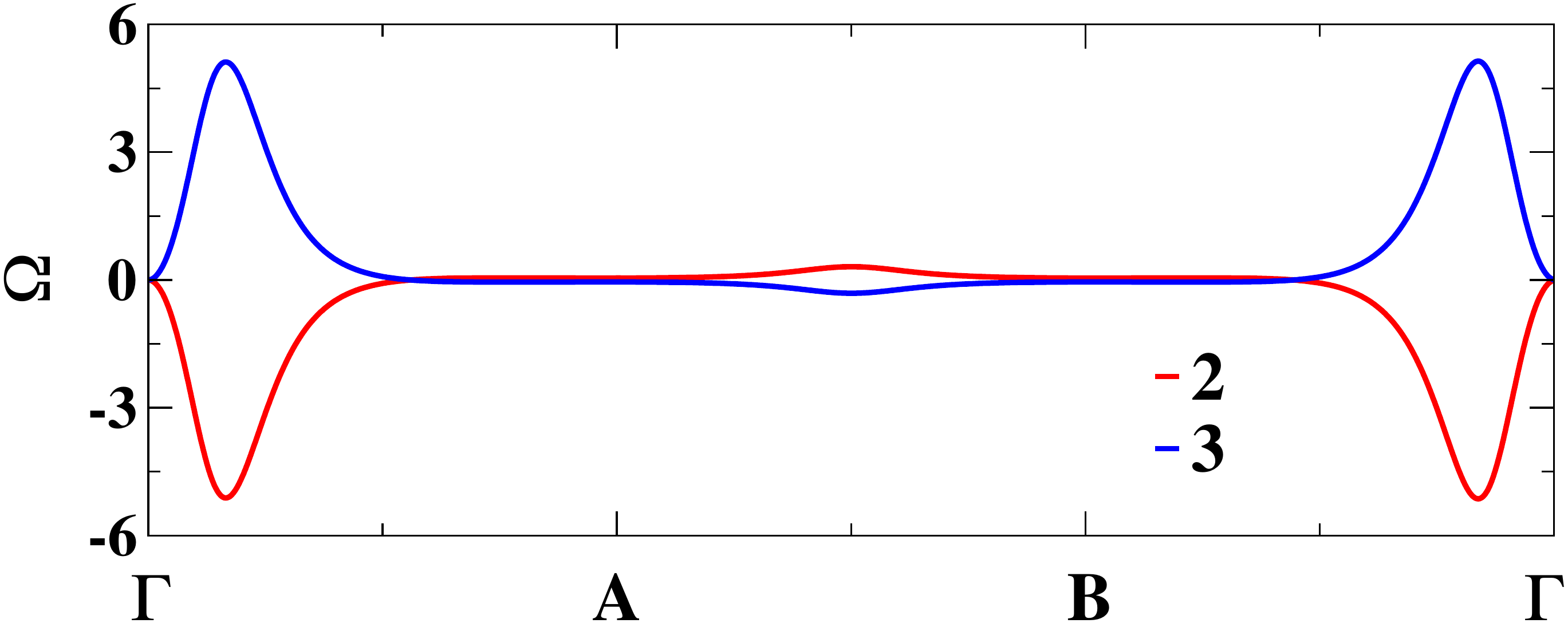}
\vspace{-0.15cm}
\caption{Red and blue lines show the Berry curvature along the high-symmetry directions of band $2$ and $3$, respectively. The calculation is performed with $A_0=0.3$, $r=0.5$.}
\label{Berry_curvature}
\end{center}
\end{figure}

\begin{figure}
\begin{center}
\includegraphics[width=0.65\textwidth]{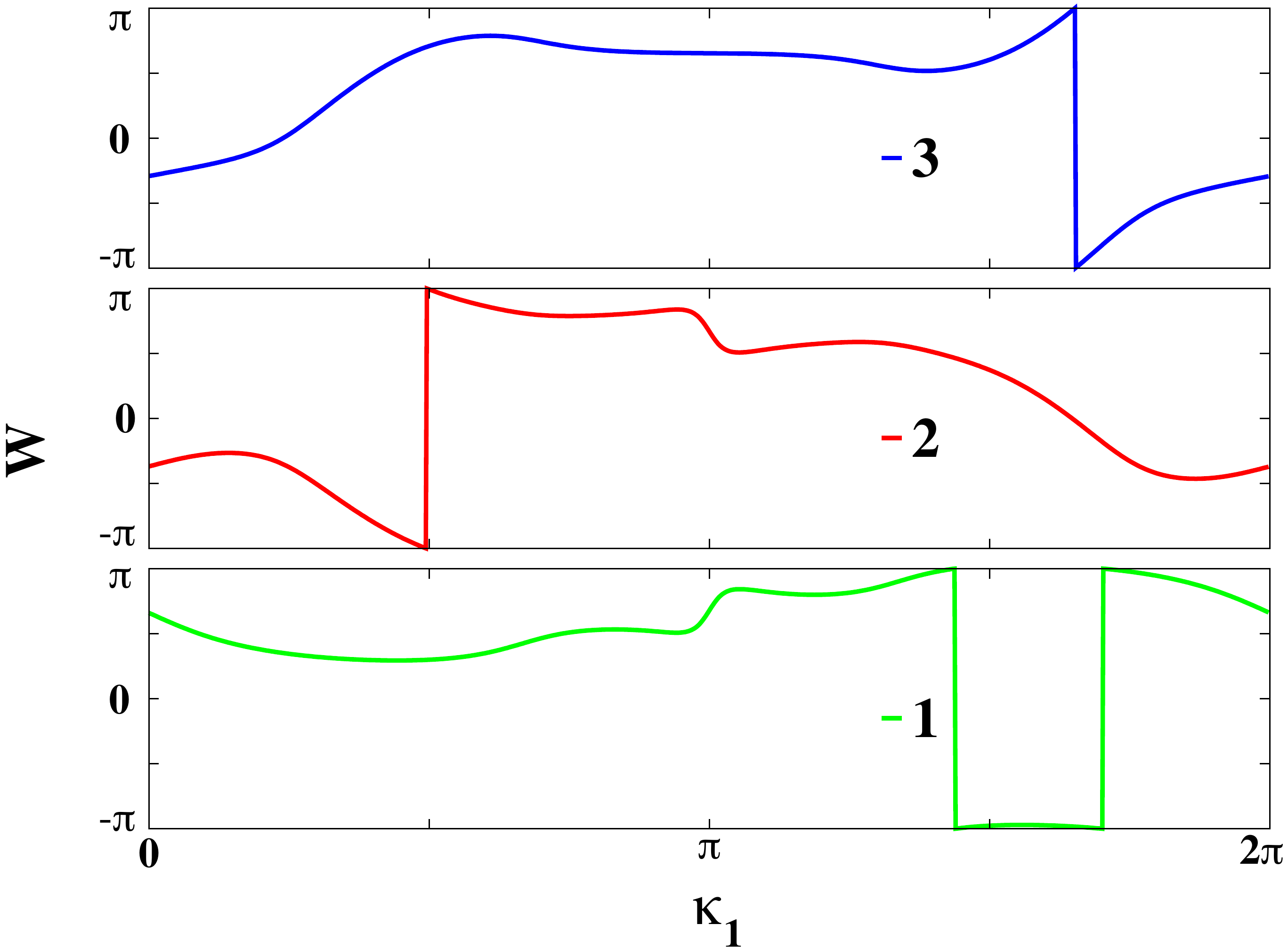}
\vspace{-0.15cm}
\caption{Wilson loop as a function of $\kappa_1=\bk\cdot\bm{c}_1$ for the bands with $E<0$. 
From the winding of $\mathcal{W}(\kappa_1)$ we deduce that  the Chern number are $C_{1}=0$, $C_{2}=-1$ and $C_{3}=+1$. The calculation is performed with $A_0=0.3$, $r=0.5$.}
\label{WLoop}
\end{center}
\end{figure}

The dipole matrix element \eqn{dipole_g_T_graphene}, that couples the zero-energy points $\bm{K}$ and $\bm{K}^\prime$, opens a gap at the Fermi level and gives rise to an electronic insulating state, see the band structure in Fig. 2(a) of the manuscript. The TRS breaking results in topological bands with non-zero Chern numbers and therefore in a topological superradiant phase. The model exhibits a chiral symmetry, $\gamma=\text{diag}(-1,1,-1,1,-1,1)$, anti-commuting with $h(\bk,\bm{A})$ that imposes bands of opposite energy to have the same Chern number.
The Berry curvature $\Omega_n(\bk)$ of the topological bands is shown in Fig. \ref{Berry_curvature} along the high-symmetry directions. Finally, we report the Wilson loops of the bands with negative energy in Fig. \ref{WLoop} (due to the symmetry $\gamma$, the Wilson loops of the bands with opposite energy are the same).

\subsection{Superradiant edge states and band structure for ribbon boundary conditions}
For ribbon boundary conditions, i.e. periodic along $\bm{c}_2$ while open in the $\bm{c}_1$ direction (Armchair BC see \cite{Bernevig_book} for more details), the spectrum (Fig. 2(b) of the main text) presents topological edge states connecting the bands $2$ and $3$.  
In addition, we observe non-topological edge states close to the Fermi energy.

\bibliographystyle{apsrev}
\bibliography{SMbiblio}